\documentclass[11pt]{article}
\usepackage{csquotes}
\usepackage{amssymb}
\usepackage{amsmath}
\usepackage{amsfonts}
\usepackage{amsthm}
\usepackage{float}
\usepackage{rotating}
\usepackage{subcaption}
\usepackage{caption}
\usepackage{geometry}
\usepackage[onehalfspacing]{setspace}
\usepackage{graphicx}
\usepackage{tabularx}
\usepackage{xcolor}
\usepackage{booktabs}
\usepackage{color}
\usepackage{threeparttable}
\usepackage{enumitem}
\usepackage{multicol}
\usepackage{multirow}
\usepackage{makecell}
\usepackage{natbib}
\usepackage{hyperref}
\usepackage{xargs}
\usepackage[colorinlistoftodos,prependcaption,textsize=tiny]{todonotes}

\hypersetup{hypertex=true,
                    colorlinks=true,
                    linkcolor=blue,
                    citecolor=blue,
                    filecolor=blue,
                    urlcolor=blue,
                    anchorcolor=blue}
\newcommandx{\vnote}[2][1=]{\todo[linecolor=red,backgroundcolor=white,bordercolor=red,#1]{#2}}
\newtheorem{definition}{Definition}[section]
\newtheorem{model}{Model}[section]
\newtheorem{theorem}{Theorem}[section]
\newtheorem{lemma}{Lemma}[section]
\newtheorem{corollary}{Corollary}[section]
\newtheorem{remark}{Remark}[section]
\newtheorem{example}{Example}[section]
\newtheorem{assumption}{Assumption}[section]

\allowdisplaybreaks

\usepackage{xr}
\begin{document}
\title{Unlocking the Regression Space}

\author{Liudas Giraitis$^1$, George Kapetanios$^2$, Yufei Li$^2$, Alexia Ventouri$^2$ \\
{\footnotesize {$^1$Queen Mary University of London} }\\
{\footnotesize {$^2$King's College London} }}
\maketitle

\begin{singlespacing}
\begin{abstract}
This paper introduces and analyzes a framework that accommodates general heterogeneity in regression modeling.
It demonstrates that regression models with fixed or time-varying parameters can be estimated using the OLS and time-varying OLS methods, respectively, across a broad class of regressors and noise processes not covered by existing theory.
The proposed setting facilitates the development of asymptotic theory and the estimation of robust standard errors.
The robust confidence interval estimators accommodate substantial heterogeneity in both regressors and noise.
The resulting robust standard error estimates coincide with White’s (1980) heteroskedasticity-consistent estimator but are applicable to a broader range of conditions, including models with missing data. They are computationally simple and perform well in Monte Carlo simulations. Their robustness, generality, and ease of implementation make them highly suitable for empirical applications. Finally, the paper provides a brief empirical illustration.
\vspace{2mm}


\vspace{1.5mm}
\end{abstract}
\end{singlespacing}

\vfill

\thispagestyle{empty} \pagebreak

\section{Introduction}
Regression analysis is the cornerstone of statistical theory and practice. Ordinary least squares (OLS) has been applied, within various regression contexts, to build an extensive toolkit, for the exploration of economic and financial datasets. The basic theory underlying OLS estimation and inference in regression models has been largely established  for over half of  a century (see e.g. \cite{lai1982}). The  problem  of   robust  estimation  has  long been a focus of  empirical work in economics,
beginning  with the seminal  work  by  \cite{white1980}. Its importance is  well understood in applied  econometrics.
 At the  same time, several important  concerns have been raised   by applied researchers. \cite{angrist2010}  noted   that
``\cite{leamer1983} diagnosed his contemporaries’ empirical work as suffering from a distressing lack of robustness to changes in key assumptions'', and \cite{leamer2010} later reflected that ``sooner or later, someone articulates the concerns that gnaw away in each of us and asks if the Assumptions are valid.''
Similarly, \cite{karmakar2022} observed, that the  assumption of parameter  constancy, or ``stationarity is often an oversimplified assumption that ignores systematic deviations of parameters from constancy". Clearly, this concern extends beyond parameter stability to encompass the stability of regressors, regression noise, and the underlying modelling assumptions.

In this paper, we focus on the inherent capacity of   regression modelling to accommodate the effects of structural change in settings with both fixed and time-varying parameters. Many such structural changes influence not only the model parameters but also the regression space itself. This space comprises both the regressors and regression noise, and improper treatment of  these  components may result in incorrect inferences, misinterpretations, and forecasting distortions. We therefore examine which specifications of regression space  can flexibly account for structural change while still enabling  estimation of both fixed and time-varying regression parameters, the construction of confidence intervals, and the computation of standard errors.

Among  recent  developments,  \cite{wu2007}, \cite{boldea2012}, and others, have proposed  advanced theoretical methods  for the estimation of the fixed  parameters, while
\cite{jansson2018}, \cite{jochmans2019} developed procedures to estimate both  fixed parameters and  standard errors in  regression models  with an increasing number of covariates  and  heteroscedasticity. Meanwhile, \cite{degui2020},
\cite{sun_hong2021} and \cite{linton2019}   introduced  new  modelling frameworks that explicitly account for structural change. A  common  response  to  concerns about  heteroskedasticity in the recent  literature is the   use of heteroscedasticity-robust variance and standard error estimators  for linear regression models, see \cite{eicker1963}, \cite{white1980},  \cite{MacKinnon(2012)} and  \cite{jansson2018},  among others.

There is  also a sizeable and growing literature on the estimation of time-varying coefficient regression models, including works of \cite{fan1999},  \cite{vogt2012},  among others. This literature further explores tests for different types of parameter variation, see  e.g. \cite{bai1998}, \cite{zhangwu2012},  \cite{zhangwu2015}, \cite{HKW2024}. In addition, specification tests and tests for parameter instability have received significant attention, with important
contributions by  \cite{hansen2000},   \cite{georgiev2018}, \cite{hidalgo2019}, \cite{boldea2019}, \cite{hong2023}, and others.

The modelling of deterministic, smooth parameter evolution
has a long history in statistics. Early examples include linear processes with time-varying spectral densities, introduced by \cite{pri1965}. This framework is  essentially nonparameteric and it has been further developed  by \cite{rob1989}, \cite{rob1991}, \cite{dah1997}, \cite{dahlhaus2019}, \cite{dahlhaus2023}, some  of  whom refer to these processes as
{\it locally stationary}. The estimation of time-varying   parameters, as well as fixed  parameters under heteroskedasticity in time  series models, has been studied  in \cite{dahlhausG1998}, \cite{xu_phillips2008}, \cite{dgr2020}, among other. Nonlinear time-varying  time series models  have also  been   developed  by  \cite{doukhan2008}, \cite{bardet2009},  \cite{vogt2012} and \cite{karmakar2022}.
Despite these advances, such approaches have not  been
not been widely adopted in applied economics,  where random coefficient models remain more prevalent.

Various methods have been proposed over the years to identify and handle structural change. Early contributions assumed  that changes were  deterministic, rare, and  abrupt. Testing for  parameter breaks dates back to the pioneering work by \cite{cho1960}, with further contributions by \cite{bro1974}, \cite{plo1992}, among others. More recent approaches
allow for random evolution of parameters, where changes may be discrete, as in Markov Switching models by \cite{ham1989} or threshold models by \cite{ton1990}, or continuous as in smooth transition models by \cite{ter1998}, or those driven by unobservable shocks, as in random coefficient models by \cite{nyb1989}. For example, \cite{cog2005} use random coefficient models to study stochastic volatility,
while \cite{pri2005} examines whether  changes in parameters or in the variance of shocks - policy-induced or otherwise -  contributed  to the period of macroeconomic calmness known as the ``Great Moderation" after 1985. In these frameworks, parameters typically evolve as random walks or autoregressive
processes.

Building on this literature, \cite{gky2014}, \cite{gky2018},
\cite{dgk2022}, and others have developed a theoretical time series  framework for random coefficient models and their estimation using kernel-based methods, which  performs well in finite samples. These methods are computationally simple and straightforward to implement in applied research. For  example,
\cite{cgk2022} demonstrated the empirical prevalence of persistent volatility, suggesting that GARCH-type volatility structures may be less common than previously thought.  Nevertheless, a full treatment of estimation and inference within a general regression framework has, surprisingly, not yet been provided.

In this paper, we provide a rigorous validation of the asymptotic normality of the feasible $t$-statistic for the estimation of both fixed and time-varying parameters in linear regression models within an extended regresion space of regressors and regression disturbances. Our main objective is to describe, in transparent terms, the extended regression space under which such normality is preserved.
The class of admissible regressors and regression noises is broad. Regressors are obtained by rescaling and shifting stationary short-memory sequences, while regression errors are generated by arbitrary rescaling of a stationary martingale difference sequence. The restrictions imposed on the scale factors and mean processes are weak, allowing these to be either deterministic or stochastic, and to vary over time, possibly abruptly or through non-stationary (e.g. unit-root) dynamics. Some assumptions on the scale factors are necessary and resemble the Lindeberg condition in the classical Lindeberg–Feller central limit theorem. Importantly, the robust feasible $t$-statistic retains the same form and limiting distribution as in the standard setting. The infeasible robust standard errors 
coincide with the heteroskedasticity-consistent standard error estimator of \cite{white1980}. Our assumptions do not rely on mixing or near-epoch dependence conditions, which prevail throughout the existing literature. Given the generality of the regression space, these assumptions typically require no additional empirical verification.

The estimation framework for fixed regression parameters is developed in Section~\ref{s:OLS}, which introduces the extended regression space, the underlying assumptions, and the main theoretical results. Section~\ref{s:FTV} establishes the estimation theory for time-varying regression parameters within the same framework. The proofs highlight how the results for the fixed-parameter case naturally extend to time-varying settings, with only negligible additional terms.

Our results are complementary to existing frameworks.
The novelty lies in providing a methodological foundation that confirms the validity of robust regression estimation in an extended regression space. The fundamental theory in this area traces back to \cite{lai1982}, who studied regression models with heteroskedastic martingale difference noise under eigenvalue-based assumptions. Alternative methods, such as bootstrap procedures, see \cite{boldea2012}; \cite{boldea2019}, are widely used in regression analysis but may not be directly applicable to such a general class of regressors and regression noises. In contrast, we demonstrate that White-type standard errors remain applicable and computationally straightforward.

All theoretical results are supported by detailed, rigorous proofs. Monte Carlo simulations confirm that the proposed robust regression estimators perform well in finite samples. Overall, the framework developed in this paper is particularly suited to modelling economic and financial data, where heterogeneity, structural change, and dependence are inherent features.

The remainder of this paper is organised as follows. Section \ref{s:OLS} presents the regression setting with the  extended regression space, accommodating  heterogeneity and  dependence,
and  outlines  the theoretical  results for infeasible and feasible $t$-statistics in the case of fixed parameters.
Section \ref{s:FTV} extends the analysis to the time-varying regression parameters.  Section \ref{s:missing}  addresses  regression modelling with missing  data patterns.
Section \ref{sss:ARp}  illustrates the flexibility of our robust estimation method by its application to the  estimation of  an AR(p) model generated  by a  stationary martingale  difference noise. Sections~\ref{MC} presents Monte Carlo simulation results. In Section \ref{Ch2:empirical}, we provide  an empirical application of  the robust regression framework
to modelling asset returns. Finally, Section \ref{s:concl}  concludes. Proofs and additional simulations are provided in the Supplemental Material.

\section{OLS estimation in general regression space}\label{s:OLS}

In this section, we focus on ordinary least squares (OLS) estimation in an environment that permits general heterogeneity in regression modelling. We analyze the model
\begin{eqnarray}  \label{e:r1}
y_t=\beta^\prime z_t +u_t, 
\end{eqnarray}
where $\beta$ is a $p$-dimensional parameter vector, $z_t=(z_{1t}, ...., z_{pt})^\prime$ is a stochastic regressor and $u_t$ is an uncorrelated noise term. To include an intercept, the first component can be set as $z_{1t}=1$. We refer to  the  collection of $\{z_t, u_t\}$  jointly as  ``the regression space".

An applied  researcher may want to  work  within a regression space that accommodates a  wide range of regressors and regression noises, without being hindered  by restrictive technical assumptions. Ideally, such a setting should permit
regressors exhibiting non-stationarity and  undefined generic structural change, while enabling estimation and inference under weak theoretical constraints that do not require empirical verification.

Our goal is to extend the OLS estimation procedure to  a broad regression framework defined by baseline assumptions aligned  with empirical research practice. These assumptions cover a wide variety of  types of  potentially non-stationary regression variables encountered in applied work. The framework achieves a level of generality comparable to that in \cite{glp2024}, which addresses   testing for absence of correlation and cross-correlation under general heterogeneity.

We begin with specifying the structure of an uncorrelated regression noise $u_t$.  Suppose that
\begin{equation}
u_t =h_t\varepsilon_t,  \label{e:r2}
\end{equation}
where $\{\varepsilon_t\}$ is a zero mean stationary uncorrelated martingale difference noise, and $\{h_t\}$ is a deterministic or stochastic scale factor independent of $\{\varepsilon_t\}$. 
The following assumption formalizes these conditions.
\begin{assumption}
\label{a:r0}  $\{\varepsilon_t\}$ is a stationary martingale difference (m.d.)
noise with respect to some $\sigma$-field filtration $\mathcal{F}_t
$, such that
\begin{equation*}
\mathbb{E}[\varepsilon_t|\mathcal{F}_{t-1}]=0,\quad \mathbb{E}
\varepsilon_t^8<\infty, \quad \mathbb{E}\varepsilon_t^2=1.
\end{equation*}
The sequence $\{\varepsilon_t\}$ is independent of $\{h_t\}$. Moreover, variable $\varepsilon_1$ has a probability  density function $f(x)$ satisfying $f(x)\le
c<\infty$ for  all $|x|\le x_0$, for some $x_0>0$.
\end{assumption}

\noindent The information set $\mathcal{F}_t$ is generated by the past history $\mathcal{F}_t=\sigma(\varepsilon_s, \, s\le t)$ and possibly other variables.

A typical example of an m.d. noise in applied work is provided by the ARCH/GARCH family and the  class of stochastic volatility processes. The specification (\ref{e:r2}) therefore allows for
conditional heteroskedasticity in $u_t$.

We next specify the regressors $z_t=(z_{1t}, ..., z_{pt})^\prime$ which form the key structural component of our regression space. For $k=1, ..., p$,  the regressors  can  be  written as
\begin{eqnarray}  \label{e:rz2}
z_{kt}=\mu_{kt}+g_{kt}\eta_{kt},\quad t=1, ..., n,
\end{eqnarray}
where $\eta_t=(\eta_{1t}, ..., \eta_{pt})^\prime$ is a stationary sequence, $g_t=(g_{1t},...., g_{pt})^\prime$ are deterministic or stochastic scale
factors, and $\mu_t=(\mu_{1t}, ..., \mu_{pt})^\prime$ is a vector of deterministic or stochastic means. We assume that $\{\mu_t, g_t,h_t\}$ are independent of $\{\varepsilon_t, \,\eta_t\}$. To include an intercept in model (\ref{e:r1}), we set
\begin{eqnarray}  \label{e:inter}
z_{1t}\equiv 1= \mu_{1t}+g_{1t}\eta_{1t}, \quad \mu_{1t}=0, \quad
g_{1t}=\eta_{1t}=1.
\end{eqnarray}
We further suppose that in (\ref{e:rz2}) $E\eta_{kt}=0$ except for the intercept (\ref{e:inter}), where $\eta_{1t}=1$.

In summary, the admissible regressors $\{z_t\}$ in our  setting  are  obtained  by  shifting and  rescaling a  short-memory stationary  process  $\{\eta_t\}$
by the  mean process  $\mu_t$  and the scale  factor $g_t$:
$$
z_t=\mu_t+I_{gt}\eta_t,  \quad I_{gt}= {\rm diag}(g_{1t}, ..., g_{pt})^\prime.
$$

The underlying stationary sequence $\{\eta_t\}$ is the fundamental component structuring  regressors $z_t$. Estimation of the  regression parameter $\beta$
requires only mild assumptions  on $\{\mu_t,g_t\}$, and short-memory dependence  assumption on  $\eta_t$, satisfied by ARMA and related stationary time series models. This framework eliminates the need for additional empirical validation.

\begin{definition}
\label{d:SM} \textrm{A (univariate) covariance stationary
sequence $\{\xi_t\}$ has short memory (SM) if $\sum_{h=-\infty}^\infty|
{\rm cov}(\xi_h,\xi_0)|<\infty. $ }
\end{definition}

\begin{assumption}
\label{a:ETA}  $\eta_t=(\eta_{1t}, ..., \eta_{pt})^\prime$ is an $\mathcal{F}_{t-1}$ measurable sequence with $E[\eta_{kt}^2]=1$ and $E[\eta_{kt}^8]<\infty$.

\vskip.1cm \noindent (i) For $k,j=1, ..., p$, the sequences $\{\eta_{kt}\}$ and $
\{\eta_{jt}\eta_{kt}\}$ are covariance stationary  and have short memory (SM).

\vskip.1cm \noindent (ii) The matrix $E[\eta_1\eta_1^\prime]$ is  positive definite.
\end{assumption}

The novelty of this regression framework lies in the structural specification (\ref{e:rz2}), which accommodates  regressors $z_t=(z_{1t}, ...., z_{pt})^\prime$
that may be  deterministic or stochastic, and  stationary or non-stationary.
This flexibility arises from allowing a broad class of scale factors and mean processes $\{h_t, g_t, \mu_t\}$ which  brings the OLS  estimation closer to empirical practice.

The estimation framework also accommodates   triangular arrays of  means and scale
factors: $\big(\mu_t, g_{t}, h_{t},\,\,\,$ $t=1,..., n\big)=\big(\mu_{nt},
g_{nt}, h_{nt}, \,\,\,t=1,..., n\big)$. Throughout the paper, we assume that these quantities may depend on the   sample size $n$. For   brevity of notation, the subscript $n$ is  omitted.

The underlying stationary noise component $\eta_t$ in the regressors $z_t$ in (\ref{e:rz2}) is weakly exogenous with respect to the stationary m.d. noise  $\varepsilon_t$ in  $u_t = h_t \varepsilon_t$. The mean  and scale factors $\{\mu_t,g_t\}$ are independent of  $\{\varepsilon_t\}$,  though they may be dependent on  $\{h_t\}$. Overall,  $\{\mu_t, g_t, h_t\}$ are mutually independent of $\{\eta_t, \varepsilon_t\}$, while  potential dependence among  $\{\mu_t\}$, $\{g_t\}$ and $\{h_t\}$ is unrestricted.

The processes $\mu_{kt}$  and $g_{kt}$  can  be interpreted  as  conditional  mean  and   variance, $\mu_{kt}=E[z_{kt}\, |\mathcal{F}_n^*]$, and $g_{kt}^2=\mathrm{var}(z_{kt}|\mathcal{F}_n^*) $ of $z_{kt}$, where  $\mathcal{F}_n^*=\sigma\big(\mu_t, g_t,h_t, t=1, ..., n\big)$  denotes the information set generated by the means and scale factors.

\noindent Denote for $k=1, ..., p$,
\begin{eqnarray}  \label{e:Dk}
v_k^2&=&\sum_{t=1}^n g_{kt}^2h_t^2,\quad v_{gk}^2=\sum_{t=1}^n g_{kt}^2,\\
D&=&\mathrm{diag}(v_1,..., v_p), \quad D_g=\mathrm{diag}(v_{g1},..., v_{gp}).\nonumber
\end{eqnarray}
We write $a_n\asymp_p b_n$ if $a_n=O_p(b_n)$ and $b_n=O_p(a_n)$.  

\begin{assumption}
\label{a:r3} The scale factors $h_t\ge 0$ and $g_t\ge 0$ are  deterministic
or stochastic non-negative variables such that, for $k=1, ..., p$,
\begin{eqnarray}  \label{e:hkj}
&&\frac{\max_{1\le t \le n}g_{kt}^2}{v_{gk}^2}= o_p(1),\quad \frac{\max_{1\le t \le n}\mu_{kt}^2}{v_{gk}^2}= o_p(1), \\
&&\frac{\sum_{t=1}^n \mu_{kt}^2}{v_{gk}^2}= O_p(1),\quad \frac{\sum_{t=1}^n
\mu_{kt}^2h_t^2}{v_{k}^2}= O_p(1), \quad v_k^2\asymp_pv^2_{gk}, \quad v_k \rightarrow_p\infty.
\label{e:hkj3}
\end{eqnarray}
\end{assumption}

\noindent 
Assumptions (\ref{e:hkj})–(\ref{e:hkj3}) impose only mild restrictions on the means $\mu_t$ and scale factors $g_t$.
In particular, condition (\ref{e:hkj}) resembles the Lindeberg condition in the classical Lindeberg–Feller central limit theorem, as it excludes the possibility that the OLS estimation is dominated by a single extreme observation of $z_t$.

The  first  restriction on $g_{kt}$ in (\ref{e:hkj}) is  necessary. For  example,  consider the  regressor $z_t=g_{t}\eta_{t}, t=1, ..., n$, with  scale  factors
 $g_{1}=1$ and  $g_{2}=g_{3}=...=g_{n}=0$, so  that $z_{2}=z_{3}=...=z_{n}=0$.
In this case, the OLS estimator of $\beta$  is inconsistent, and such a scale factor $g_t$  does not satisfy  (\ref{e:hkj}).

The second  condition (\ref{e:hkj3}) ensures that OLS estimation is driven by the stochastic component  $g_t\eta_t$  of the regressor $z_t$, rather than by deterministic or stochastic drift in $\mu_t$.

In the presence of  an  intercept,  condition (\ref{e:hkj3}) further implies that $\sum_{t=1}^n h_t^2\asymp_pn$, since $v_1^2\asymp_pv^2_{g1}$, $g_{1t}=1$,  $v_{g1}^2=n$, and $v_{1}^2=\sum_{t=1}^n h_t^2$.

\noindent To estimate $\beta=(\beta_1, ..., \beta_p)^\prime$,
we use the standard OLS estimator
\begin{eqnarray}  \label{e:OLSr}
\widehat \beta=\big( \sum_{j=1}^n z_j z_j^\prime\big)^{-1}\big( \sum_{j=1}^n
z_j y_j\big)
\end{eqnarray}
computed  from the  sample $y_j, z_j, \,j=1, ..., n$.

\vskip.2cm
\noindent\textbf{Consistency}. We first establish the consistency of the OLS
estimator $\widehat \beta$. 
\begin{theorem}
\label{t:r1} Suppose that $(y_1, ..., y_n)$ is a sample from the regression model
(\ref{e:r1})  and Assumptions \ref{a:r0}, \ref{a:ETA}
and \ref{a:r3} are satisfied. Then, the OLS estimator $\widehat\beta$ is consistent, i.e.
\begin{eqnarray}  \label{e:hatb1}
D( \widehat \beta -\beta) = \big(v_1(\widehat \beta_1
-\beta_1),...,v_p(\widehat \beta_p -\beta_p)\big)^\prime=O_p(1).
\end{eqnarray}
\end{theorem}

\vskip.2cm \noindent
\noindent
This result implies that the $k$-th component $\widehat \beta_k$ of the OLS estimator is $v_k$-consistent, that is,
$\widehat \beta_k -\beta_k = O_p(v_k^{-1})$.
The convergence rate $v_k$ may deviate from the conventional $\sqrt{n}$ rate and may differ across components. From the definition of $v_k$ and $v_{gk}$, it follows that
\begin{equation}
\label{e:raf1}
\text{if } g_{kt},h_t\ge c>0 \text{ for all } t,n, \text{ then } v_k, v_{gk}\ge c\sqrt n.
\end{equation}

\noindent\textbf{Asymptotic normality}. The asymptotic normality of an element $\widehat \beta_k$ of the OLS estimator, as well as the computation of its standard errors, requires additional  assumptions on the scale factors and the stationary processes $\{\eta_t, \varepsilon_t\}$.

\begin{assumption}
\label{a:4R} \noindent (i) For $k,j=1, ..., p$, the  sequences $\{\varepsilon_t^2\}$, $\{\eta_{jt}\eta_{kt}\varepsilon_t^2\}$ and $\{\eta_{jt}\varepsilon_t^2\}$
are covariance stationary and  have  short memory  (SM). \noindent\textrm{(ii)} For $k=1,..., p,$
\begin{eqnarray}  \label{e:zzuu}
&&\frac{\max_{1\le t \le n}g_{kt}^2h_t^2}{v_{k}^2}= o_p(1),\quad \frac{\max_{1\le t \le n}\mu_{kt}^2h_t^2}{v_{k}^2}= o_p(1).
\end{eqnarray}
\end{assumption}
\noindent
Assumption \ref{a:4R} is not  required  when  $\varepsilon_t$  is i.i.d.
Together, Assumptions \ref{a:r3} and \ref{a:4R}(ii)  exclude cases in which the mean process $\mu_t$ or  a  few  extreme observations  of  $z_t$ or  $u_t$, dominate the estimation of the regression parameter. Overall, these assumptions are mild. They accommodate both deterministic and  stochastic means $\mu_t$ and  scale  factors $h_t, g_{t}$ that   may change  abruptly over time unlike other theoretically rigorous treatments which restrict structural change to be deterministic and  smooth. This flexibility makes the  framework particularly suitable  for modelling  financial  data, as it allows for volatility jumps, commonly observed  in empirical finance (see, e.g., \cite{eraker2003}). In modern macroeconomic VAR models, the scale factor $h_t$  in the uncorrelated noise representation $u_t=h_t\varepsilon_t$ is typically assumed to be stochastic (see, e.g., \cite{chan2024}, \cite{carriero2024}), which our framework naturally encompasses.

Lemma \ref{l:le} below shows that Assumptions \ref{a:r3} and \ref{a:4R}(ii) holds for regressors $z_t$ and noises $u_t$ with bounded $4+\delta$ moments satisfying (\ref{e:raf1}). The following example provides additional sufficient conditions.

\begin{example}\label{ex:1}{\rm
Assumptions \ref{a:r3} and \ref{a:4R}(ii)
are satisfied by regressors $z_t$ and noises
$u_t$ whose scale  factors $h_t,g_t$  and  means $\mu_t$  satisfy
$0<c\le h_t, g_{kt}\le C,  \,\,\,||\mu_t||\le C,$
where  $0<c,\,C<\infty$ do not  depend on $t, n$  or $k=1, ..., p$.

When $0<c\le h_t\le C$, $||\mu_t||\le C$
for all $t, n$,  Assumptions \ref{a:r3} and \ref{a:4R}(ii) hold for scale  factors   $g_{kt}$  satisfying
\begin{eqnarray*}
\frac{\min_{t=1, ..., n}g^2_{kt}}{\sum_{t=1}^ng^2_{kt}}&=&o_p(1), \quad  k=1, ..., p.
\end{eqnarray*}
This  condition is, for example,   satisfied  when $g_{kt}$  follows  a  unit   root  process defined by $g_{kt}=\sum_{j=1}^n \xi_j$,  where  $\{\xi_j$\}  is  a sequence of  i.i.d  $(0, \sigma^2)$
random  variables  with  finite  moments of order   $\theta>2$.
The idea of modelling parameters as unit root processes was discussed, for example, in \cite{nyblom1989}.
}
\end{example}

\noindent We now describe the infeasible standard errors $\sqrt {\omega_{kk}}$  using the notation:
\begin{eqnarray}
&&\mbox{$S_{zz}=\sum_{t=1}^n z_tz^\prime_t, \quad S_{zzuu}=\sum_{t=1}^n
z_tz^\prime_t u_t^2,$}  \notag \\
&&\Omega_n=(E[S_{zz} \,|\mathcal{F}^*_n])^{-1}E[S_{zzuu}\,|\mathcal{F}
^*_n](E[S_{zz}\,|\mathcal{F}^*_n])^{-1}=(\omega_{jk}),   \label{e:OmegaR}
\end{eqnarray}
where $\omega_{jk}$ denotes the  $(j,k)$-th element of the matrix  $\Omega_n$.
The  infeasible standard error of $\widehat \beta_k$ is  defined as  $\sqrt {\omega_{kk}}$, i.e.,  the square root of the corresponding diagonal element of  $\Omega_n$.

\noindent The generality of our regression setting limits the multivariate asymptotic theory that can be  established for $\widehat\beta_t$. While a full joint distribution of  $\widehat\beta_t$ is not available, we can derive asymptotic normality for linear combinations $a^\prime \widehat \beta$ and then construct feasible inference for individual component $\beta_k$. 

\noindent
Existing  literature typically  imposes   stronger   assumptions on regressors and errors such as  mixing  regressors \cite[Theorem 3.78]{white2014},
locally stationary regressors in \cite[eq.~(2.3)]{zhangwu2012}, or near-epoch dependent errors in \cite[Assumption 8]{boldea2012}.

\begin{theorem}
\label{t:r1-R} Suppose that the assumptions of Theorem \ref{t:r1}
and Assumption \ref{a:4R} hold. Then, for any $a=(a_1, ..., a_p)^\prime\ne 0$, the OLS estimator $\widehat\beta$ satisfies
\begin{eqnarray}  \label{e:hatb1-R}
\frac{a^\prime D( \widehat \beta -\beta)}{\sqrt{a^\prime D\Omega_n D a }}
\rightarrow _d \mathcal{N}(0, 1).
\end{eqnarray}

\vskip.1cm \noindent  In particular, for $k=1, ..., p$, the  $t$-statistic for  $\beta_k$  satisfies
\begin{eqnarray}  \label{e:hatb1-R13}
\frac{\widehat \beta_k -\beta_k}{\sqrt{\omega_{kk}}}\rightarrow _d \mathcal{N}(0, 1).
\end{eqnarray}
\end{theorem}

\vskip.2cm \noindent

\noindent Property (\ref{e:hatb1-R}) is difficult to implement in practice because it requires  estimation of the unknown matrices $D,\, \Omega_n$,  except in the special case $a^\prime=(0, ...,1,...0)$ with only the $k$-th element nonzero.
In this case,  (\ref{e:hatb1-R}) reduces to (\ref{e:hatb1-R13}), and the infeasible standard error $\sqrt{\omega_{kk}}$ can be consistently estimated by
\begin{eqnarray}  \label{e:OmegaR+}
&&\widehat \Omega_n=S_{zz}^{-1}S_{zz\widehat u \widehat
u}S_{zz}^{-1}=(\widehat \omega_{jk}), \quad \widehat u_t=y_t-\widehat
\beta^\prime z_t.
\end{eqnarray}
The feasible standard error $\sqrt{\widehat \omega_{kk}}$ is the square root of the diagonal element $\widehat \omega_{kk}$ of $\widehat \Omega_n$.
 
\begin{corollary}
\label{c:co1} Under the assumptions of Theorem \ref{t:r1-R}, for $k=1, ..., p$,
as $n \rightarrow \infty$,
\begin{eqnarray}  \label{e:hatb1-Rno}
&& \frac{\widehat \beta_k -\beta_k}{\sqrt{\widehat \omega_{kk}}}\rightarrow
_d \mathcal{N}(0, 1), \quad \frac{\widehat \omega_{kk}}{ \omega_{kk}}
=1+o_p(1),\quad \sqrt{ \omega_{kk}}\asymp_pv_k^{-1}.
\end{eqnarray}
\end{corollary}

\vskip.2cm \noindent This result
is  the  main contribution of  Section \ref{s:OLS}.
It enables straightforward computation of standard errors and the construction of confidence intervals for  $\beta_k$
in the  extended  regression framework. Notably, the estimator $\widehat \Omega_n$ 
coincides with the heteroskedasticity-consistent standard error estimator of \cite{white1980}.

\vskip.2cm
\noindent \begin{remark}\label{r:2.1}
{\rm
The consistency rate $v_k =(\sum_{k=1}^n   g_{kt}^2 h_t^2)^{1/2}$
for the  parameter $\beta_k$ may take the form $v_k\sim c n^\alpha$  for  any $\alpha>0$, ranging from super-slow $(0<\alpha<1)$ to  super-fast $(\alpha>1)$ convergence. To illustrate this, consider the regression model
\begin{eqnarray*}
y_t&=&\beta_1+\beta_2 z_{2t}+\beta_3 z_{3t}+u_t,\quad \mbox{ $u_t=h_t\varepsilon_t$ with  $h_t=1$,}\\
  & & z_{kt}=g_{kt}\eta_{kt}, \quad g_{kt} =t^{(\alpha_k-1)/2} \,\,\mbox{for $k=2,3,$}
\end{eqnarray*}
where $\alpha_2>1$, $0<\alpha_3<1$, and  $\{\eta_{2t}\}$, $\{\eta_{3t}\}$,  $\{\varepsilon_t\}$ are   i.i.d. ${\cal N}(0,1)$.  Then $v_1=\sqrt n$ and $v_k\sim \alpha_k^{-1/2}n^{\alpha_k/2}$ for $k=2,3$, producing different convergence rates across parameters. Even in this simple case, the usual multivariate asymptotic normality for $\sqrt n(\widehat\beta-\beta)$ does not hold.
 }
\end{remark}

\noindent Corollary~\ref{c:co1} allows us to establish the asymptotic power and consistency of a  test for testing the hypothesis
\[
H_0: \beta_k = \beta_k^0, \quad {\rm vs.} \quad H_1: \beta_k \ne \beta_k^0,
\]
i.e., whether the $k$-th element of the regression parameter $\beta=(\beta_1,\ldots,\beta_p)^\prime$ is equal to a specific value $\beta_k^0$.

\begin{corollary}
\label{c:co1Power}
Suppose that $\beta_k^0 \ne \beta_k$. Then, under the assumptions of Corollary \ref{c:co1},
\begin{eqnarray}  \label{e:hatb1-RnoPower}
t = \frac{\widehat \beta_k - \beta_k^0}{\sqrt{\widehat \omega_{kk}}} \ \asymp_p \ v_k \ \rightarrow_p \ \infty.
\end{eqnarray}
\end{corollary}

\noindent We conclude this section with a   lemma that  provides simple sufficient moment-type conditions for  the validity  of Assumptions \ref{a:r3} and \ref{a:4R}(ii). In  particular,  condition (\ref{e:raf1})  implies  (\ref{e:lec1++}).

\begin{lemma}
\label{l:le} Suppose that  for $k=1, ..., p$,
\begin{eqnarray}
Ez_{kt}^4&\le& c,\quad E|u_t|^{4+\delta}\le c \,\,\,\, \mbox{for some $\delta>0$},
\label{e:lec1}\\
n/v_k^2&=&O_p(1), \quad n/v_{gk}^2=O_p(1),\label{e:lec1++}
\end{eqnarray}
where $c<\infty$ does not depend on $t,n$. Then  Assumptions \ref{a:r3} and 
\ref{a:4R}(ii) hold.

\noindent In particular, (\ref{e:lec1++}) is  satisfied  if $\min_{t=1, ..., n}h_t^{-1}=O_p(1)$,  $\min_{t=1, ..., n}g_{kt}^{-1}=O_p(1).$

\end{lemma}

\noindent The regular estimator of standard errors in OLS regression estimation is given by
\begin{equation}  \label{e:clRFTx}
\widehat \Omega_{n}^{(st)}=S_{zz}^{-1} \,\widehat \sigma ^2_u, \quad
\widehat \sigma ^2_u=n^{-1}\sum_{j=1}^n\widehat u_j^2.
\end{equation}
Unlike the robust standard errors $\sqrt{\widehat\omega_{kk}}$, these conventional standard errors may produce coverage distortions, particularly when heteroskedasticity or heterogeneity in $g_t$, $h_t$, or $\mu_t$ is present, see Section \ref{MC}. This underscores the robustness and strong empirical performance of the normal approximation in (\ref{e:hatb1-Rno}).

In this section, we have provided a rigorous validation of the asymptotic normality of feasible $t$-statistics for the components of the OLS estimator  in linear regression models with general heterogeneity. The assumptions imposed are mild yet flexible, allowing a wide class of (possibly nonstationary) regressors and noise processes beyond those typically considered in the existing literature.
Some conditions on scale factors are analogous to the Lindeberg condition and remain necessary. Our framework complements, rather than replaces, prior approaches; for instance, near-unit-root regressors \cite{georgiev2018} require a distinct theoretical treatment. Although bootstrap methods, see, e.g., \cite{boldea2012}, \cite{boldea2019}, are widely applied in regression analysis, they may not extend to the heterogeneous structures considered here. By contrast, we demonstrate that the heteroskedasticity-consistent standard errors of \cite{white1980} remain applicable and computationally straightforward.

\vskip.2cm
\noindent In this  paper  we  focus    on the regression model (\ref{e:r1}), where  the regression noise  $u_t$ in (\ref{e:r1})
is  uncorrelated. Extending  the asymptotic theory  to account for dependence in $u_t$ is a natural next step and is currently under consideration.

\vskip.2cm \noindent Detailed proofs of all results are provided in the Online Supplement.

\section{Time-varying OLS estimation in extended regression space}
\label{s:FTV}
This  section  demonstrates  further  advantages of the theory of  regression estimation with a fixed  parameter, developed in Section \ref{s:OLS}. Thanks to the flexible setting, estimation of time-varying parameters naturally follows  from our theory for fixed-parameter regression in the extended space, along with bounding of some negligible terms.

In the  previous  section, we discussed the estimation of the regression model (\ref{e:r1}), $y_j=\beta^\prime z_j +u_j$, with  a fixed  parameter  $\beta$.
We now extend the model by allowing the regression parameter  to vary over time.
Specifically, we consider the model
\begin{eqnarray}  \label{e:r1TV}
y_j=\beta^\prime_j z_j +u_j, \quad j=1, ..., n,
\end{eqnarray}
where the regressors $z_j$ and the regression noise $u_j$, as defined  in  (\ref{e:rz2}) and (\ref{e:r2}), remain unchanged. That is,  they belong to the  same   regression space  as  in Section \ref{s:OLS}.

The primary objective is to develop a point-wise estimation procedure for the
path $\beta_1, ..., \beta_n$ of the time-varying parameter $\beta_j$ in
model (\ref{e:r1TV}), while preserving the   same  regression space  introduced in Section \ref{s:OLS}.

The literature on estimation of  time-varying regression parameters $\beta_j$  is extensive. It primarily focuses on estimation and testing for parameter stability    under   relatively strong assumptions on the regressors and regression  noise.
For instance,  regressors  are  assumed  to be  locally stationary in (\cite{vogt2012}, model (3)), stationary and strongly mixing in (\cite{hong2023}, Assumption A.1) and strictly stationary in (\cite{HKW2024}, Assumption P$(d)$). It  is clear  that  the class of  regressors considered in our  setting is  broader,   and  they  may be  neither mixing nor  stationary.

The objective of this section is to describe the extended regression space of regressors $z_t$ and disturbances $u_t$ that ensures the asymptotic normality of the feasible $t$-statistic  estimating the components of the time-varying parameter $\beta_t$. We show that, as long as the regressors and the disturbance follow the structure $z_t=\mu_t+I_{gt}\eta_t$ and    $u_t=h_t\varepsilon_t$, the  class of  admissible means  $\mu_t$  and    scale factors $g_t, h_t$ is very broad and  characterized by weak restrictions that may not require empirical verification.

Further extensions of the regression space are possible. For example, the weakly exogenous component $\eta_t$ of the regressors $z_t$  in our paper is assumed to be a short-memory process. In contrast, \cite{HKW2024} demonstrate that estimation of the time-varying parameter $\beta_t$ also permits weakly exogenous, strictly stationary regressors $z_t$  that exhibit long-memory behavior.

While most  assumptions on the regressors $z_j$  and regression noise $u_j$  remain unchanged from Section \ref{s:OLS}, the estimator requires some modifications.  Under an additional   smoothness assumption  on $\{\beta_j\}$,  the time-varying OLS  estimator $\widehat \beta_t$ of  parameter   $\beta_t$ at time   $t$ is  the standard  OLS  estimator for a  fixed  regression  parameter, obtained by regressing
$\widetilde y_j= b_{n,tj}^{1/2}y_j$  on $\widetilde z_j=b_{n,tj}^{1/2}z_j$:
\begin{eqnarray}
\widehat \beta_t=\big( \sum_{j=1}^n \widetilde z_j \widetilde z_j^\prime\big)^{-1}
\big( \sum_{j=1}^n \widetilde z_j  \widetilde y_j\big)
=\big( \sum_{j=1}^nb_{n,tj} z_j z_j^\prime\big)^{-1}\big( 
\sum_{j=1}^nb_{n,tj} z_j y_j\big).\label{e:OLS10}
\end{eqnarray}
The weights $b_{n, tj}$ are generated as follows:
\begin{eqnarray}  \label{e:vnt}
b_{n, tj}=K(\frac{|t-j|}{H}), \,\, t,j=1, ..., n,
\end{eqnarray}
where $H=H_n$ is a bandwidth parameter such that $H\rightarrow \infty$ and $H=o(n)$. The kernel function $K$ is bounded and there exist $a_0, \, \delta>0$ and $\theta>3$ such that
\begin{eqnarray}  \label{e:kernel}
K(x)&\ge& a_0>0, \,\,\,0\le x\le \delta, \\
K(x)&\le& Cx^{-\theta}, \,\,\ x> \delta.  \notag
\end{eqnarray}

\noindent For example, (\ref{e:kernel}) is satisfied by functions $K(x)=I(x\in [0,1])$ and $K(x)=p(x)$ where $p(x)$ is the probability density function of
the standard normal distribution.

\noindent We impose a smoothness assumption on the time-varying parameter $\beta_j$, which may be either deterministic or stochastic.

\begin{assumption}
\label{a:sm} For some $\gamma\in (0,1]$ and for  $t,j=1, ..., n$,
\begin{eqnarray}  \label{e:bsm1}
E ||\beta_t-\beta_j||^2\le c\big(\frac{|t-j|}{n}\big)^{2\gamma},
\end{eqnarray}
where $c<\infty$ does not depend on $t,j,n$.
\end{assumption}

\noindent Next, we briefly outline how our asymptotic theory for the time-varying robust estimator
 builds on  the results from Section \ref{s:OLS} on fixed-parameter regression estimation and the smoothness assumption (\ref{e:bsm1}).
 To demonstrate this, we
 introduce  the following regression model  with a fixed parameter $\beta=\beta_t$:
 \begin{eqnarray}  \label{e:r1TV1TV}
 y_j^*&=&\beta^\prime \widetilde z_j + \widetilde u_j, \,\,\, \widetilde u_j =b_{n,tj}^{1/2}u_j, \quad j=1,
..., n.
\end{eqnarray}
Notice  that the  OLS   estimator $\widehat \beta$ of the  fixed  parameter  $\beta$ satisfies:
\begin{eqnarray}\label{e:OLSrTV+TV+}
\widehat \beta &=&\big( \sum_{j=1}^n \widetilde z_j \widetilde z_j^\prime\big)^{-1}
\big( \sum_{j=1}^n \widetilde z_j  y_j^*\big)=\beta+\big( \sum_{j=1}^n \widetilde z_j \widetilde z_j^\prime\big)^{-1}
\big( \sum_{j=1}^n \widetilde z_j  \widetilde u_j\big).
\end{eqnarray}
Since $\widetilde  y_j =y_j^*+(\beta_j-\beta_t)^\prime \widetilde z_j$,
 the  time-varying estimator $\widehat \beta_t $ given in (\ref{e:OLS10})   satisfies:
\begin{eqnarray}
\widehat \beta_t-\beta_t&=&
\big( \sum_{j=1}^n \widetilde z_j \widetilde z_j^\prime\big)^{-1}
\big( \sum_{j=1}^n \widetilde z_j \{y_j^*+(\beta_j-\beta_t)^\prime \widetilde z_j\}\big)-\beta_t\nonumber \\
&=&\widehat \beta-\beta+R_t,\quad R_t=\big( \sum_{j=1}^n \widetilde z_j \widetilde z_j^\prime\big)^{-1}
\big( \sum_{j=1}^n \widetilde z_j \widetilde z_j^\prime (\beta_j-\beta_t)\big).\label{e:OLSrTV+DD}
\end{eqnarray}
Notice that $\widehat \beta -\beta$
in (\ref{e:OLSrTV+DD})  does not depend  on $\beta$.
Additionally,  the  regression  space  in estimation of   the  fixed  parameter  in Section \ref{s:OLS} permits rescaling, so premultiplying by the kernel  weights $b_{n,tj}^{1/2}$ does not change the structure of regressors  $\widetilde z_{j} =( \widetilde z_{1j}, ...,\widetilde z_{pj})^\prime$  and $\widetilde u_{j} $:
they still satisfy the settings (\ref{e:rz2}) and (\ref{e:r2}).
Consequently, the model (\ref{e:r1TV1TV}) is  covered by the  regression  model (\ref{e:r1}) with a fixed  parameter, and the asymptotic results for $\widehat \beta -\beta$ follow from  Section  \ref{s:OLS}. The  main technical task  in this   section is to show  that the remainder term $R_t$ in (\ref{e:OLSrTV+DD}) is negligible,  which follows from the smoothness assumption (\ref{e:bsm1}).

\vskip.2cm
The regressors $z_j$ and regression noise $u_j$  belong to the  same  regression space as defined in as in Section  \ref{s:OLS}. While the assumptions on the  stationary process $\{\eta_j\}$ and the martingale  difference
noise $\{\varepsilon_j\}$ remain unchanged, for simplicity,  we replace the previous conditions  on the scale factors $g_j, h_j$ and the means $\mu_j$ with simple sufficient assumptions similar to  those used  in Lemma \ref{l:le}. As before,  the scale factors $\{h_j, g_j, \mu_j\}$ can be deterministic or stochastic, may vary with $n$, and are independent of $\{\eta_j, \varepsilon_j\}$.

Denote
\begin{eqnarray*} 
\mbox{$ v_{kt}^2=\sum_{j=1}^n b_{n,tj}^2g_{kj}^2h_j^2, \,\,\,\,\,\,
v_{gk,t}^2=\sum_{j=1}^n b_{n,tj}^2g_{kj}^2,\quad k=1, ..., p.$}
\end{eqnarray*}

\begin{assumption}
\label{a:tv5} $z_t$  and  $u_t$  are such that, for $k=1, ..., p$,
\begin{eqnarray}  \label{e:lecTV}
Ez_{kt}^4&\le& c,\quad E|u_t|^{4+\delta}\le c \,\, \mbox{for some $\delta>0$},\\
\label{e:lecTV++}
H/v_{kt}^2&=&O_p(1), \qquad H/v_{gk,t}^2=O_p(1),
\end{eqnarray}
where $c<\infty$ does not depend on $t,n$.
\end{assumption}
\noindent It is straightforward to show  that (\ref{e:lecTV++})  is valid if  $g_{kt},h_t\ge c>0$ for all $t,n$.

To describe the infeasible standard errors $\sqrt{\omega_{kk,t}}$, we   use:
\begin{eqnarray*} 
&&\mbox{$S_{zz,t}=\sum_{j=1}^n b_{n, tj}z_jz^\prime_j, \quad
S_{zzuu,t}=\sum_{j=1}^nb^2_{n, tj} z_jz^\prime_j u_j^2,$} \\
&&\Omega_{nt}=E[S_{zz,t}|{\cal F}_n^*]^{-1}E[S_{zzuu,t}|{\cal F}_n^*]E[S_{zz,t}|{\cal F}_n^*]^{-1}=(\omega_{jk,t}),
\notag
\end{eqnarray*}
where $\omega_{jk,t}$ denotes the  $(j,k)$-th element of the matrix  $\Omega_{nt}$.
The  infeasible standard error $\sqrt {\omega_{kk,t}}$ is  defined  by the diagonal element  $ \omega_{kk,t}$  of  the  matrix $ \Omega_{nt}$.

\noindent The next theorem establishes the consistency rate and asymptotic
normality property  for the components  of the time-varying OLS estimator $
\widehat \beta_t=(\widehat\beta_{1t}, ..., \widehat\beta_{pt})^\prime$,
 and  allows for arrays  of  integers $t=t_n\in [1, ..., n]$,  which may  depend on $n$.

\begin{theorem}
\label{t:r1FT} Suppose that $(y_1, ..., y_n)$ is a sample from a regression
model (\ref{e:r1TV}). Assume that  Assumptions \ref{a:r0}, \ref{a:ETA}, \ref{a:4R}(i), \ref{a:sm} and \ref{a:tv5} hold. Then, for $1\le t=t_n\le n$ and $k=1,
..., p$:
\begin{eqnarray}  \label{e:hatb1FT}
&&\widehat \beta_{kt} -\beta_{kt}=O_p\big( H^{-1/2}+ (H/n)^\gamma\big), \\
&&\frac{\widehat \beta_{kt} -\beta_{kt}}{\sqrt{\omega_{kk,t}}}\rightarrow _d
\mathcal{N}(0, 1)\quad \mbox{
if $H=o(n^{2\gamma/(2\gamma+1)})$},  \label{e:hatb1FTCLT}
\end{eqnarray}
and $\sqrt{ \omega_{kk,t}}\asymp_p H^{-1/2}$.
\end{theorem}

\noindent
The   consistency rate in (\ref{e:hatb1FT})  is  determined  by the   bandwidth parameter $H$  and the smoothness  parameter $\gamma\in (0,1)$   in (\ref{e:bsm1}).
The condition $H=o(n^{2\gamma/(2\gamma+1)})$ ensures
that in (\ref{e:hatb1FTCLT})
 the bias term remains negligible.

As  in the
fixed-parameter case, for $(z_j, u_j)$  from the  extended regression   space,  the asymptotic  normality
can    be  established in point-wise estimation  for   each individual component     $\widehat \beta_{kt}$ of  $\widehat \beta_{t}$.

\noindent The unknown
standard error $\sqrt{\omega_{kk,t}}$ can be consistently
estimated by:
\begin{eqnarray}  \label{e:OmegaRFT1+}
\widehat \Omega_{nt}=S_{zz,t}^{-1}S_{zz\widehat u \widehat
u,t}S_{zz,t}^{-1}=(\widehat \omega_{jk,t}), \quad \widehat u_j=y_j-\widehat
\beta_j ^\prime z_j.
\end{eqnarray}
The  feasible standard error $\sqrt {\widehat \omega_{kk,t}}$ is  defined  by the diagonal element  $\widehat \omega_{kk,t}$  of  $\widehat \Omega_{nt}$.

\begin{corollary}
\label{c:co1TV}Under assumption of Theorem \ref{t:r1FT}, for $k=1, ..., p$,
and $H=o(n^{2\gamma/(2\gamma+1)})$ it holds:
\begin{eqnarray}  \label{e:hatb1-RnoTV}
&& \frac{\widehat \beta_{kt} -\beta_{kt}}{\sqrt{\widehat \omega_{kk,t}}}
\rightarrow _d \mathcal{N}(0, 1), \quad \frac{\widehat \omega_{kk,t}}{
\omega_{kk,t}}=1+o_p(1).
\end{eqnarray}
\end{corollary}

 \noindent  Corollary \ref{c:co1TV}  allows us to establish the asymptotic power of the test of the hypothesis
$$H_0: \beta_{kt}=\beta_{kt}^0,\quad {\rm vs.} \quad H_1: \beta_{kt} \ne \beta_{kt}^0, $$
 based  on the  $t$-statistics
$(\widehat \beta_{kt} -\beta_{kt}^0)/\sqrt{\widehat \omega_{kk,t}}$.
\begin{corollary}
\label{c:co1TVPower} Suppose   that $|\beta_{kt}^0-  \beta_{kt}|\ge a>0$ for $t=t_n \in [1,...,n]$  as $n \rightarrow  \infty$.  Then, under  assumption of   Corollary \ref{c:co1TV},
\begin{eqnarray}  \label{e:hatb1-RnoPowerTV}
\frac{\widehat \beta_{kt} -\beta_{kt}^0}{\sqrt{\widehat \omega_{kk,t}}}\asymp_p H^{1/2}  \rightarrow  _p\infty.
\end{eqnarray}
\end{corollary}

The estimator $\widehat \Omega_{nt}$ used  to  obtain
 robust standard errors in (\ref{e:OmegaRFT1+}) is a time-varying version
of  heteroskedasticity-consistent estimator of standard errors by \cite{white1980}.  Simulation results  confirm that it does not produce coverage distortions in the estimation of $\beta_t$ under the settings considered in this section.

In conclusion, we provide examples of smoothly varying  deterministic and  stochastic parameters  $\beta_t$  that satisfy Assumption \ref{a:sm}.

\begin{example}
\label{ex:2}\textrm{A standard example of a deterministic time-varying
parameter $\beta_t$ which satisfies Assumption \ref{a:sm}, is $\beta_t=\beta_{t,n}=g(t/n)$, $t=1, ..., n$, where $g(\cdot)$ is a
deterministic smooth function  that has property $|g(x)-g(y)|\le C|x-y|$.
Such $\beta_t$ satisfies (\ref{e:bsm1}) with $\gamma=1$. }

\textrm{A standard example of a stochastic smooth parameter $\beta_t$ is a
re-scaled random walk $\beta_t=\beta_{t,n}=n^{-1/2}\sum_{j=1}^te_j$, $t=1,
..., n$, where $\{e_j\}$ is an i.i.d. sequence with $E[e_t]=0$ and $E[e^2_j]<\infty$. It satisfies (\ref{e:bsm1}) with $\gamma=1/2$, that is for
$t>s$,
\begin{eqnarray*}
E (\beta_t-\beta_s)^2 &=& \mbox{$n^{-1}E(\sum_{j=s+1}^te_j)^2\le C(t-s)/n.$}
\end{eqnarray*}
}
\end{example}

 \noindent The  above results  are  equipped with  thorough and mathematically rigorous proofs, which can be  found in the Online  Supplement.

\vskip.2cm

\noindent The  key  new features  in the estimation  of  time-varying parameter  $\beta_t$
are similar to those highlighted
in the estimation of the fixed  parameter  in Section \ref{s:OLS}.
 Although the computation is straightforward, establishing the validity of the robust standard errors $\sqrt{\widehat \omega_{kk, t}}$ in the   extended  regression space of
$(z_t, u_t)$ is challenging because the scale factors
 $h_t,g_t, \mu_t$ in model (\ref{e:r1TV}) are unknown and potentially random, and highly general, while the asymptotic behaviour of
the $\omega_{kk,t}$ may  not  be  well-defined.
The asymptotic normality of a single component of the estimator can still be established, even though a full multivariate asymptotic theory is not available.
Unlike most existing literature, $\beta_t$
is permitted to evolve as a smoothly varying stochastic process.

\section{Regression with missing data}
\label{s:missing}
\noindent In the previous  sections,  we showed  that the extended  regression space  enables the estimation  of both  fixed  and time-varying regression parameters.
It  offers several theoretical advantage, in particular, the ability to  estimate  regression models   in the presence of missing data. Given the importance in empirical regression analysis in situations where some observations $y_t$ or regressors $z_t$ are missing, see, e.g., \cite{enders2022}, we now present new and somewhat unexpected results on regression estimation with missing data. We
show  that the  foundational assumptions underlying the constriction of regression space also allow us to  accommodate an  a broad range of  missing data  patterns.

\noindent In this  section we  suppose that instead  of the full sample  $(y_1, z_1), ..., (y_n,z_n)$, we  observe  a  subsample
\begin{equation}\label{e:subs}
(y_{k_1},z_{k_1}), ..., (y_{k_N},z_{k_N}), \quad N\le n,
\end{equation}
of dependent  variable  $y_t$  and  regressor  $z_t$.
Our primary interest is to estimate  both fixed and  time-varying  regression parameters
using  the subsample (\ref{e:subs}).

To that end, we   represent the  observed  data  as partially  observed  sample
\begin{equation}\label{e:part1}
(\widetilde y_j, \widetilde z_j)= (\tau_jy_j,\, \tau_jz_j), \quad  j=1, ..., n
\end{equation}
where $\tau_j$ is  missing-data  indicator. In (\ref{e:subs}) it is  defined  as
\begin{equation}\label{e:part2}
\tau_j=
\begin{cases}
1  \quad \text{for} \quad j=k_1, k_2,\dots,k_N, \quad\text{where} \quad k_1<k_2<\dots<k_N\leq n, \\
0  \quad \text{otherwise}.
\end{cases}
\end{equation}
We  set $\tau_j=1$  if  both $y_{j}$  and  $z_{j}$ are observed, otherwise   $\tau_j=0$. Throughout  this  section, $\tau_j$   is  treated as a sequence of   random  or  deterministic variables, allowing for regularly  missing,  block-wise missing,  or randomly missing data patterns.

In order for the theoretical results of the previous section to apply, we  impose  the  following  assumptions on the  missing data indicator $\tau_t$, the regressors $z_{kt}=\mu_{kt}+g_{kt}\eta_{kt}$  in (\ref{e:rz2})  and the regression noise  $u_t=h_t\varepsilon_t$ in (\ref{e:r2}).

\begin{assumption}
\label{a:tau}
The  missing-data  indicator  $\{\tau_t\}$  is assumed to be independent of  $\{\varepsilon_t, \eta_t\}$ in (\ref{e:r2})  and  (\ref{e:rz2}). 
\end{assumption}
\begin{assumption}
\label{a:mis}
{\rm (i)} $Ez_{kt}^4\le c $ and $ E|u_{t}|^{4+\delta}\le c$  for some  $\delta>0$,
where  $c>0$  does not depend  on $k,t,n$.

{\rm (ii)}  $g_{kt} \ge  c>0$  and $h_{t} \ge  c>0$, where  $c$  does not depend  on $k,t,n$.

{\rm (iii)} $\varepsilon_t, \eta_t$  satisfy Assumptions \ref{a:r0}, \ref{a:ETA}, and \ref{a:4R}(i).
\end{assumption}

\vskip.2cm
\noindent{\bf Estimation of  a fixed  parameter}.
Suppose  that $y_t= \beta^\prime z_t+u_t$  follows the regression model   (\ref{e:r1}) with a fixed  parameter  $\beta$ as  in Section \ref{s:OLS}.
Our primary interest is to estimate the parameter $\beta$ using  subsample (\ref{e:subs}). In view  of  (\ref{e:r1}), we can write the partially  observed  regression  model  as
\begin{eqnarray}
\widetilde y_t&=& \tau_ty_t=\tau_t (\beta^\prime z_t +u_t),\nonumber \\
 \label{e:r1mis}
\widetilde y_t&=&
\beta^\prime \widetilde z_t +\widetilde u_t, \quad \widetilde u_t=\tau_tu_t=\{\tau_th_t\}
\varepsilon_t.
\end{eqnarray}
In (\ref{e:r1mis}), the regressors
 $\widetilde z_t$  and the noise $\widetilde u_t$ can be   represented  as
\begin{eqnarray}  \label{e:r1tilde}
\widetilde z_{kt}&=& \widetilde \mu_{kt}+\widetilde g_{kt}\eta_{kt}, \quad
\widetilde \mu_{kt}=\tau_t \mu_{kt}, \quad \widetilde g_{kt}=\tau_t g_{kt},\\
\widetilde u_{t}&=& \widetilde h_{t}\varepsilon_{t}, \quad \widetilde h_{t}=\tau_t h_{t}. \nonumber
\end{eqnarray}
They   belong to the  regression space  described  in
(\ref{e:r2})  and  (\ref{e:rz2}). Therefore,
 parameter $\beta$  and the  correspondent  standard errors in model  (\ref{e:r1mis}) can be  estimated  using  the OLS  estimator $\widehat \beta$ and
 $\widehat \omega_{kk}$:
\begin{eqnarray} \label{e:OLSrTILDE}
\widehat \beta&=&\big( \sum_{t=1}^n \widetilde z_{t} \widetilde z_{t}^\prime\big)^{-1}\big( \sum_{t=1}^n
\widetilde z_{t}\widetilde  y_{t}\big),\quad \widehat  \Omega_n= S_{\widetilde z\widetilde z}^{-1} S_{\widetilde z\widetilde z\widehat u \widehat
u}  S_{\widetilde z\widetilde z}^{-1}=(\widehat \omega_{jk}), \\
& &\mbox{$ S_{\widetilde z\widetilde z}=\sum_{t=1}^n\widetilde z_{t}\widetilde z^\prime_{t}, \quad      S_{\widetilde  z\widetilde  z\widehat u \widehat
u}=    \sum_{t=1}^n
\widetilde  z_{t}\widetilde  z^\prime_{t} \widehat u_{t}^2,   \quad \widehat u_{t}=\widetilde y_{t}-\widehat
\beta^\prime \widetilde  z_{t}.$}
\nonumber
\end{eqnarray}

\begin{theorem}\label{t:prop1}  The OLS  estimator $\widehat \beta $  of  parameter   $\beta$ in  regression model  (\ref{e:r1tilde}) with missing  data  has  the  following  asymptotic properties.
If  Assumptions   \ref{a:tau}  and  \ref{a:mis}  hold and   $n/N=O_p(1)$, then,
for $k=1, ..., p$, as $n \rightarrow \infty$,
\begin{eqnarray}  \label{e:hatb1-RnoMD}
&& \frac{\widehat \beta_k -\beta_k}{\sqrt{\widehat \omega_{kk}}}\rightarrow
_d \mathcal{N}(0, 1), \qquad
\quad \sqrt{ \widehat \omega_{kk}}\asymp_p n^{-1/2}.
\end{eqnarray}
\end{theorem}

\begin{remark} \label{r:1}
{\rm
Theorem  \ref{t:prop1}  shows that ignoring  missing data does not affect the estimation of the fixed  parameter. That is, the researcher can  compute the  estimators $\widehat \beta$  and  $\sqrt {\widehat \omega_{kk}}$
directly using subsample $y_{k_j}, z_{k_j}, \, j=1, ..., N$:
\begin{eqnarray*}
\widehat \beta&=&\big( \sum_{j=1}^N z_{k_j} z_{k_j}^\prime\big)^{-1}\big( \sum_{j=1}^N
z_{k_j} y_{k_j}\big),\quad \widehat \Omega_n= S_{*,zz}^{-1} S_{*,zz\widehat u \widehat
u} S_{*,zz}^{-1}=(\widehat \omega_{jk}), \\
& &\mbox{$ S_{*,zz}=\sum_{j=1}^Nz_{k_j}z^\prime_{k_j}, \quad  S_{*,zz\widehat u \widehat u}= \sum_{j=1}^N z_{k_j}z^\prime_{k_j} \widehat u_{k_j}^2,  \quad \widehat u_{k_j}=y_{k_j}-\widehat \beta^\prime z_{k_j}.$}
\nonumber
\end{eqnarray*}
}
\end{remark}

\vskip.4cm
\noindent{\bf Estimation of a time-varying  parameter}.
Assume now that $y_t= \beta^\prime _tz_t+u_t $  follows the regression model   (\ref{e:r1TV}) with  time-varying  parameter  $\beta_t$, where regressors $z_t$
 and regression noise $u_t$ are as  in  (\ref{e:rz2}) and  (\ref{e:r2}). We  are interested  in estimating the  parameter $\beta_t$  in the presence of  missing data  using the subsample  (\ref{e:subs}). Similarly to  (\ref{e:r1mis}),  we base  the estimation on  the  partially observed regression model  with a time-varying parameter,
\begin{eqnarray}  \label{e:r1misTV}
\widetilde y_j=
\beta^\prime_j \widetilde z_j +\widetilde u_j, \quad j=1, ..., n,
\end{eqnarray}
where  regressors $\widetilde z_j$  and the noise $\widetilde u_j$
are  defined  as  in  (\ref{e:r1tilde}). They   belong to the regression space described by (\ref{e:r2}),  and  (\ref{e:rz2})  and thus   results of  Section \ref{s:FTV} on the estimation of  time-varying parameter $\beta_j$ apply.

We  show in the  following theorem that under  Assumptions  \ref{a:tau}  and  \ref{a:mis}, parameter   $\beta_t$ and  standard errors
can be   estimated  point-wise  at each time $t=1, ..., n$
provided that the missing data pattern satisfies the following condition:
\begin{eqnarray}  \label{e:Nt}
H/N_t=O_p(1), \quad N_t=\sum_{j=1}^n \tau_j b_{n,tj}.
\end{eqnarray}
This  condition holds,  for example,  if $\tau_j=1$  for   $|j-t|\le  \epsilon H$   for  some $\epsilon>0$.

The  estimator $\widehat  \beta_t$   and  the estimator of the robust standard errors $\widehat \omega_{kk,t}$ given in (\ref{e:OLS10}) and (\ref{e:OmegaRFT1+})  are  defined as
\begin{eqnarray}  \label{e:OLS10MV}
\widehat \beta_t&=&\big( \sum_{j=1}^nb_{n,tj} \widetilde z_j \widetilde z_j^\prime\big)^{-1}\big(
\sum_{j=1}^nb_{n,tj} \widetilde z_j \widetilde y_j\big),\\
\widehat \Omega_{nt}&=&S_{\widetilde z\widetilde z,t}^{-1}S_{\widetilde z\widetilde z\widehat u \widehat
u,t}S_{\widetilde z\widetilde z,t}^{-1}=(\widehat \omega_{jk,t}), \quad \widehat u_j=\widetilde y_j-\widehat
\beta_j ^\prime \widetilde z_j. \nonumber
\end{eqnarray}

\begin{theorem}\label{t:prop2}
The OLS  estimator $\widehat \beta_t $  of the time-varying parameter   $\beta_t$
in   regression model  (\ref{e:r1misTV}) with missing  data  has  the  following   properties. Assume  that $1\le t=t_n\le n$,  Assumptions   \ref{a:tau}, \ref{a:sm}  and  \ref{a:mis}  are  satisfied and  that the  condition $H/N_t=O_p(1)$ holds. Then, for $k=1, ..., p$, as $n \rightarrow \infty$,
\begin{eqnarray}  \label{e:hatb1FTMD}
&&\widehat \beta_{kt} -\beta_{kt}=O_p\big( H^{-1/2}+ (H/n)^\gamma\big), \\
&&\frac{\widehat \beta_{kt} -\beta_{kt}}{\sqrt{\widehat \omega_{kk,t}}}\rightarrow _d
\mathcal{N}(0, 1)\quad \mbox{
if $H=o(n^{2\gamma/(2\gamma+1)})$},  \label{e:hatb1-R13TVMD}\\
&&\widehat \omega_{kk,t}\asymp_p H^{-1}. \label{e:omegaVMD}
\end{eqnarray}
\end{theorem}

\section{Estimation of a stationary AR($p$) model with an m.d. noise}\label{sss:ARp}
In this  section we focus on another practical application of our regression framework developed in Section \ref{s:OLS}. We  show that   it covers the  estimation of parameters of a  stationary  AR($p$) model driven by
a stationary martingale  difference  noise $\varepsilon_t$:
\begin{equation}
y_t=\phi_0+\phi_1y_{t-1}+...+\phi_py_{t-p}+\varepsilon_t,  \label{e:AR2S}
\end{equation}
where parameters  $\phi_0, ..., \phi_p$ are such  that the model (\ref{e:AR2S}) has a stationary solution. \cite{xu_phillips2008} developed  estimation theory  for AR$(p)$  model $y_t=\phi_0+\phi_1y_{t-1}+...+\phi_py_{t-p}+u_t$,  when $u_t=h_t\varepsilon_t$  where  $h_t$  is  smoothly varying  deterministic sequence and a m.d. sequence $\varepsilon_t$ has  property $E[\varepsilon_t^2|{\cal F}_{t-1}]=1$ a.s. \,\cite{GTT2018} were  among the first  to  analyze  the distortions of   standard errors caused by m.d. noise in estimation of   ARMA  models.
This paper shows that the variance of the parameter vector
$\phi$ converges to a well-defined limit; however, its complex structure complicates the estimation of the limiting variance and the corresponding standard errors in empirical applications. They restricted the  estimation of standard errors
 to  AR(1)  and MA(1)  models. In the  case of  AR$(p)$ model, using   our   method
 we  are  able  to estimate  standard  errors for  any $p$  without relying on asymptotic approximations
 which is  the  main novelty  and  contribution of  this  section.
Notice   that the model (\ref{e:AR2S})
  can be written as a special case
 of the   regression model  (\ref{e:r1}),
 \begin{equation}
  \label{e:RAR1}
 y_t=\beta^\prime
z_t+u_t,  \quad u_t=\varepsilon_t.
\end{equation}
Here, the parameter
 $\beta=(\beta_1, ..., \beta_{p+1})^\prime=(\phi_0,
...., \phi_p)^\prime$ is fixed,  and the  regressors $z_t=(z_{1t},z_{2t},...,z_{p+1,t})^
\prime=(1,y_{t-1},y_{t-2}, ..., y_{t-p})^\prime $  are stationary random variables. It is straightforward to verify that the regressors
$$z_{kt}= \mu_{kt}+g_{kt}\eta_{kt},  \quad \mu_{kt}=E[y_{t-k}]=Ey_1, \quad g_{kt}=1, \quad \eta_{kt}=y_{t-k}-E[y_{t-k}]$$
 for  $k=2, ..., p+1$  satisfy  the regression  assumption   (\ref{e:rz2}).
In the theorem  below,   we assume that the standard  stationarity conditions on parameters of  the AR($p$) model  (\ref{e:AR2S}) are  satisfied,
see  e.g.  Theorem 3.1.1 in \cite{bd1991}, which ensure the existence of a  stationary  solution
\begin{equation}
  \label{e:AREE}
y_t=\mu+\sum_{j=0}^\infty  a_j\varepsilon_{t-j},  \quad  \mbox{where} \,\,\sum_{j=0}^\infty  |a_j|<\infty, \,\,\mu=Ey_t.
\end{equation}
We assume  that $\varepsilon_t$  satisfies  Assumption  \ref{a:r0} and  $\eta_t=(y_{t-1},y_{t-2}, ..., y_{t-p})^\prime $  satisfy  Assumptions  \ref{a:ETA} and  \ref{a:4R}(i). These  assumptions  impose only mild  restrictions on the m.d. noise   $\varepsilon_t$, and  their validity  can  be   verified  for  typical  examples of  uncorrelated  m.d. noise, such as ARCH-type processes.

\noindent The   OLS  estimator $\widehat \beta $  of  $\beta$ in   regression model  (\ref{e:RAR1}) is  defined  as  in  (\ref{e:OLSr})  and  $ \widehat \omega_{kk}$  as in (\ref{e:OmegaR+}).

\begin{theorem}
\label{t:AR} Suppose that
 AR($p$) model (\ref{e:AR2S}) with  m.d. noise $\varepsilon_t$ has a stationary  solution as in (\ref{e:AREE}), that $E\varepsilon_t^8<\infty$ and that  $(\varepsilon_t, \eta_t)$ satisfy  Assumptions   \ref{a:r0}, \ref{a:ETA}, and  \ref{a:4R}(i). Then the OLS  estimator $\widehat \beta $  of  parameter   $\beta$
in  regression model  (\ref{e:RAR1})   has  the  following   properties: for $k=1, ..., p+1$, as $n \rightarrow \infty$,
\begin{eqnarray}  \label{e:hatb1-RnoMDAR}
&& \frac{\widehat \beta_k -\beta_k}{\sqrt{\widehat \omega_{kk}}}\rightarrow
_d \mathcal{N}(0, 1), \qquad
\quad \sqrt{ \widehat \omega_{kk}}\asymp_p n^{-1/2}.
\end{eqnarray}
\end{theorem}
\vskip.2cm
\noindent The Monte Carlo results presented in Section \ref{s:ARp} demonstrate that the robust OLS estimation produces correct $95\%$ confidence intervals for $\beta_k$, whereas the standard OLS method  exhibits coverage distortions, when the noise $\varepsilon_t$ is not i.i.d. This finding indicates that the robust OLS estimator has a broader range of applicability than merely addressing heteroscedasticity, and that it can also be effectively used in regression settings not covered by the standard OLS estimation and inference theory.

It is  worth noting  that  the papers  by  \cite{doukhan2008}, \cite{bardet2009} and \cite{karmakar2022} provide advanced theoretical results on the modelling and  estimation of general  nonlinear time-varying  time series models; however, they address  the linear AR$(p)$  model (\ref{e:AR2S}) only in the trivial  case  of an i.i.d. noise $\varepsilon_t$.

\section{Monte Carlo Simulations}
\label{MC}

In this section, we explore the finite sample performance of the robust and standard OLS
estimation methods in regression settings, outlined in Sections \ref{s:OLS} and \ref{s:FTV}.
We examine the  impact of time-varying deterministic and stochastic parameters, means, scale factors and heteroskedasticity of the regression noise on estimation.
Comparison of simulation results for  standard and  robust  estimation  methods
shows that, despite the generality of our regression setting, estimation based  on the robust standard errors produces well-sized coverage intervals for  fixed and time-varying regression parameters $\beta$ and $\beta_{t}$, while application of the standard  confidence intervals  leads to severe distortion of coverage  rates.

\subsection{Estimation of a fixed parameter} 
We generate arrays of samples of regression model  with fixed  parameter and an intercept:
\begin{equation}
y_t=\beta_1+\beta_{2}z_{2t}+\beta_{3}z_{3t}+u_t, \quad u_t=h_t \varepsilon_t,
\quad \beta=(\beta_1,\beta_2,\beta_3)^\prime= (0.5, 0.4 , 0.3)^\prime.
\label{MC:OLSy}
\end{equation}
We set the sample size to $n=1500$ and conduct $1000$ replications and  set  the nominal coverage  probability at $0.95$. (Estimation results for $n=200, 800$ are available upon request). We also include  a more complex example in the online supplement.

This model  includes three parameters and three regressors. We set $z_{1t}=1$ and define
\begin{eqnarray}  \label{MC:rz2}
z_{kt}&=&\mu_{kt}+g_{kt}\eta_{kt},\,\,\,k=2,3, \\
& &\mu_{kt}=0.5\sin(\pi t/n)+1, \quad \eta_{kt}= 0.5\eta_{k,t-1}+\xi_{kt},
\notag
\end{eqnarray}
where $\xi_{2t}=\varepsilon_{t-1}$ and $\xi_{3t}=\varepsilon_{t-2}$.
The stationary martingale difference noise $\varepsilon_t$ in $u_t$ is generated by a GARCH($1,1$) process
\begin{eqnarray}  \label{GARCHnoise}
\varepsilon_t =\sigma_t e_t, \quad \sigma^2_t =
1+0.7\sigma^2_{t-1}+0.2\varepsilon^2_{t-1}, \quad e_t\sim i.i.d.\,\mathcal{N}
(0,1).
\end{eqnarray}

\noindent \begin{model}
\label{MC:OLSmodel1} $y_t$ follows (\ref{MC:OLSy}) with deterministic scale  factors. We set: $h_t=0.3(t/n)$ and $g_{2,t}=g_{3,t}=0.4(t/n)$.
\end{model}

\begin{model}
\label{MC:OLSmodel2} $y_t$  follows (\ref{MC:OLSy}) with stochastic scale factors.
We  set
\begin{eqnarray*}
h_{t}=\Big|\dfrac{1}{2\sqrt{n}}\sum\limits^t_{j=1} \zeta_{j}\Big|+0.25,
\quad g_{2t}=g_{3t}=\Big|\dfrac{1}{2\sqrt{n}}\sum\limits^t_{j=1} \nu_{kj}
\Big|+0.25.
\end{eqnarray*}
The generating noises $\{\zeta_{j},\, \nu_{2j}, \nu_{3j}\}$
are i.i.d. $\mathcal{N}(0,1)$ and independent of $\{\varepsilon_{j}\}$.
\end{model}

\begin{figure}[]
\centering
\begin{minipage}[t]{0.32\linewidth}
    \centering
\includegraphics[width=\textwidth]{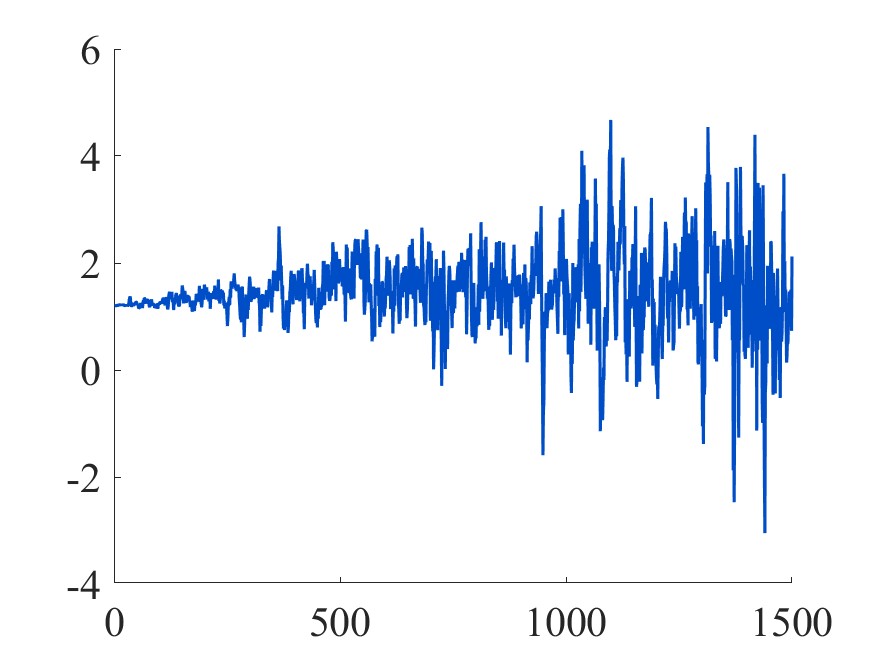}
 \subcaption{$y_t$}
	  \end{minipage}
\begin{minipage}[t]{0.32\linewidth}
    \centering
\includegraphics[width=\textwidth]{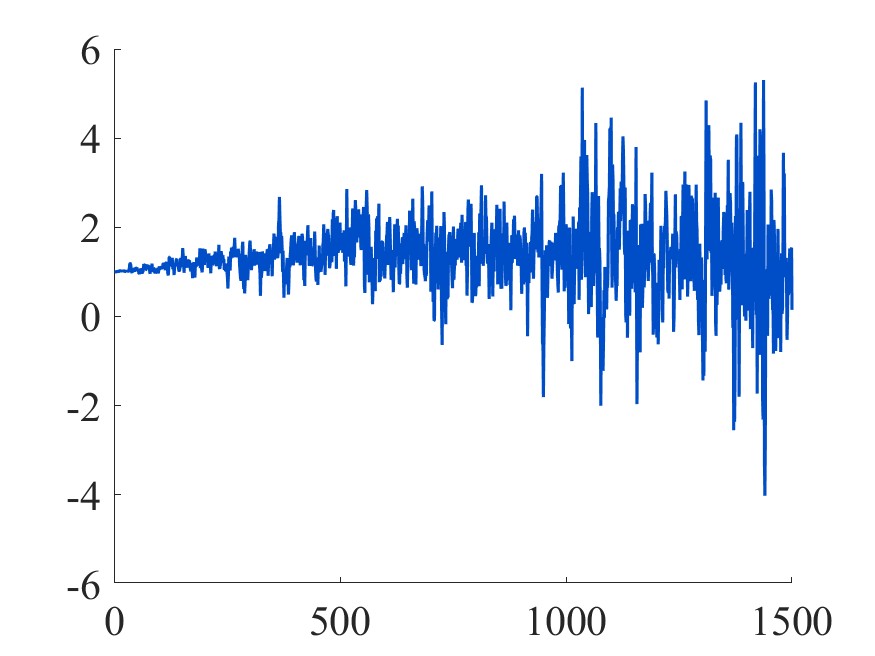}
\subcaption{{$z_{2t}$}}
	  \end{minipage}
\begin{minipage}[t]{0.32\linewidth}
    \centering
\includegraphics[width=\textwidth]{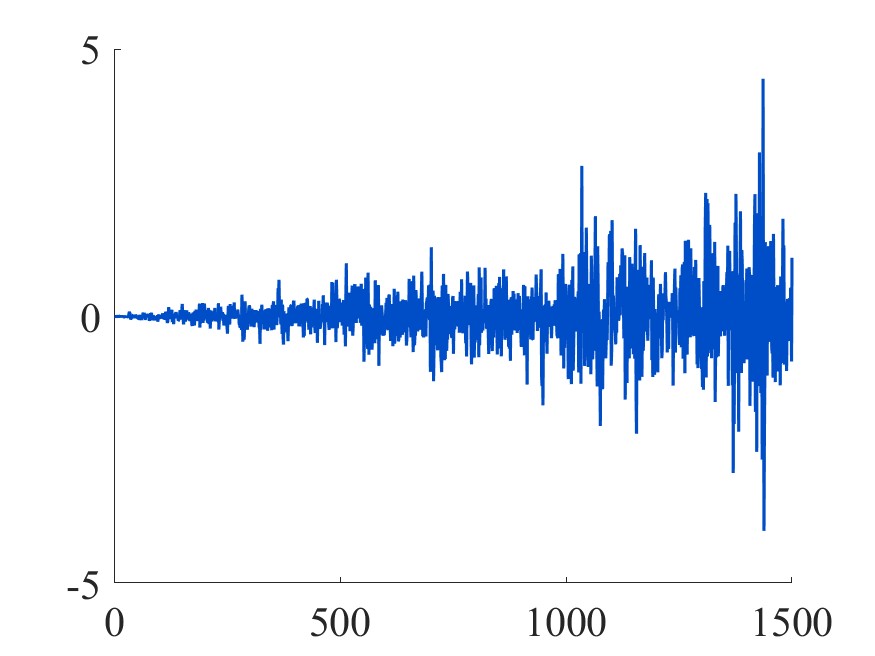}
 \subcaption{$u_t$}
	  \end{minipage}
\caption{Plots of $y_t$, $z_{2t}$, $u_t$ in Model
\ref{MC:OLSmodel1}.}
\label{WDTA2YZ}
\end{figure}

\begin{figure}[]
\centering
\begin{minipage}[t]{0.32\linewidth}
    \centering
\includegraphics[width=\textwidth]{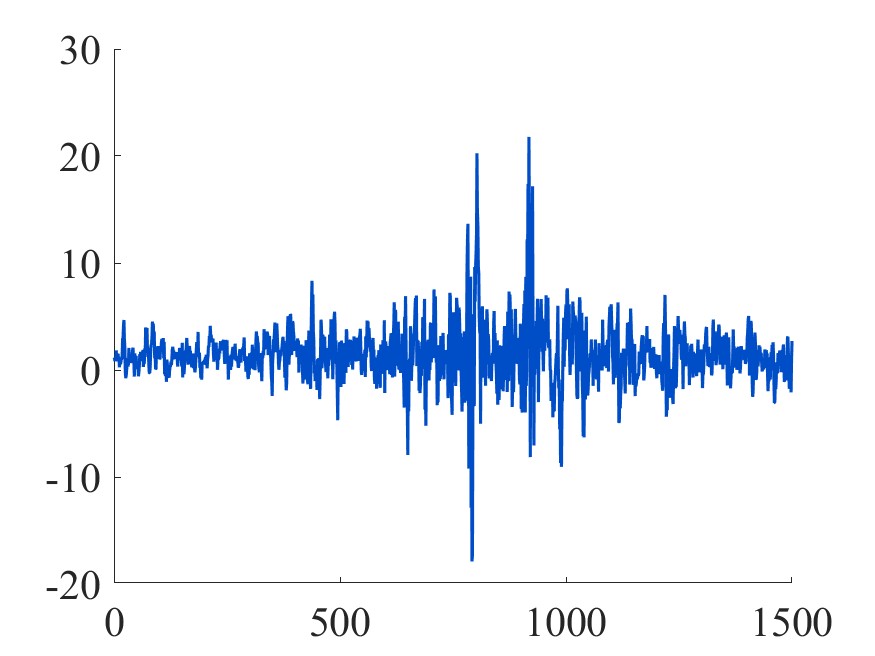}
 \subcaption{$y_t$}
	  \end{minipage}
\begin{minipage}[t]{0.32\linewidth}
    \centering
\includegraphics[width=\textwidth]{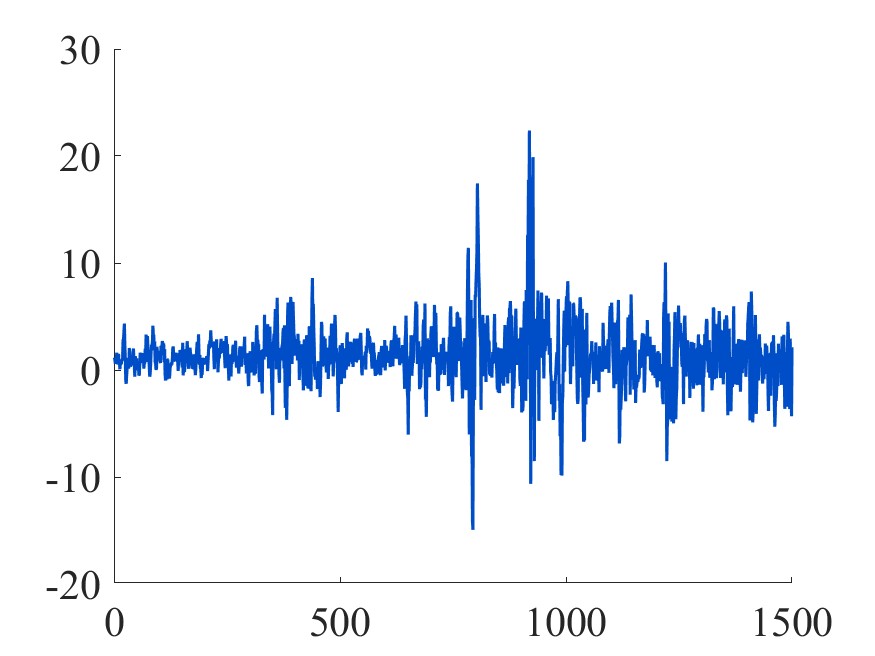}
\subcaption{{$z_{2t}$}}
	  \end{minipage}
\begin{minipage}[t]{0.32\linewidth}
    \centering
\includegraphics[width=\textwidth]{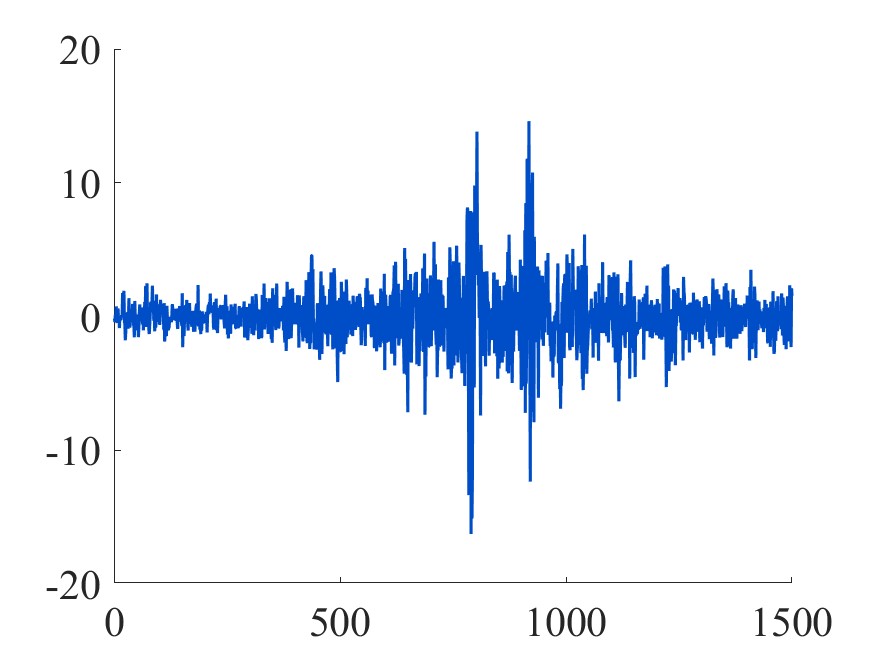}
\subcaption{{$u_{t}$}}
	  \end{minipage}
\caption{Plots of $y_t$, $z_{2t}$, $u_t$ in Model
\ref{MC:OLSmodel2}.}
\label{WSUA2YZ}
\end{figure}

Models  \ref{MC:OLSmodel1}   and  \ref{MC:OLSmodel2}   are regression models  with   fixed  parameters. Examples of plots of the simulated dependent variable, regressor and regression noise are shown in Figure \ref{WDTA2YZ} and \ref{WSUA2YZ} ($z_{2t}$ and $z_{3t}$ have similar patterns). To verify the validity of the asymptotic normal approximation of Corollary \ref{c:co1}  in finite samples, we compute empirical coverage  rates (CP)  for $95\%$ confidence intervals used  in robust OLS  estimation, for parameter $\beta$. For comparison, we compute the coverage rates CP$_{st}$ for standard confidence intervals based on the standard errors (\ref{e:clRFTx}) used in   standard OLS estimation. The robust and standard OLS procedures share the same estimator $\widehat \beta$, and whence Bias, root mean square error (RMSE) and standard deviation (SD). Their confidence intervals differ because the variances (and standard errors) in their normal approximations are different.

Table \ref{MC:tab:OLSm1}  reports  estimation results  for  Model \ref{MC:OLSmodel1}
which  contains  determinist scale  factors. It  shows  that coverage rate CP for robust  confidence intervals is close  to  the  nominal  $95\%$, while the coverage  rate CP$_{st}$ of the standard  confidence intervals drops  below  $80\%$. The Bias, RMSE, and SD are small.

Table \ref{MC:tab:OLSm2} shows
estimation results  for Model \ref{MC:OLSmodel2} which  includes  stochastic scale  factors. It  shows  that the  coverage  rate CP  for robust   confidence intervals  is close  to  the  nominal  $95\%$, whereas the standard estimation method produces coverage distortions for
parameters $\beta_2$ and $\beta_3$.

\begin{table}[]
\caption{
Robust  OLS estimation
in Model \protect\ref
{MC:OLSmodel1}. }
\label{MC:tab:OLSm1}\centering
\begin{tabular}{cccccc}
\hline Parameters & Bias & RMSE & CP & CP$_{st}$ & SD \\
\hline $\beta_1$ & -0.00570 & 0.04579 & 95.0 & 79.2 & 0.04544 \\
$\beta_2$ & 0.00206 & 0.03407 & 95.4 & 72.7 & 0.03401 \\
$\beta_3$ & 0.00204 & 0.03495 & 94.0 & 72.9 & 0.03489 \\
\hline &  &  &  &  &
\end{tabular}
\end{table}

\begin{table}[]
\caption{Robust OLS estimation in Model \protect\ref{MC:OLSmodel2}. }
\label{MC:tab:OLSm2}\centering
\begin{tabular}{cccccc}
\hline  Parameters & Bias & RMSE & CP & CP$_{st}$ & SD \\
\hline  $\beta_1$ & -0.00420 & 0.05117 & 94.6 & 92.2 & 0.05100 \\
$\beta_2$ & 0.00208 & 0.03205 & 94.6 & 87.4 & 0.03199 \\
$\beta_3$ & 0.00071 & 0.01542 & 94.8 & 85.3 & 0.01541 \\
\hline  &  &  &  &  &
\end{tabular}
\end{table}

\begin{figure}[H]
\begin{minipage}[t]{0.32\linewidth}
    \centering
\includegraphics[width=\textwidth]{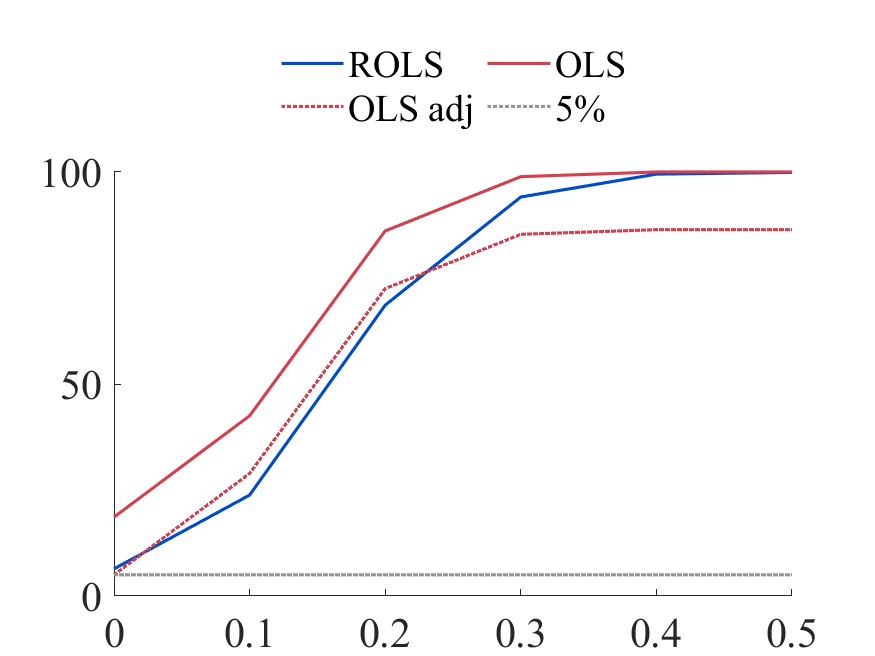}
 \subcaption{$n=200$}
	  \end{minipage}
\begin{minipage}[t]{0.32\linewidth}
    \centering
\includegraphics[width=\textwidth]{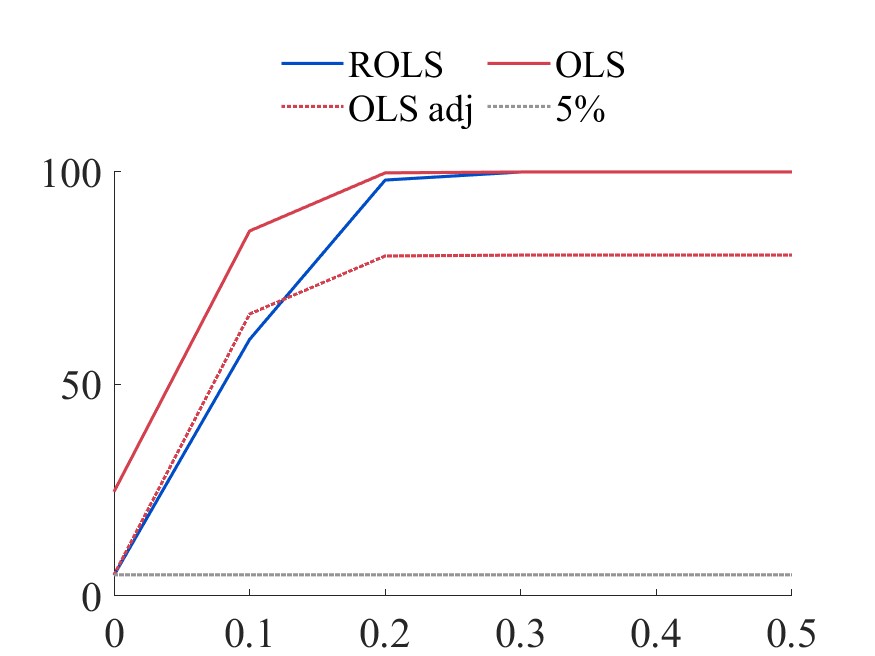}
\subcaption{{$n=800$}}
	  \end{minipage}
\begin{minipage}[t]{0.32\linewidth}
    \centering
\includegraphics[width=\textwidth]{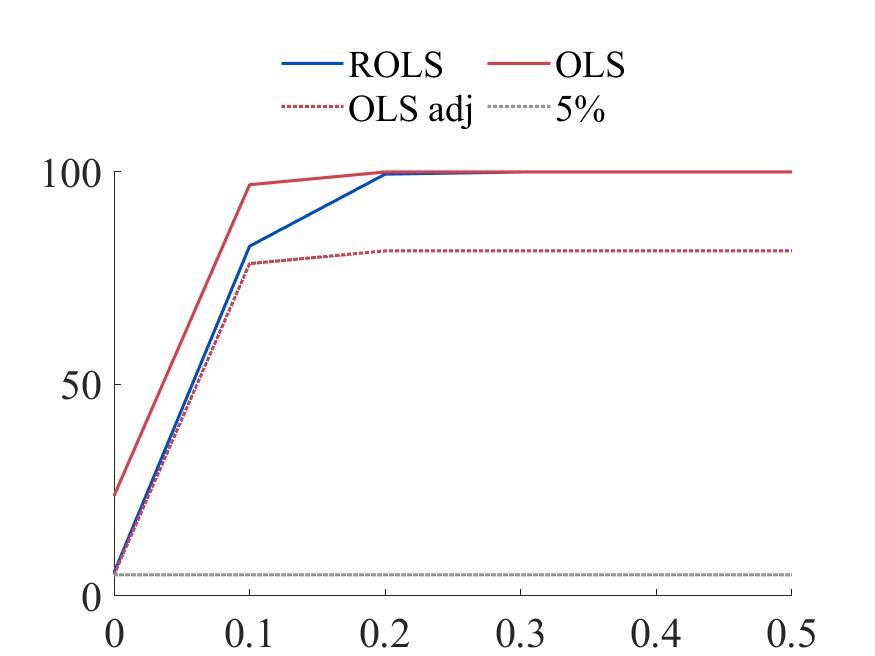}
\subcaption{{$n=1500$}}
	  \end{minipage}
\caption{Size, power, and adjusted power ($\%$) for test $H_0: \beta_3=0$ in Model \protect
\ref{MC:OLSmodel1}: $\beta_3=0,\cdots,0.5$, $n=200,\,800,\, 1500$.}
\label{WDTA2adjpower}
\end{figure}

To assess power, we vary $\beta_3$ in Model \ref{MC:OLSmodel1} from $0$ to $0.5$ and record how often the test rejects $H_0: \beta_3=0$. Figure \ref{WDTA2adjpower} reports results for ROLS and OLS at sample sizes  $n=200,\,800,\, 1500$. When $\beta_3=0$, ROLS achieves a good size close to the nominal $5\%$, while the size based on OLS results starts around $20\%$ and remains heavily oversized even as $n$ increases. For $\beta_3\neq 0$, power rises monotonically with $\beta_3$ for both methods. In Figure \ref{WDTA2adjpower}, the blue solid lines represent power based on ROLS, and the red solid lines correspond to standard OLS.
Considering the OLS estimation has large size distortion, we compute its adjusted power, shown by the red dotted lines. With small sample size $n=200$, OLS appears more powerful for $\beta_3 \le 0.2$, whereas   ROLS catches up and achives good power when $\beta_3 \ge 0.3$. For $n=800$ and  $1500$, both methods already achieve good power around $\beta_3=0.2$. Overall, ROLS provides reliable size and competitive power across different sample sizes. Similary results are observed for Model \ref{MC:OLSmodel2}.

\subsection{Estimation of a time-varying parameter}
\label{MC:FTVP}
In this  section we   examine  the   validity
of the normal  approximation for   the  estimator $\widehat  \beta_t$, (\ref{e:OLS10}), of   time-varying parameter   $\beta_t$, as
established  in  Corollary  \ref{c:co1TV}   of  Section \ref{s:FTV}.
We replace the fixed regression parameter $\beta$ in the
 model (\ref{MC:OLSy}) by a time-varying parameter $
\beta_t=(\beta_{1t},\beta_{2t},\beta_{3t})^\prime$:
\begin{equation}
y_t=\beta_{1t}+\beta_{2t}z_{2t}+\beta_{3t}z_{3t}+u_t, \quad u_t=h_t
\varepsilon_t,  \label{MC:modelTV}
\end{equation}
where
$z_{1t}=1$ and $z_{2t}, z_{3t}$ are defined
using $\mu_{2t},\mu_{3t}$ and $\eta_{2t}, \eta_{3t}$ as in (\ref{MC:rz2}).

We  consider  two simulation  models.   Model \ref{MC:FTVmodel1} assumes deterministic parameters and scale factors, while Model \ref{MC:FTVmodelmixb} combines  deterministic and  stochastic  parameters and scale factors.

\begin{model}\label{MC:FTVmodel1}$y_t$ follows (\ref{MC:modelTV}) with $\varepsilon_t$ as
 in (\ref{GARCHnoise}). The scale factors $h_t,  g_{2t},g_{3t}$   and parameters
 $\beta_{1t}, \beta_{2t},\beta_{3t}$
 are deterministic:
 \begin{eqnarray*}
h_{t}&=&0.5 \sin(2\pi t/n)+1, \quad g_{2t}=g_{3t}=0.5\sin(\pi t/n)+1.
\\ \beta_{1t}&=&0.5\sin(0.5\pi t/n)+1, \quad 
\beta_{2t}=0.5\sin(\pi t/n)+1,\quad \beta_{3t}= 0.5\sin(2\pi t/n)+1.
\end{eqnarray*}
\end{model}
\begin{model}
\label{MC:FTVmodelmixb}$y_t$ follows (\ref{MC:modelTV}) with $\varepsilon_t
\sim i.i.d. \,\mathcal{N} (0,1) $
and  scale factors:
\begin{equation*}
h_{t}= 0.5 \sin(2\pi t/n)+1, \quad g_{2t}=\Big|n^{-\gamma}\sum\limits_{i=1}^{t}\zeta_{j}\Big|+0.2, \quad g_{3t}=0.5 \sin(\pi t/n)+1.
\end{equation*}
Parameters $\beta_{1t},\beta_{2t}$ are   the  same as in Model \ref%
{MC:FTVmodel1}, while $\beta_{3t}$ is  stochastic:
\begin{equation*}
 \beta_{3t}=\Big|n^{-\gamma}\sum\limits_{i=1}^{t}\nu_{j}\Big|+0.3(t/n),
\end{equation*}
\end{model}
\noindent where $\{\zeta_{j}\}, \{\nu_{j}\}$  are  stationary  ARFIMA$(0, d, 0)$ processes  with memory parameter $d=0.4$.

\noindent We estimate $\beta_t$ using the estimator $\widehat \beta_t$, (\ref{e:OLS10}), where the weights $b_{n,tj}= K(|t-j|/H) $ are computed
with the Gaussian kernel function $K(x)= (2\pi )^{-1/2}\exp(-x^2/2)$ with
 bandwidth $H=n^h$, $h=0.4, 0.5, 0.6, 0.7$.

\begin{figure}[]
\begin{minipage}[t]{0.32\linewidth}
    \centering
\includegraphics[width=\textwidth]{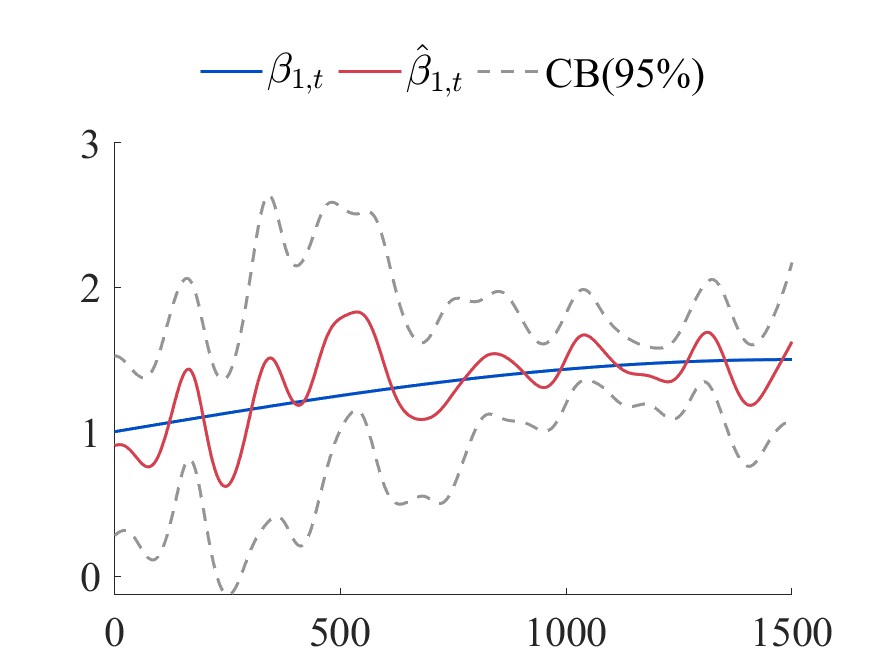}
 \subcaption{$\beta_{1,t}$}
	  \end{minipage}
\begin{minipage}[t]{0.32\linewidth}
    \centering
\includegraphics[width=\textwidth]{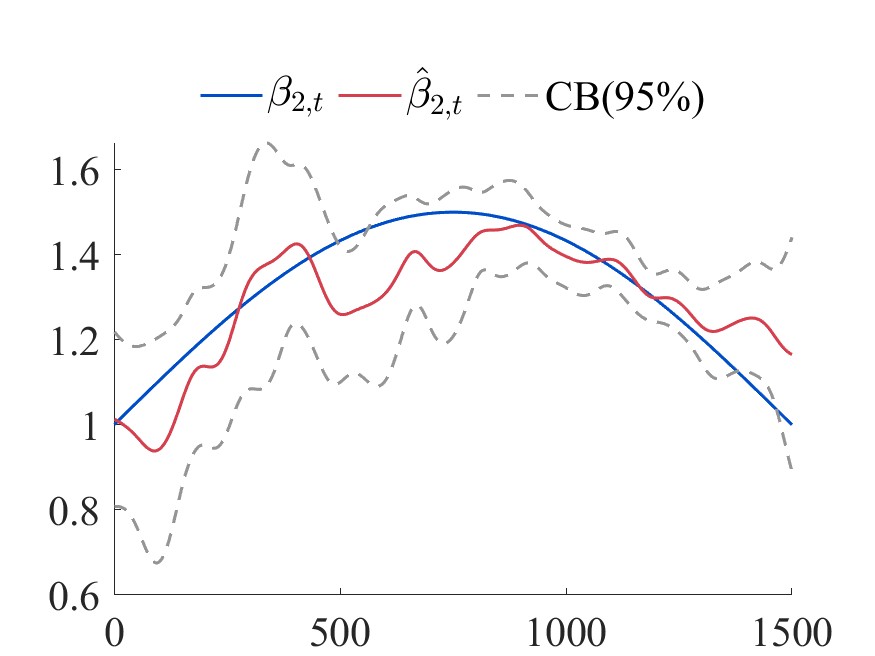}
\subcaption{{$\beta_{2,t}$}}
	  \end{minipage}
\begin{minipage}[t]{0.32\linewidth}
    \centering
\includegraphics[width=\textwidth]{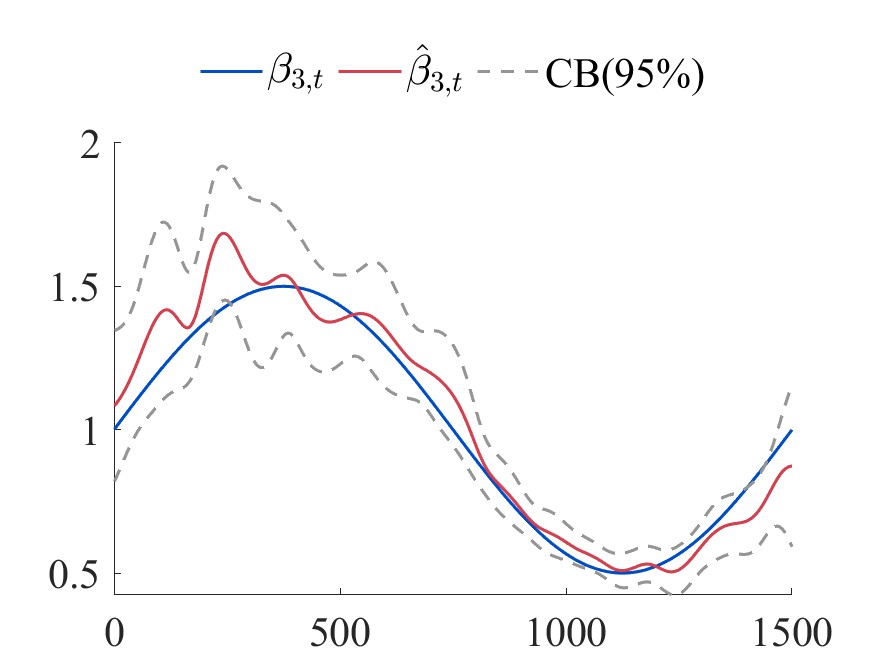}
\subcaption{{$\beta_{3,t}$}}
	  \end{minipage}
\caption{ Robust 95$\%$ confidence intervals for time-varying parameters $\protect\beta_{1t}, \protect\beta_{2t}, \protect\beta_{3t}$ in Model \protect
\ref{MC:FTVmodel1}: $n=1500$, bandwidth $H=n^{0.5}$. Single   replication.}
\label{WDDA2estimator5}
\end{figure}

\begin{figure}[]
\begin{minipage}[t]{0.32\linewidth}
    \centering
\includegraphics[width=\textwidth]{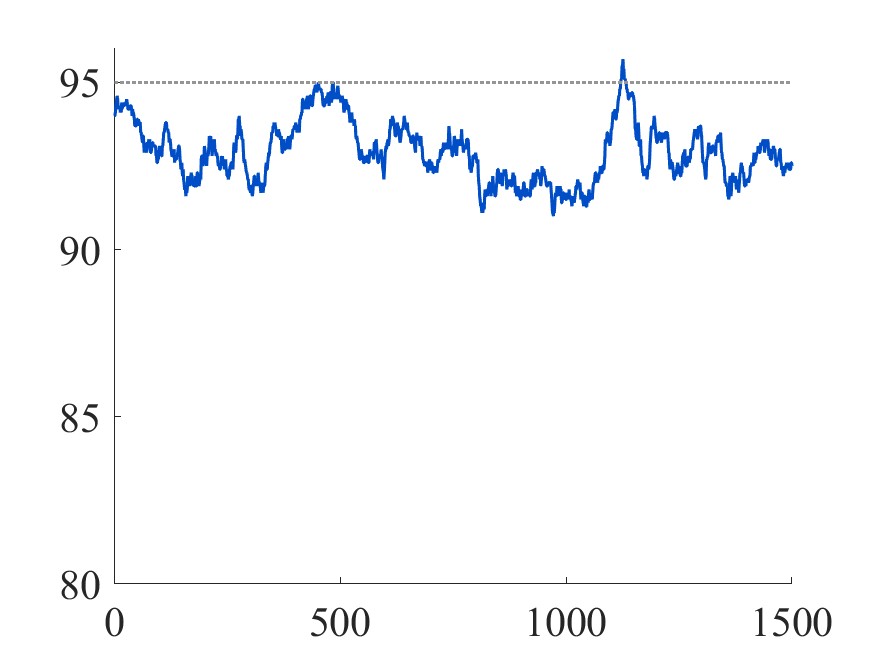}
\subcaption{$\beta_{1,t}$}
	  \end{minipage}
\begin{minipage}[t]{0.32\linewidth}
    \centering
\includegraphics[width=\textwidth]{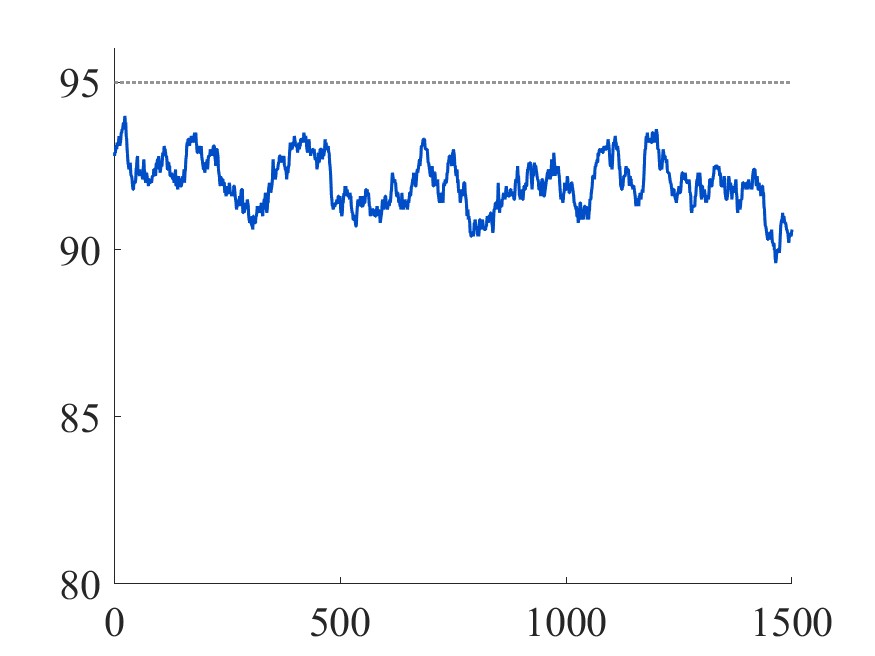}
\subcaption{{$\beta_{2,t}$}}
	  \end{minipage}
\begin{minipage}[t]{0.32\linewidth}
    \centering
\includegraphics[width=\textwidth]{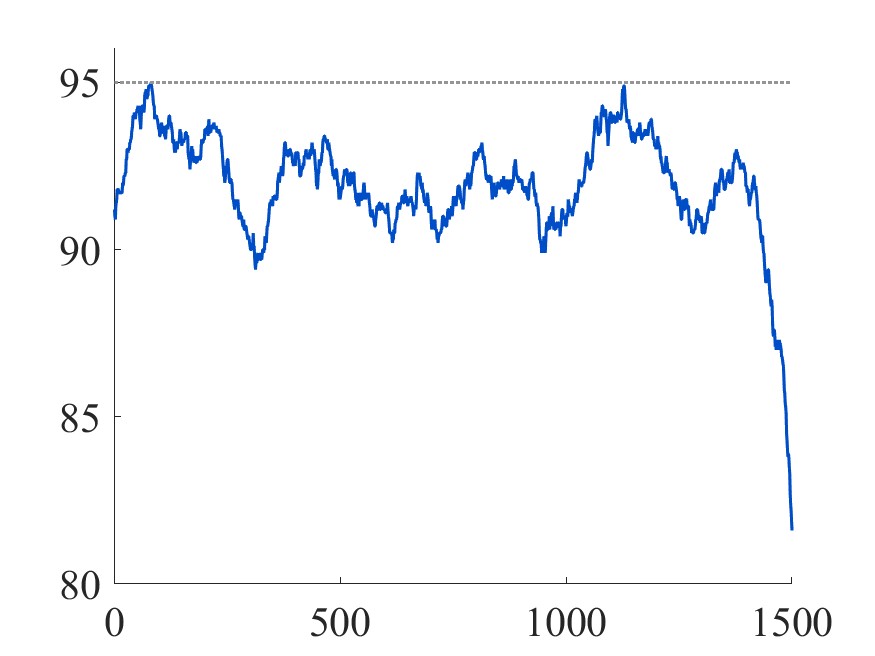}
\subcaption{{$\beta_{3,t}$}}
	  \end{minipage}
\caption{ Coverage rates (in \%) of robust confidence intervals for time-varying parameters $\protect\beta%
_{1t}, \protect\beta_{2t}, \protect\beta_{3t}$ in Model \protect\ref{MC:FTVmodel1}: $n=1500$, bandwidth $H=n^{0.5}$. }
\label{WDDA2CP5}
\end{figure}

\begin{figure}[]
\begin{minipage}[t]{0.32\linewidth}
    \centering
\includegraphics[width=\textwidth]{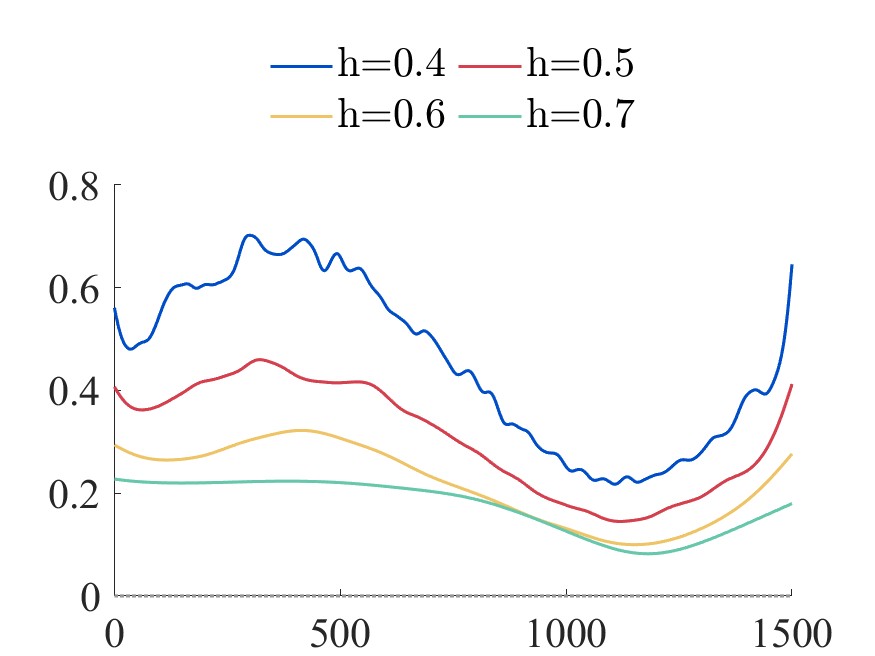}
\subcaption{$\beta_{1,t}$}\label{WDDA2RMSEb1}
	  \end{minipage}
\begin{minipage}[t]{0.32\linewidth}
    \centering
\includegraphics[width=\textwidth]{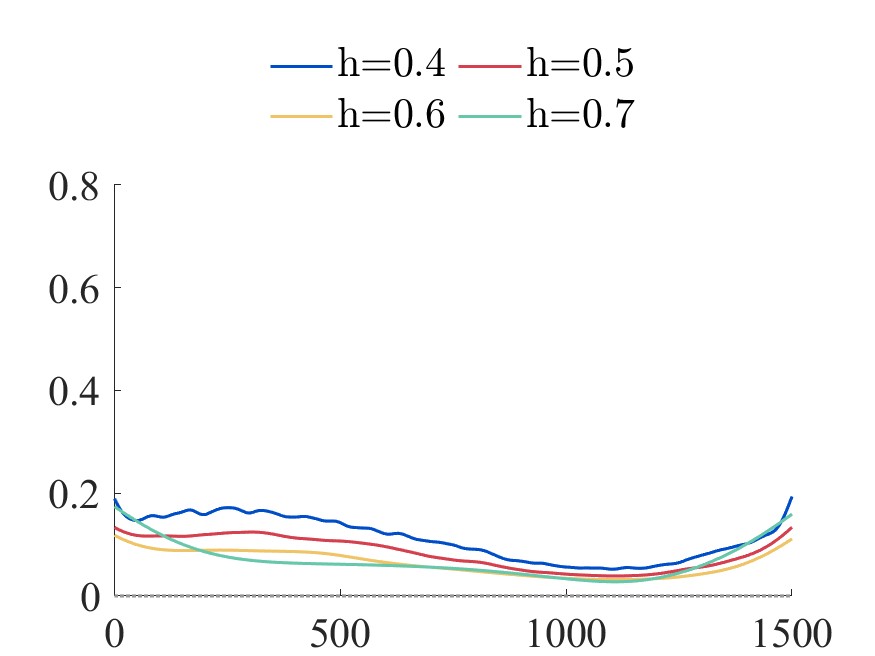}
\subcaption{{$\beta_{2,t}$}}\label{WDDA2RMSEb2}
	  \end{minipage}
\begin{minipage}[t]{0.32\linewidth}
    \centering
\includegraphics[width=\textwidth]{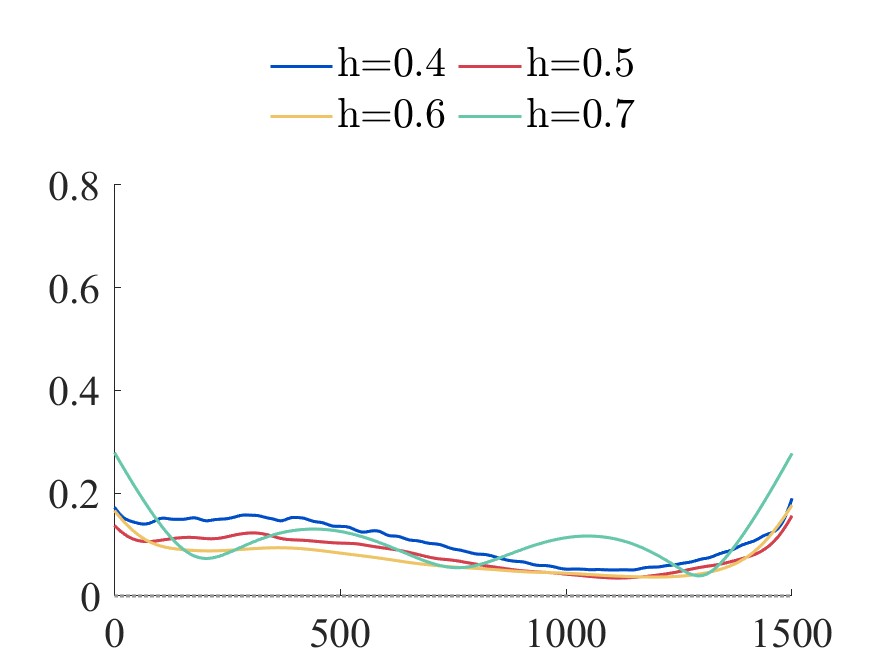}
\subcaption{{$\beta_{3,t}$}}\label{WDDA2RMSEb3}
	  \end{minipage}
\caption{ RMSE for time-varying parameters $\protect\beta_{1t}, \protect\beta_{2t}, \protect\beta_{3t}$ in Model \protect\ref{MC:FTVmodel1}: $n=1500$,
bandwidth $H=n^{h}$, $h={0.4, 0.5, 0.6, 0.7}.$}
\label{WDDA2RMSE}
\end{figure}

Figure \ref{WDDA2estimator5} displays parameter estimation results for a  single
simulation from  Model \ref{MC:FTVmodel1}. It depicts the estimates $\widehat \beta_{k1}, ..., \widehat \beta_{kn}$ (red line) against the true parameters $\beta_{kt}$ (blue line), $k=1,2,3$ obtained with the bandwidth $H=n^{0.5}$, and their point-wise $95\%$  confidence intervals (grey dashed lines),
computed using the robust standard errors. The robust time-varying confidence intervals cover the true parameters $\beta_{kt}$, $t=1,...,n$, for
most of the time points.

Figure \ref{WDDA2CP5} reports  the point-wise empirical coverage rates (blue line) in time-varying robust  estimation of parameters  $\beta_{kt}, k=1,2,3$ which
are close to the nominal  $95\%$  for most of the time points. Figure \ref{WDDA2RMSE} shows the RMSE's  for different choices of the bandwidth $H=n^h$, $h=0.4, 0.5, 0.6, 0.7$. As expected,  the RMSE depends on the smoothness of the parameter $\beta_{kt}$ and often is minimized by moderately large values of $H$,
for  example, $H=n^{0.6}$.

\begin{figure}[]
\begin{minipage}[t]{0.32\linewidth}
    \centering
\includegraphics[width=\textwidth]{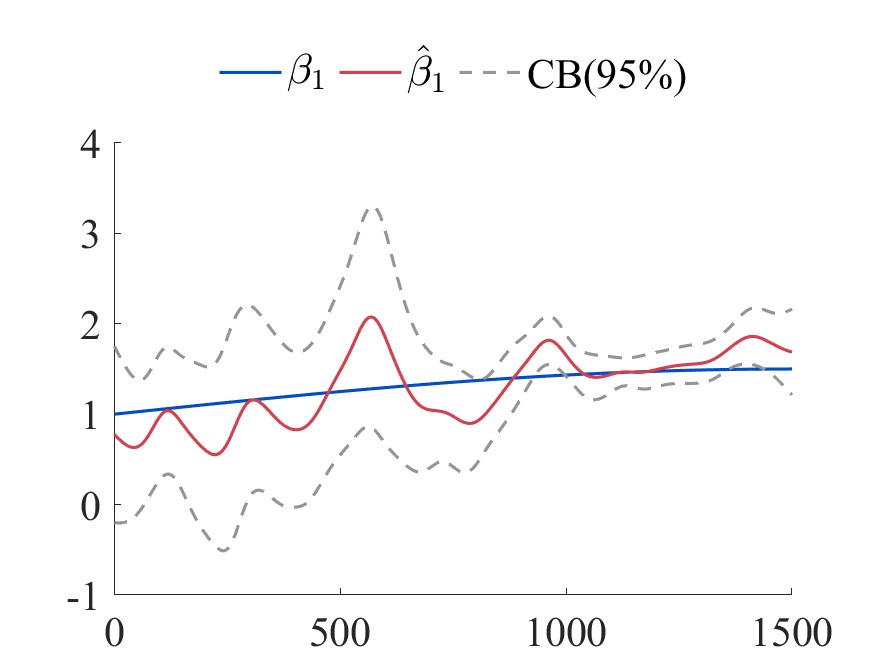}
\subcaption{$\beta_{1,t}$}
	  \end{minipage}
\begin{minipage}[t]{0.32\linewidth}
    \centering
\includegraphics[width=\textwidth]{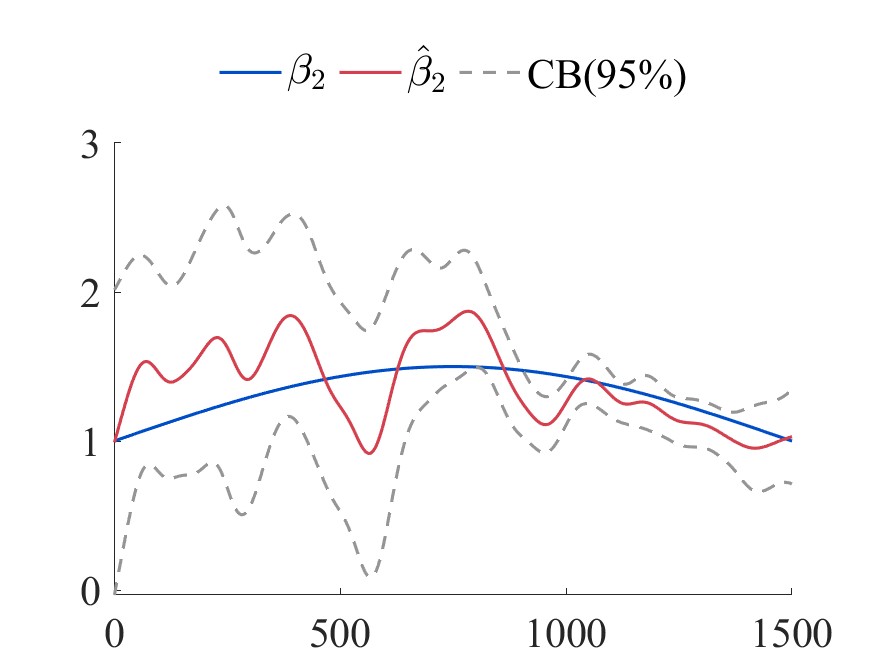}
\subcaption{{$\beta_{2,t}$}}
	  \end{minipage}
\begin{minipage}[t]{0.32\linewidth}
    \centering
\includegraphics[width=\textwidth]{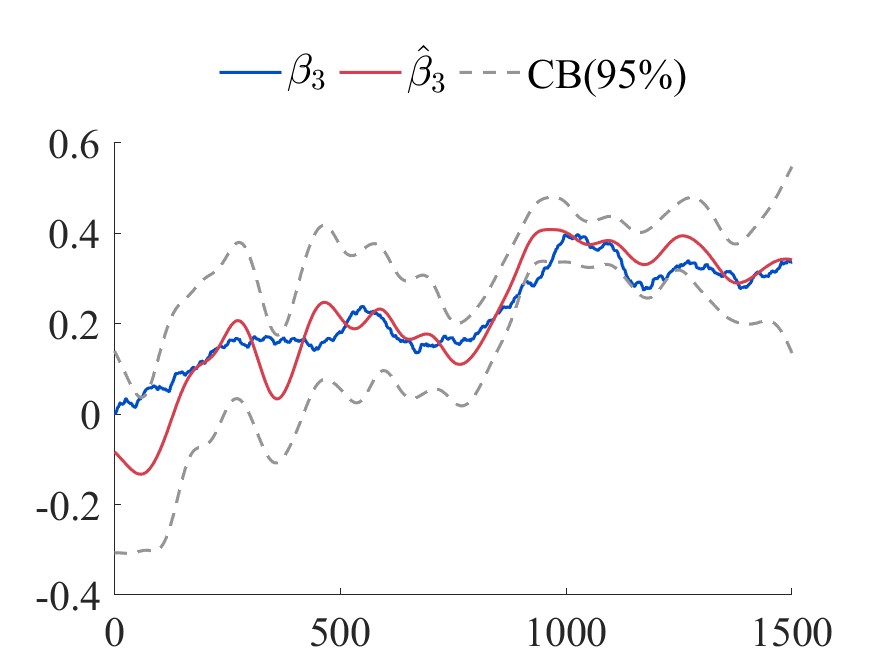}
\subcaption{{$\beta_{3,t}$}}\label{MIXE5estimator5b3}
	  \end{minipage}
\caption{ Robust 95$\%$ confidence bands for time-varying parameters $\protect\beta_{1t}, \protect\beta_{2t}, \protect\beta_{3t}$ in Model \protect
\ref{MC:FTVmodelmixb}: $n=1500$, bandwidth $H=n^{0.5}$. Single  replication.}
\label{MIXE5estimator5}
\end{figure}

\begin{figure}[]
\begin{minipage}[t]{0.32\linewidth}
    \centering
\includegraphics[width=\textwidth]{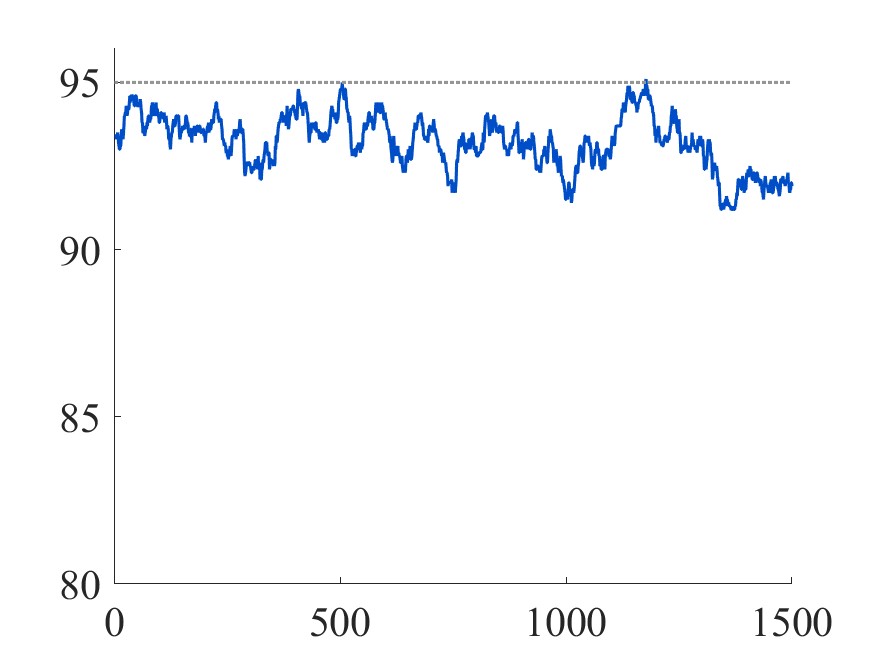}
\subcaption{$\beta_{1,t}$}
	  \end{minipage}
\begin{minipage}[t]{0.32\linewidth}
    \centering
\includegraphics[width=\textwidth]{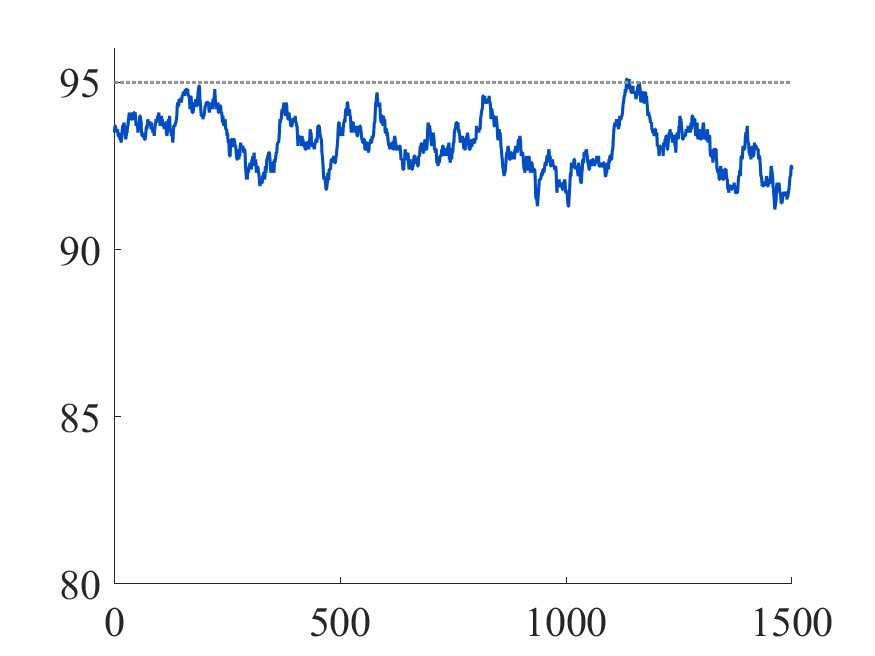}
\subcaption{{$\beta_{2,t}$}}
	  \end{minipage}
\begin{minipage}[t]{0.32\linewidth}
    \centering
\includegraphics[width=\textwidth]{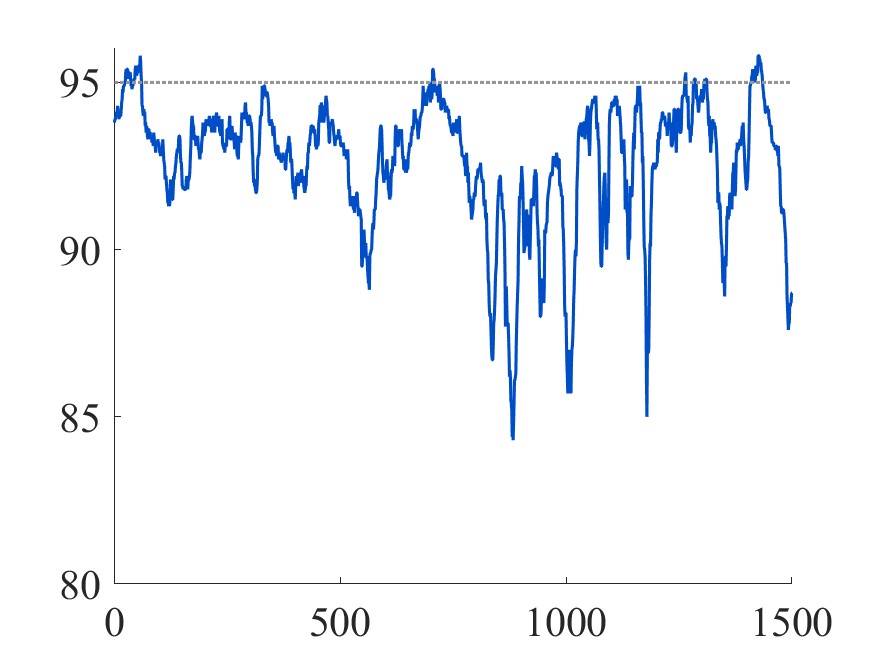}
\subcaption{{$\beta_{3,t}$}}\label{MIXE5CP5b3}
	  \end{minipage}
\caption{ Coverage rates (in \%) of robust confidence intervals for time-varying parameters $\protect\beta%
_{1t}, \protect\beta_{2t}, \protect\beta_{3t}$ in Model \protect\ref{MC:FTVmodelmixb}: $n=1500$, bandwidth $H=n^{0.5}$.
}
\label{MIXE5CP5}
\end{figure}

Figure \ref{MIXE5estimator5} reports estimation results for a single simulation from
Model \ref{MC:FTVmodelmixb}, and Figure \ref{MIXE5CP5} displays point-wise
empirical coverage rates for  robust $95\% $ confidence intervals. For deterministic
parameters $\beta_{1t}$ and $\beta_{2t}$, estimation quality is good and
results are similar to those obtained for Model \ref{MC:FTVmodel1}. For the
stochastic parameter $\beta_{3t}$, the  robust point-wise confidence intervals 
cover  the path of stochastic parameter $\beta_{3t}$ for most of  the  time points, see Figure \ref{MIXE5estimator5}(\subref{MIXE5estimator5b3}). Figure \ref{MIXE5CP5}(\subref{MIXE5CP5b3})  shows  that coverage rates of robust time-varying confidence intervals for $\beta_{3t}$ might  be slightly affected by stochastic variation  in the parameter and scale factors. Nevertheless, they are still satisfactory and reasonably close to the nominal $95\%$ coverage.

\subsection{Estimation of regression parameter with missing data}\label{s:midata}

\begin{table}[]
\caption{Robust OLS estimation
in Model \protect\ref{MC:OLSmodel1} with  block missing data (Type 1).}
\label{tab:OLSm1missblk}\centering
\begin{tabular}{cccccc}
\hline Parameters & Bias & RMSE & CP & CP$_{st}$ & SD \\
\hline $\beta_{1}$ & -0.00818 & 0.04983 & 94.60 & 74.60 & 0.04915 \\
$\beta_{2}$ & 0.00356 & 0.03875 & 94.00 & 67.90 & 0.03859 \\
$\beta_{3}$ & 0.00246 & 0.03840 & 93.80 & 70.00 & 0.03832 \\
\hline &  &  &  &  &
\end{tabular}
\end{table}

\begin{table}[]
\caption{Robust OLS estimation in Model \protect\ref{MC:OLSmodel1} with randomly missing data (Type 2).}
\label{tab:OLSm1missrnd}\centering
\begin{tabular}{cccccc}
\hline Parameters & Bias & RMSE & CP & CP$_{st}$ & SD \\
\hline $\beta_{1}$ & -0.00567 & 0.05732 & 94.30 & 66.60 & 0.05704 \\
$\beta_{2}$ & 0.00144 & 0.04251 & 95.20 & 63.50 & 0.04249 \\
$\beta_{3}$ & 0.00289 & 0.04128 & 94.80 & 64.70 & 0.04118 \\
\hline &  &  &  &  &
\end{tabular}
\end{table}

To examine the impact of missing data on  the robust and  standard OLS
estimation
based on partially observed data
$(y_{j_1},z_{j_1}), (y_{j_2},z_{j_2}), ...., (y_{j_N},z_{j_N}),$
we  use two types of missing data patterns over the time period $1,
..., 1500$.
\vskip.2cm
\noindent  {\it Type 1}.
The block of data $j\in [650, \,850]$ is missing.

\noindent {\it Type 2}. $500$ single observations are missing at randomly selected times.

\vskip.2cm
Tables \ref{tab:OLSm1missblk} and \ref{tab:OLSm1missrnd} report  robust  and standard estimation results for Model \ref{MC:OLSmodel1} with fixed  parameter.
Table \ref{tab:OLSm1missblk} shows  that block missing data (Type 1) do not lead to noticeable changes in Bias, RMSE and SD, and the coverage rate for
robust confidence intervals remains around $95\%$. At  the  same time, the coverage rate CP$_{st}$ of the  standard confidence intervals is   substantially distorted.

Table \ref{tab:OLSm1missrnd} shows that randomly missing data do  not affect  the coverage rate of robust confidence intervals which remains to the nominal $95\%$, while the coverage rate of  the  standard confidence intervals drops to  around  $65\%$. This emphasises the flexibility of the robust OLS estimation of the fixed parameter in the presence of block or randomly missing data.

\begin{figure}
\begin{minipage}[t]{0.32\linewidth}
    \centering
\includegraphics[width=\textwidth]{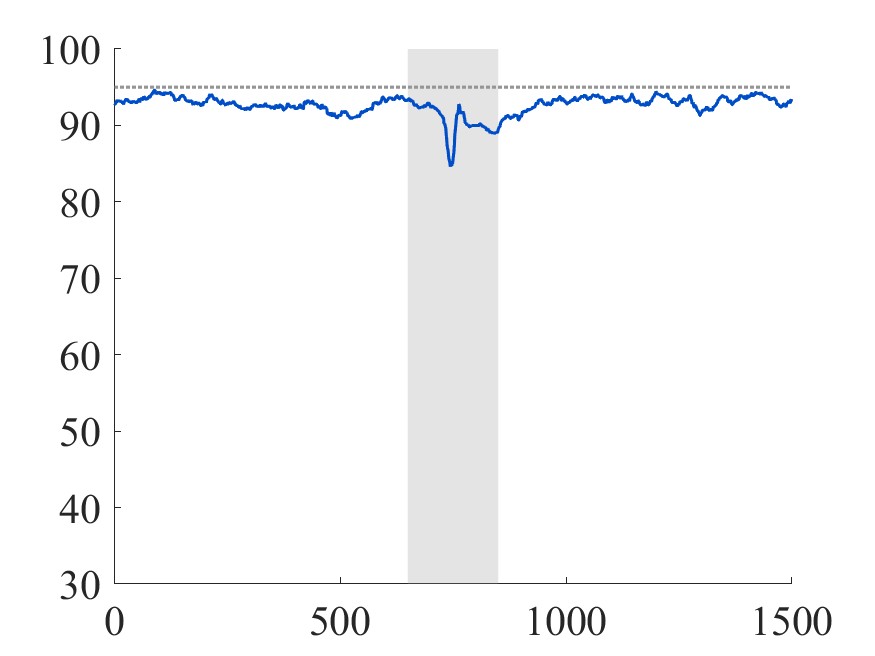}
\subcaption{$\beta_{1,t}$}
	  \end{minipage}
\begin{minipage}[t]{0.32\linewidth}
    \centering
\includegraphics[width=\textwidth]{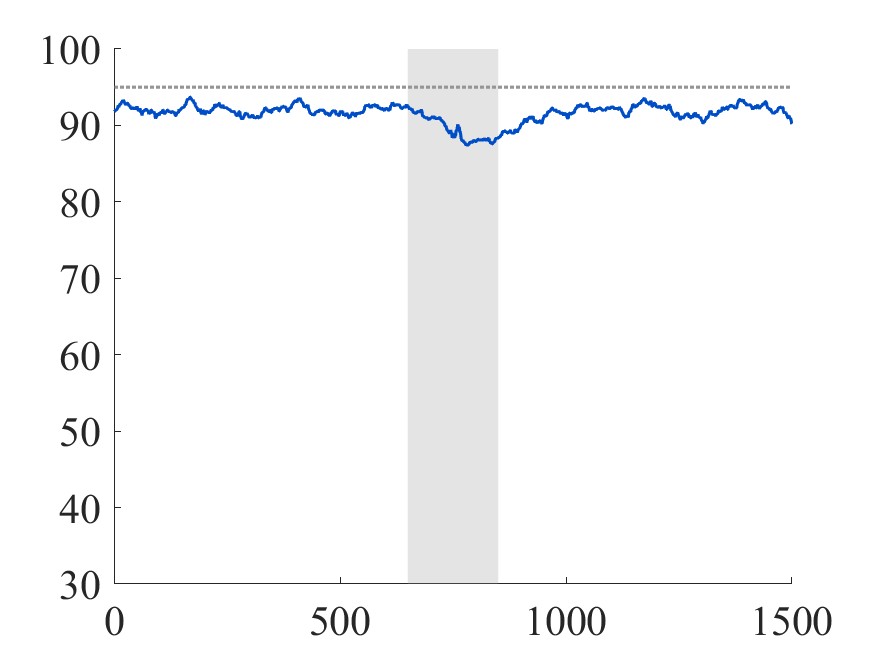}
\subcaption{{$\beta_{2,t}$}}
	  \end{minipage}
\begin{minipage}[t]{0.32\linewidth}
    \centering
\includegraphics[width=\textwidth]{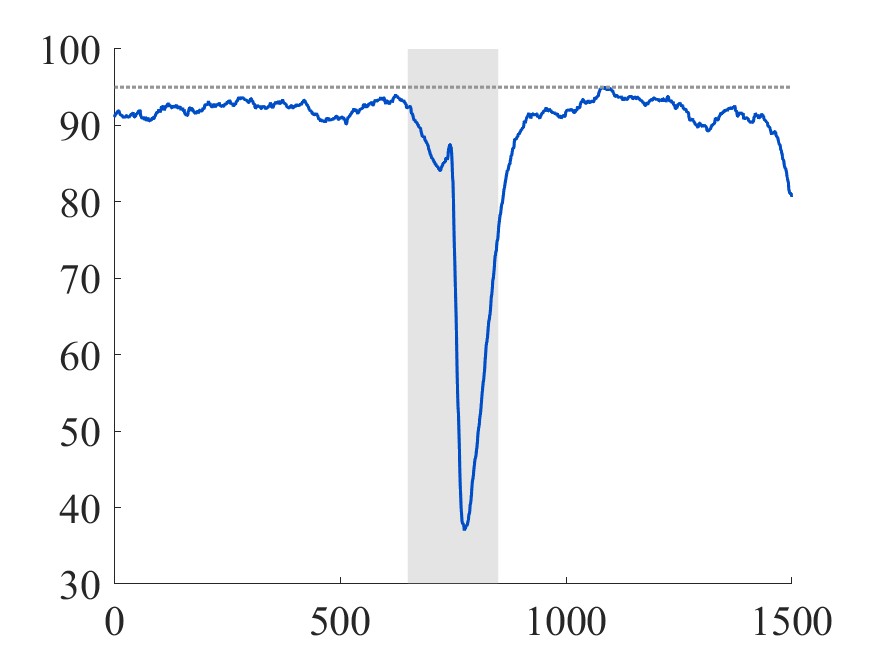}
\subcaption{{$\beta_{3,t}$}}
	  \end{minipage}
\caption{Coverage rates  (in $\%$)
 of robust confidence intervals
for time-varying parameters $\protect\beta%
_{1t}, \protect\beta_{2t}, \protect\beta_{3t}$ in Model \protect\ref{MC:FTVmodel1} with  block missing data (Type 1), $n=1500$, bandwidth $H=n^{0.5}$.  }
\label{WDDA2missblkCP}
\end{figure}

\begin{figure}[]
\begin{minipage}[t]{0.32\linewidth}
    \centering
\includegraphics[width=\textwidth]{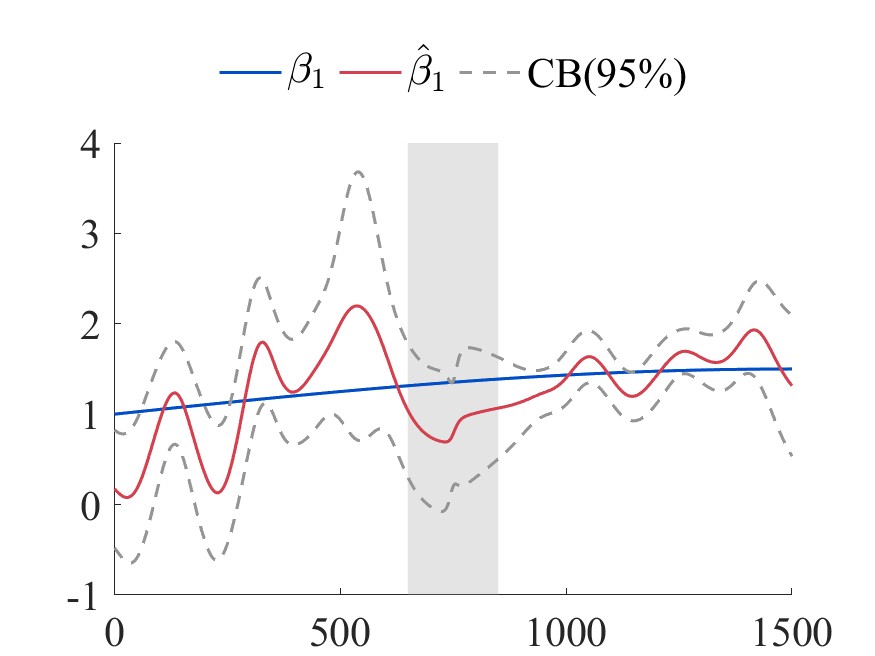}
\subcaption{$\beta_{1,t}$}
	  \end{minipage}
\begin{minipage}[t]{0.32\linewidth}
    \centering
\includegraphics[width=\textwidth]{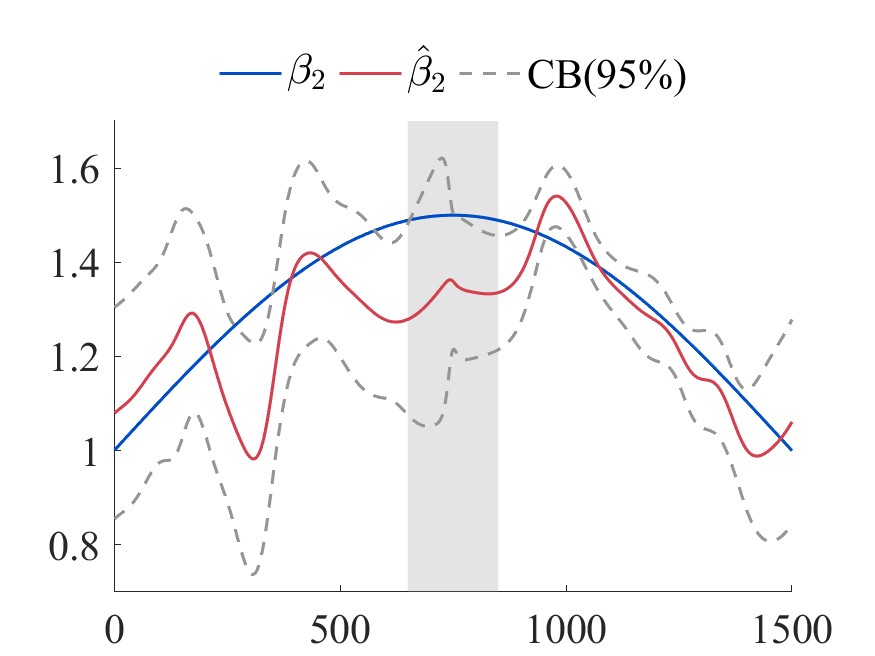}
\subcaption{{$\beta_{2,t}$}}
	  \end{minipage}
\begin{minipage}[t]{0.32\linewidth}
    \centering
\includegraphics[width=\textwidth]{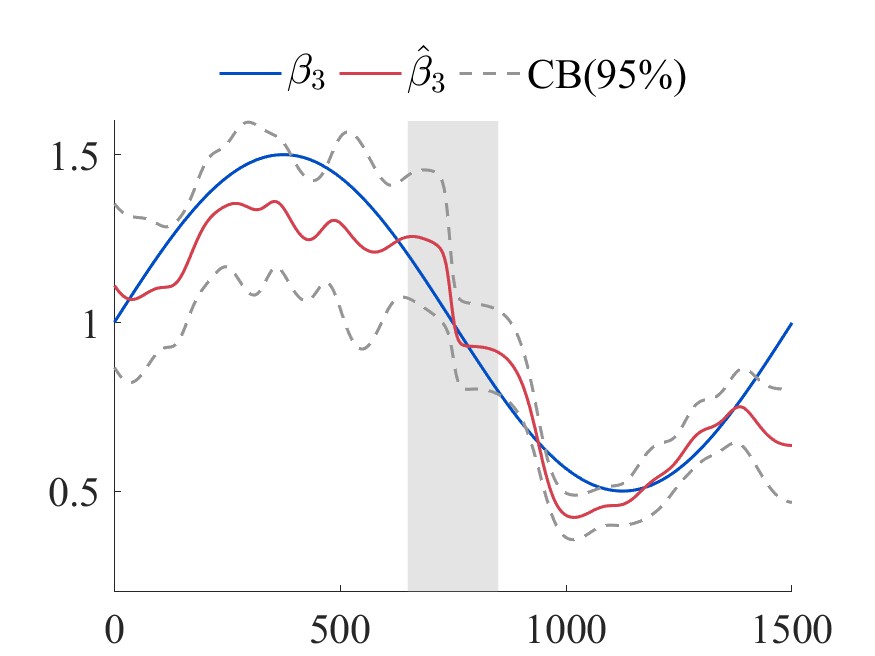}
\subcaption{{$\beta_{3,t}$}}
	  \end{minipage}
\caption{Robust 95$\%$ confidence bands for time-varying parameters $\protect%
\beta_{1t}, \protect\beta_{2t}, \protect\beta_{3t}$ in Model \protect\ref{MC:FTVmodel1} with  block missing data (Type 1), $n=1500$, bandwidth $H=n^{0.5}$. Single  replication.}
\label{WDDA2missblkestimator}
\end{figure}

\begin{figure}[]
\begin{minipage}[t]{0.32\linewidth}
    \centering
\includegraphics[width=\textwidth]{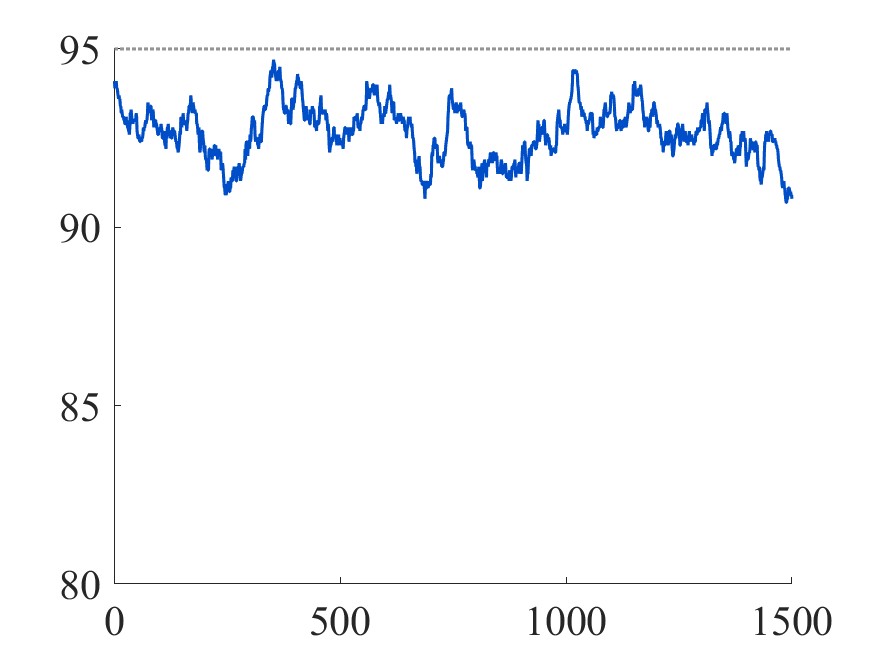}
\subcaption{$\beta_{1,t}$}
	  \end{minipage}
\begin{minipage}[t]{0.32\linewidth}
    \centering
\includegraphics[width=\textwidth]{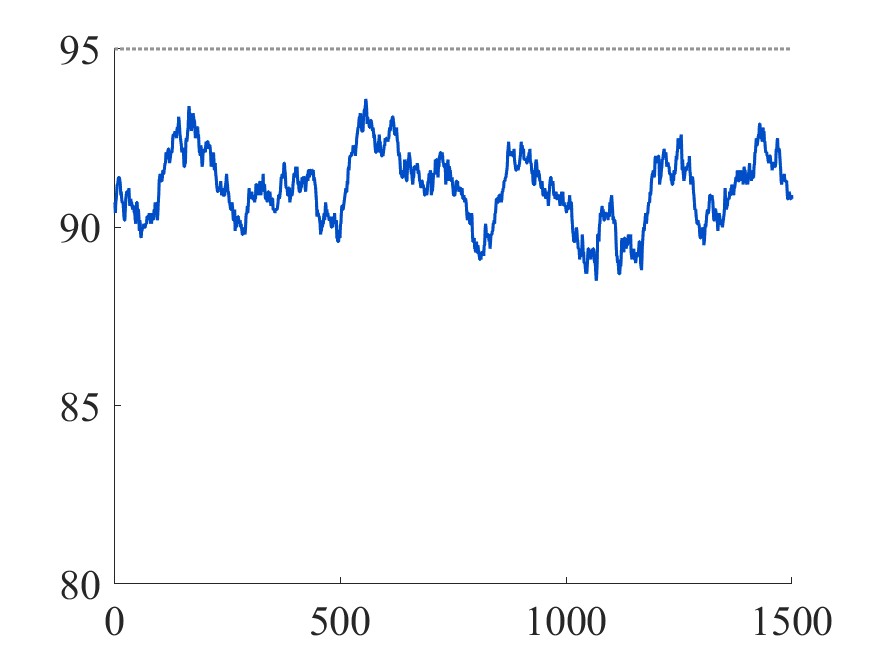}
\subcaption{{$\beta_{2,t}$}}
	  \end{minipage}
\begin{minipage}[t]{0.32\linewidth}
    \centering
\includegraphics[width=\textwidth]{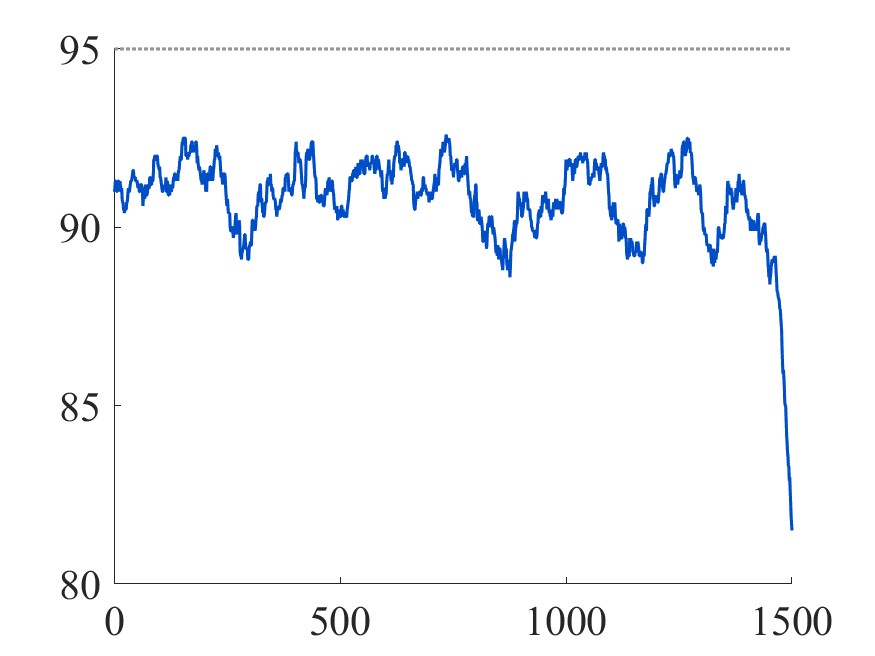}
\subcaption{{$\beta_{3,t}$}}
	  \end{minipage}
\caption{Coverage rates  (in $\%$)
of robust confidence intervals for time-varying parameters $\protect\beta_{1t},
\protect\beta_{2t}, \protect\beta_{3t}$ in Model \protect\ref{MC:FTVmodel1},
 $500$ randomly missing data, $n=1500$, bandwidth $H=n^{0.5}$.}
\label{WDDA2missbrndCP}
\end{figure}

\noindent Figures \ref{WDDA2missblkCP} -- \ref{WDDA2missbrndCP} report
 estimation results for Model \ref{MC:FTVmodel1} with time-varying parameter $\beta_t$.

Figure \ref{WDDA2missblkCP} shows the  coverage rates  in time-varying robust estimation with  block missing data (Type 1, shaded region)  for $t=1,  ..., 1500$.
The coverage is close to the nominal $95\%$, with some distortion for parameters $\beta_{1,t}$ and $\beta_{2,t}$ and a larger distortion for parameter $\beta_{3,t}$ within the  shaded region. The  distortion peaks at the centre of the block, as expected. Although the width of missing data block, $200$, exceeds the bandwidth $H=n^{0.5}=39$ used in  estimating  $\beta_t$, the coverage distortion 
seems to be  offset by the smooth down-weighting of the data,  and the performance of the robust time-varying OLS estimation exceeds  expectations.

Figure \ref{WDDA2missblkestimator} reports the path of the estimator $\widehat\beta_{kt}$ and the point-wise robust confidence intervals, for a single simulation. The robust confidence intervals become wider in the shaded region, which likely explains the satisfactory coverage performance during that period.

Figure \ref{WDDA2missbrndCP} shows that  randomly missing data  (Type 2) do not  distort the robust time-varying OLS  estimation. For all three parameters and time periods $t$, the coverage rate is close to the nominal. Overall, robust estimation of  time-varying parameter does not appear  be  affected  by randomly missing data.

\subsection{Estimation of a stationary AR($p$) model}
\label{s:ARp}
We assess the performance of  the  robust and   standard   procedures in the case  of a stationary AR($2$) model:
\begin{equation}
y_t=\beta_1+\beta_{2}y_{t-1}+\beta_{3}y_{t-2}+\varepsilon_t,  \quad \beta=(\beta_1,\beta_2,
\beta_3)^\prime= (0.5, 0.4 , 0.3)^\prime, \label{e:AR2}
\end{equation}
where  $\varepsilon_t=e_te_{t-1}$,
 $e_t\sim i.i.d. \,\mathcal{N}(0,1)$ is a stationary martingale difference noise.
The regressors $z_t=(z_{1,t},z_{2,t},z_{3,t})^\prime=(1,y_{t-1},y_{t-2})^\prime $ include an intercept and the two past lags of $y_t$. By Theorem \ref{t:AR},  the parameter  $\beta$  can be  estimated  by using the robust  estimation method.

Table \ref{tab:OLSm6} shows  that the coverage rate for the robust OLS estimation   is close to the nominal $95\%$, while   the standard   OLS estimation   exhibits  extensive  coverage  distortion for $\beta_2$ and $\beta_3$.

\begin{table}[]
\caption{Robust OLS estimation in $AR(2)$ model (\protect\ref{e:AR2}).}
\label{tab:OLSm6}\centering
\begin{tabular}{cccccc}
\hline Parameters & Bias & RMSE & CP & CP$_{st}$ & SD \\
\hline $\beta_1$ & -0.00808 & 0.05250 & 94.9 & 92.3 & 0.05187 \\
$\beta_2$ & 0.00104 & 0.04183 & 94.5 & 75.0 & 0.04182 \\
$\beta_3$ & 0.00356 & 0.03091 & 94.8 & 88.8 & 0.03070 \\
\hline &  &  &  &  &
\end{tabular}
\end{table}

\section{Empirical experiment}\label{Ch2:empirical} 
In this section, we analyze the structure and dynamics of daily S$\&$P 500 log returns, $r_t$, from 02/01/1990 to 31/12/2019, (sample size $n=7558$). We employ robust regression estimation to assess whether the returns
$r_t$ can  be modelled  using  a time-varying regression model
of the  form
\begin{equation}
r_t=\mu_t+u_t,\quad u_t=h_t\varepsilon_t,  \label{SP500TVOLSreturn}
\end{equation}
where $\{\varepsilon_t\}$ is an i.i.d.$(0,1)$ noise, and the time-varying
mean and scale factor  $\mu_t,h_t$ are independent of $\{\varepsilon_t\}$. Our objective is to estimate the time-varying mean $\mu_t$, the scale factor $h_t$,
 and to test for the absence of autocorrelation in the  absolute residuals $|u_t|=h_t|\varepsilon_t|$, thereby assessing the fit of the model (\ref{SP500TVOLSreturn}) to the data.

It returns $r_t$  follows the model (\ref%
{SP500TVOLSreturn})   with i.i.d. noise $\varepsilon_t$, then  the  absolute residuals $|u_t|$'s are uncorrelated  then for $t\ne s$:
\begin{eqnarray*}
\mathrm{cov}(|u_t|, |u_s|)&=&\mathrm{cov}(h_t|\varepsilon_t|,
h_s|\varepsilon_s|)=
E\big[h_th_s\mathrm{cov}(|\varepsilon_t|,|\varepsilon_s|)\big]=0.
\end{eqnarray*}
Conversely, if the noise
$\varepsilon_t$ exhibits ARCH effects (stationary conditional
heteroskedasticity), the  sequence  $|u_t|$ becomes  autocorrelated, and the null hypothesis of  uncorrelated absolute residuals $|u_t|$ would be rejected.

\begin{figure}[]
\begin{minipage}[t]{0.48\linewidth}
    \centering
\includegraphics[width=\textwidth,height=5cm]{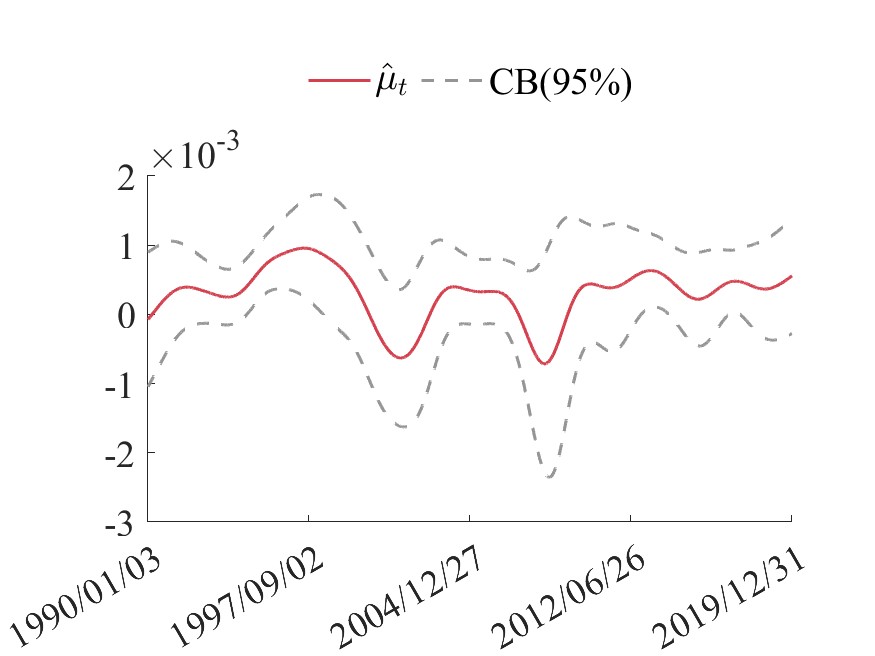}
\subcaption{Confidence bands for
 $\mu_t$}\label{ep:CI_u}
	  \end{minipage}
\begin{minipage}[t]{0.48\linewidth}
    \centering
\includegraphics[width=\textwidth,height=5cm]{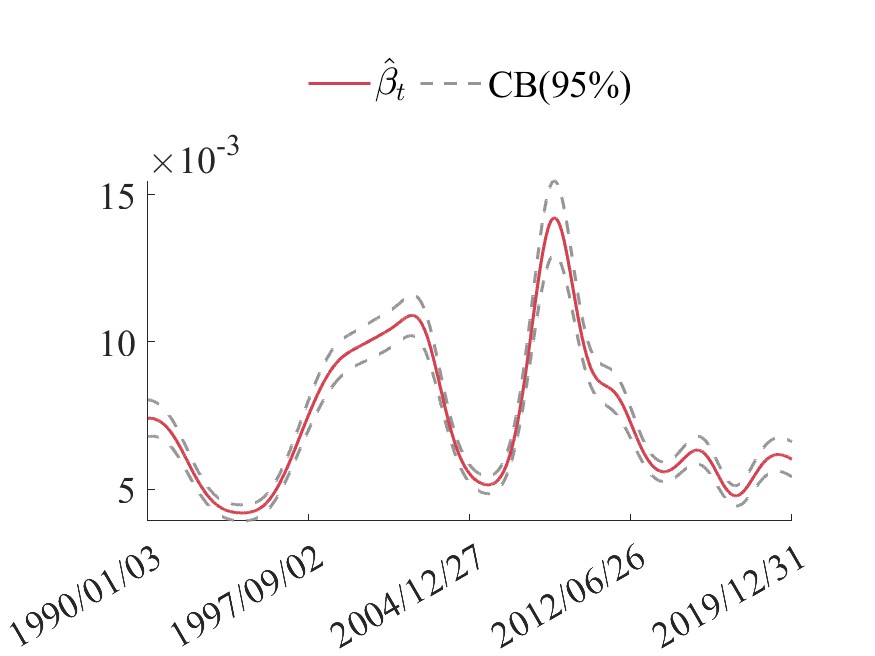}
\subcaption{Confidence bands for
$\beta_{1t}=h_tE|\varepsilon_t|$}\label{ep:CI_b}
	  \end{minipage}
\caption{ Robust $95\%$ confidence bands for $\protect\mu_t$ in model (
\protect\ref{SP500TVOLSreturn}) and $\protect\beta_{1t}=h_tE|\protect
\varepsilon_t|$ in model (\protect\ref{SP500TVOLSrest2}), $n=7558$, $
H=n^{0.6}$.}
\label{SP500mu}
\end{figure}

We estimate the the time varying mean $\mu_{t}$ using the time-varying OLS estimator with bandwidths $H=n^{0.4}, n^{0.5}, ..., n^{0.7}$.
Figure \ref{SP500mu}(\subref{ep:CI_u}) shows  the estimated path of $\widehat \mu_{t}$ and  the associated $95\%$ confidence intervals for bandwidth $H=n^{0.6}$ indicating that $\mu_{t}$ is very likely to change over time.

Assumption (\ref{SP500TVOLSreturn}) implies that
\begin{equation*}
|u_t|=|r_t- \mu_t| = h_t|\varepsilon_t|=
h_tE|\varepsilon_t|+h_t(|\varepsilon_t|-E|\varepsilon_t|).
\end{equation*}
Therefore, $|\widehat u_t|=|r_t-\widehat \mu_t| \sim
h_tE|\varepsilon_t|+h_t(|\varepsilon_t|-E|\varepsilon_t|)$ and thus $
y_t=|\widehat u_t|$ follows a time-varying regression model of the  form
\begin{equation}
y_t=\beta_{1t}+\widetilde u_t, \quad \widetilde u_t=g_t\eta_t,
\label{SP500TVOLSrest2}
\end{equation}
where $\beta_{1t}=h_tE|\varepsilon_t|$ represents a time-varying intercept,  $
g_{t}=h_t $ denotes the  scale factor, and $\eta_t=|\varepsilon_t|-E|\varepsilon_t|$
is an i.i.d. noise. Hence $\beta_{1t}$ can be consistently estimated using the time-varying OLS estimator $\widehat\beta_{1t}$. Figure \ref{SP500mu}(\subref{ep:CI_b}) displays the estimated path of $\widehat{\beta}_{1t}$ and the corresponding $95\%$ confidence intervals  for $\beta_{1t}=h_tE|\varepsilon_t|$ with bandwidth $H=n^{0.6}$, revealing pronounced time variation in the scale factor $h_t$.

\begin{figure}[]
\begin{minipage}[t]{0.48\linewidth}
    \centering
\includegraphics[width=\textwidth,height=5cm]{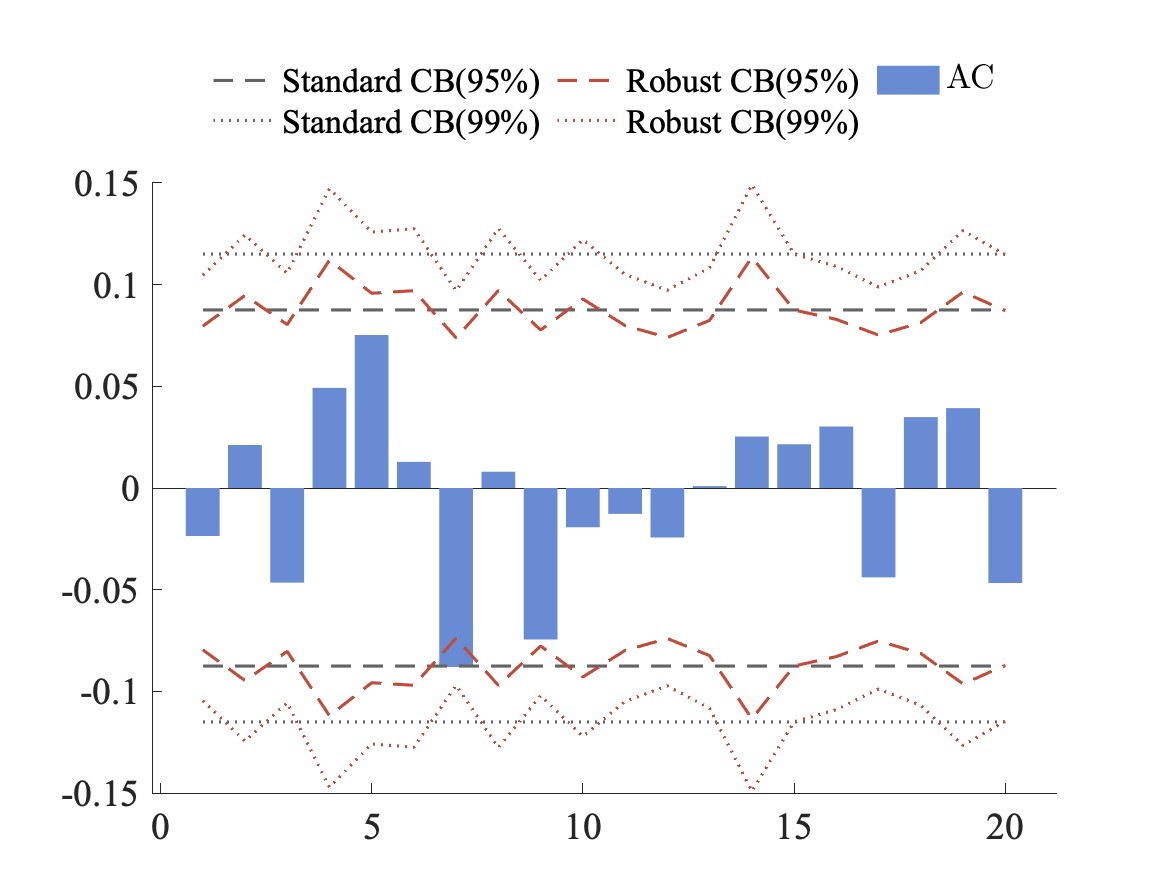}
\subcaption{Testing at individual lag: $\widehat{\widetilde{u}}_j$}\label{SP500ResCorr_u1_CH}
	  \end{minipage}
\begin{minipage}[t]{0.48\linewidth}
    \centering
\includegraphics[width=\textwidth,height=5cm]{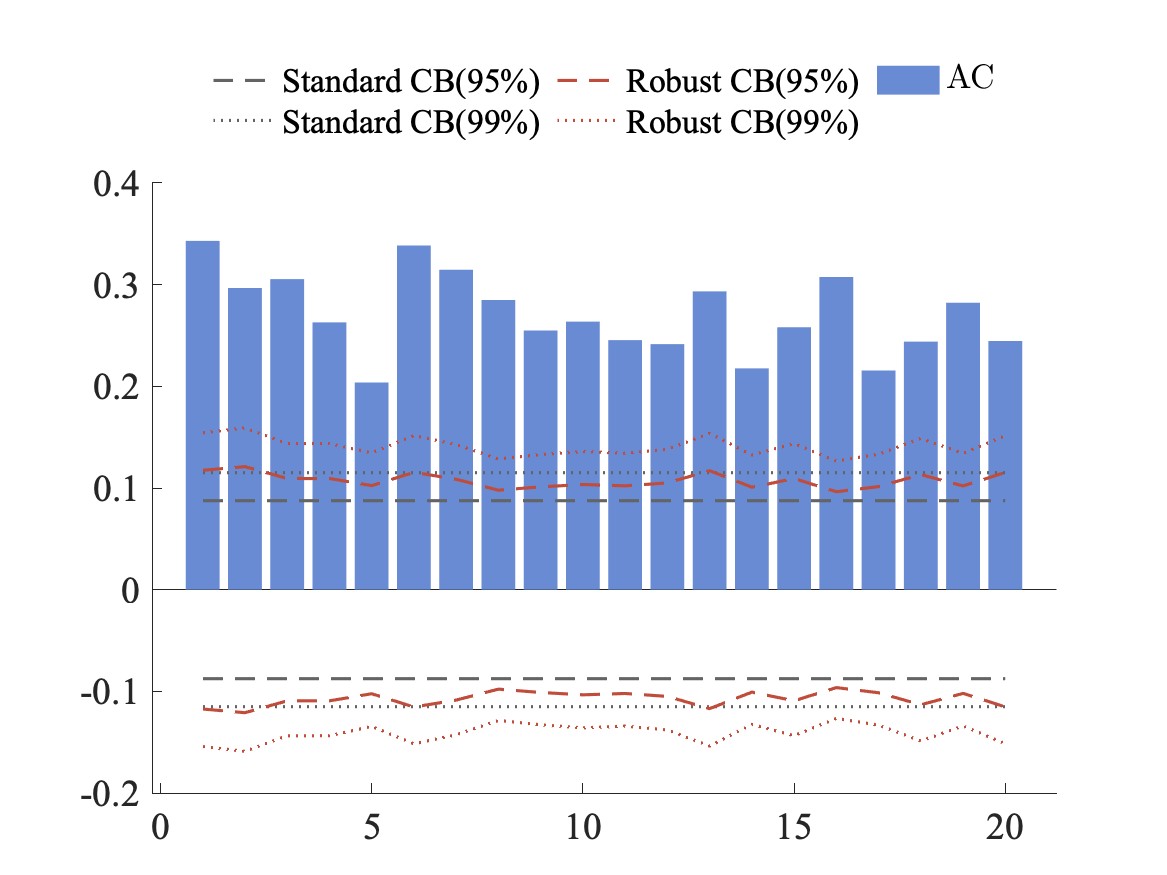}
\subcaption{Testing at individual lag: $\widehat u ^*_j$ }\label{SP500ResCorr_u2_CH}
	  \end{minipage}
\caption{ Robust and standard tests for absence of correlation in subsample
of residuals $\protect\widehat{\protect\widetilde{u}}_j$, $\protect\widehat 
u^{*}_j$, $j\in[500,1000]$, $H=n^{0.6}$, significance level $5\%$.}
\label{SP500ResCorr}
\end{figure}

Figure \ref{SP500ResCorr}(\subref{SP500ResCorr_u1_CH}) reports testing results for zero correlation at lags $k=1, ...,20$ in the residual sequence $\widehat{\widetilde{u}}_t=y_t-\widehat \beta_{1t}$.
We employ the standard test and robust test procedures developed in \cite{glp2024}. Given that the sample size  is large ($n=7558$) and $\beta_{1t}$ is
estimated non-parametrically with bandwidth $H=n^{0.6}$, we  restrict
the correlation analysis to the subsample $j\in[500,1000]$. Both tests provide no evidence of significant correlation within this subsample,  suggesting that
the model (\ref{SP500TVOLSreturn}) fits the returns $r_t$ well during this time period.

The same is not likely to be true  if $r_t^*=r_t-\widehat \mu_t$ follows a
GARCH(1,1) process, as confirmed by the following experiment.
We fit a GARCH(1,1) model to the  demeaned returns $r_t^*=r_t-\widehat \mu_t$,
\begin{eqnarray*}
r^*_t=\sigma_t\varepsilon_t,\quad \sigma^2_t=1.563\times 10^{-6}+0.88913
\sigma^2_{t-1}+0.096974r^{*\,2}_{t-1}.
\end{eqnarray*}
We generate a simulated GARCH(1,1) sample $r^*_{g 1}, ...., r^*_{g n}$,
apply the regression model (\ref{SP500TVOLSrest2})  to the absolute  values
 $y_t^*=|r^*_{g t}|$, and compute the residuals, $\widehat{u}^*_t=y_t^*-\widehat \beta_{1t}$.
Figure \ref{SP500ResCorr}(\subref{SP500ResCorr_u2_CH}) shows that both standard and robust tests detect significant correlation in residuals $\widehat u^*_t$, confirming the presence of conditional heteroskedasticity in the simulated GARCH data.

\section{Conclusion}\label{s:concl}
The robust OLS and time-varying OLS  estimation and inference methods developed in this paper offer considerable flexibility  for modelling economic and financial data. They  allow for  general heterogeneity in regression components  and for structural change of regression coefficients over time.  Moreover, the generalization of the structure of regressors and error terms further expands the range of  empirical settings to which robust OLS regression framework can be applied.  In particular, the  paper develops asymptotic theory for general
regression models with stochastic regressors possibly including a time varying mean,  and provides data-based robust standard errors that enable the  construction
of  confidence intervals for regression parameters. The Monte Carlo  analysis demonstrates the strong performance of the robust estimation approach under complex settings, and confirms the asymptotic normality property and consistency of the  proposed
estimators.

\pagebreak
 \setcounter{equation}{0}
\numberwithin{equation}{section} 

\begin{center}
{\Large Online  Supplement to }
\bigskip

{\Large ``Unlocking the Regression Space" \bigskip\bigskip}

{\large Liudas Giraitis$^1$, George Kapetanios$^2$, Yufei Li$^2$, Alexia Ventouri$^2$ }

\vskip.2cm
{\small   $^1$Queen Mary
University of  London
\\$^2$King's College London
\qquad}

\bigskip
\end{center}

This  Supplement  provides  proofs  of  the results given in the  text  of  the  main paper. It is organised as follows: Section \ref{s:pro}, \ref{s:proTV}, \ref{s:proTVmiss}   provide proofs  of the  main theorems. Section \ref{AuxLem} contains  auxiliary technical  lemmas used in the proofs.

Formula numbering in this  supplement includes the section number, e.g. $(8.1)$,
and references to lemmas are  signified as ``Lemma 10.\#", e.g. Lemma 10.1.
Theorem references to the main paper include section number and are signified, e.g. as  Theorem $2.1$,  while equation references do not  include section number, e.g. $(1)$, $(2)$.

In the proofs, $C$ stands for a generic positive constant which may assume different values in different contexts.

\section{Proofs  of Theorems  \ref{t:r1} and \ref{t:r1-R}, Corollaries \ref{c:co1} and  \ref{c:co1Power}, and Lemma  \ref{l:le}} \label{s:pro}
\noindent {\bf Proof of Theorem \ref{t:r1}}.
Notice  that in view  of   (\ref{e:r1}),
\begin{eqnarray*} 
\widehat \beta-\beta &=&\big(  \sum_{j=1}^n z_j z_j^\prime\big)^{-1}\big(  \sum_{j=1}^n z_j (z_j^\prime \beta+u_j)\big)-\beta
\\&=&S_{zz}^{-1}S_{zu},\quad
S_{zz}= \sum_{j=1}^n z_j z_j^\prime, \quad S_{zu} = \sum_{j=1}^n z_j u_j. \nonumber
\end{eqnarray*}
Recall  definition (\ref{e:Dk}) of $D$ and  $D_g$.
Then
\begin{eqnarray}
D(\widehat \beta-\beta) &=&
(DS_{zz}^{-1}D)(D^{-1}S_{zu})\nonumber \\
&=&(DD_g^{-1})(D_gS_{zz}^{-1}D_g)(D_g^{-1}D)(D^{-1}S_{zu}) =O_p(1),\label{e:R1OLS1}
\end{eqnarray}
since $DD_g^{-1}=O_p(1)$  by (\ref{e:hkj3}) of  Assumption \ref{a:r3},
$D^{-1}S_{zu}=O_p(1)$  by  (\ref{e:daz2-re+})  of  Lemma \ref{l:RegR}.
Moreover, by  (\ref{e:daz2-re}) and (\ref{e:norm-re}),
$$D_gS_{zz}^{-1}D_g=D_gE[S_{zz}\, |\mathcal{F}_n^*]^{-1}D_g+o_p(1)=O_p(1).$$
This  completes the proof of the  consistency claim  (\ref{e:hatb1})  of  the   theorem.  \hfill  $\Box$
\vskip.2cm
\noindent Recall that for $p\times p$ symmetric matrices $A$, $B$ and a $p\times 1$   vector  $b$  it holds:
\begin{eqnarray*}
||AB||_{sp}\le ||A||_{sp}||B||_{sp}, \quad ||AB||\le ||A||_{sp}||B||,\quad ||A||_{sp}\le ||A||,
\end{eqnarray*}
where  $||A||_{sp}$ denotes the  spectral norm  and $||A||$ the  Euclidean  norm  of the matrix~$A$.

\noindent Recall  the  definition of  the information set $\mathcal{F}_n^*=\sigma\big(\mu_t, g_t,h_t,t=1, ..., n\big)$.

\vskip.2cm
 \noindent{\bf Proof  of Theorem \ref{t:r1-R}}. 
 \noindent{\it Proof of (\ref{e:hatb1-R})}.
  By (\ref{e:R1OLS1}),
 \begin{eqnarray*}
D(\widehat \beta-\beta) &=&\{DS_{zz}^{-1}D\} \{D^{-1}S_{zu}\}.
\label{e:R1OLS18}
\end{eqnarray*}
Moreover, by the  same  argument  as  in the proof of  (\ref{e:R1OLS1}),
 \begin{eqnarray}DS_{zz}^{-1}D&=&
(DD_g^{-1})(D_gS_{zz}^{-1}D_g)(D_g^{-1}D)\nonumber\\
&=&(DD_g^{-1})(D_gE[S_{zz}\, |\mathcal{F}_n^*]^{-1}D_g+o_p(1))(D_g^{-1}D)
\nonumber\\&=&DE[S_{zz}\, |\mathcal{F}_n^*]^{-1}D   +o_p(1), \quad DS_{zz}^{-1}D=O_p(1).\label{e:dafr}
\end{eqnarray}
Hence,
\begin{eqnarray}
a^\prime\, D(\widehat \beta-\beta) &=&
a^\prime\{DE[S_{zz}\,|\mathcal{F}_n^*]^{-1}D+o_p(1)\}
\{D^{-1}S_{zu}\}\nonumber\\
&=&d_nS_{zu}+o_p(1), \quad d_n=a^\prime (DE[S_{zz}|\mathcal{F}_n^*]^{-1}).
\label{e:DA}
\end{eqnarray}
By (\ref{e:OMEGA})  of  Lemma \ref{l:RegR},
\begin{eqnarray}\label{e:DA2}
v_n^2:= (a^\prime D\Omega_n D a) \ge b_n, \quad  b_n^{-1}=O_p(1).
\end{eqnarray}
This together  with (\ref{e:DA})  implies:
\begin{eqnarray*}
 \frac{a^\prime D( \widehat \beta -\beta)}{\sqrt{a^\prime D\Omega_n D a }}=v_n^{-1}d_nS_{zu}+o_p(1).
 \end{eqnarray*}
Write
\begin{eqnarray*}
s_n&=&v_n^{-1}d_nS_{zu}=
\sum_{t=1}^n\xi_t, \quad  \xi_t=v_n^{-1}d_nz_tu_t.\nonumber
\end{eqnarray*}
To prove (\ref{e:hatb1-R}), it remains to show   that
 \begin{eqnarray}\label{e:DA4}
s_n \rightarrow _d \mathcal{N}(0, 1).
\end{eqnarray}Notice that $\{ \xi_t\} $  is   an m.d.  sequence  with
respect to the $\sigma$-field

 \noindent $ \mathcal{F}_{n,t}=\sigma(\varepsilon_1, ..., \varepsilon_t; \,\,\mu_s, h_s,g_s, s=1, ..., n)$:
 \begin{eqnarray}\label{e:MD*}
E[\xi_t \, |\mathcal{F}_{n,t-1}]=E[v_n^{-1}d_nz_th_t\varepsilon_t |\mathcal{F}_{n,t-1}]=v_n^{-1}d_nz_th_tE[\varepsilon_t |\mathcal{F}_{n,t-1}]=0.
\end{eqnarray}
The latter  follows noting that the variables
$v_n^{-1},d_n,h_t$  are   $\mathcal{F}_{n,t-1}$-measurable since they are function of  $\mu_s, h_s,g_s, s=1, ..., n$. Similarly, since  $\eta_t$'s are $\mathcal{F}_{n,t-1}$  measurable (see  Assumption  \ref{a:ETA}), the variables
$z_t=\mu_t+I_{gt}\eta$ are  also  $\mathcal{F}_{n,t-1}$-measurable. Finally,   by assumption, $\{\mu_s, h_s,g_s, s=1, ..., n\}$ and $\{\varepsilon_s, s=1, ..., n\}$  are mutually independent,  and therefore $E[\varepsilon_t |\mathcal{F}_{n,t-1}]=
E[\varepsilon_t |\mathcal{F}_{t-1}]=0$ by  Assumption \ref{a:r0}.
This   shows  that the conditional  expectation property $E[\xi_t \, |\mathcal{F}_{n,t-1}]=0$ is  preserved for $\xi_t$ and completes the argument showing  that $\xi_t$ is a martingale  difference sequence with respect to the  $\sigma$-field $\mathcal{F}_{n,t-1}$.

\noindent Therefore, by Corollary 3.1 of \cite{hal1980}, to  prove (\ref{e:DA4}), it  suffices  to show  that
\begin{eqnarray}\label{e:vas1+LB-R}
&&(a) \quad\sum_{t=1}^nE[\xi_{ t}^{ 2}\,|\mathcal{F}_{n,t-1}] \rightarrow_p 1, 
\\
&&(b) \quad\sum_{t=1}^nE[\xi_{ t}^{ 2}I(\xi_{ t}^{ 2}\ge \epsilon)\,|\mathcal{F}_{n,t-1}] =o_p(1) \quad \mbox{ for  any  $\epsilon>0$.}\nonumber
\end{eqnarray}
Observe that (a)  holds  with a  non-random  limit $\eta^2=1$. Thus,
the verification of  the condition (3.21) of Corollary 3.1,  that  the  $\sigma$-fields are nested, ${\cal F}_{n,t}\subset {\cal F}_{n+1,t}$  for $t=1, ..., n$  and  $n \ge  1$, is unnecessary; see  remark  on page 59
in  \cite{hal1980}.
To verify  (a), notice  that
\begin{eqnarray*}
\xi_{ t}^{ 2}&=&(v_n^{-1}d_nz_tu_t)^2=v_n^{-2}d_nz_tz^\prime_td^\prime _nu_t^2,\\
E[\xi_{ t}^{ 2}\,|\mathcal{F}_{n,t-1}]&=&v_n^{-2}d_nz_tz^\prime_td^\prime _nE[u_t^2\,|\mathcal{F}_{n,t-1}]=v_n^{-2}d_nz_tz^\prime_td^\prime _nh_t^2E[\varepsilon_t^2\,|\mathcal{F}_{t-1}].
\end{eqnarray*}
\noindent Then, setting  $S_{zzuu}^{(c)}=\sum_{t=1}^n z_tz^\prime_th_t^2E[\varepsilon_t^2\,|\mathcal{F}_{t-1}]$,
we can write,
\begin{eqnarray}
&&\sum_{t=1}^nE[\xi_{ t}^{ 2}\,|\mathcal{F}_{n,t-1}] =
v_n^{-2}d_n\,S_{zzuu}^{(c)}\,d^\prime _n
\nonumber\\
&&\qquad=v_n^{-2} \,a^\prime \{DE[S_{zz}|\mathcal{F}_n^*]^{-1}D\}\{D^{-1}
S_{zzuu}^{(c)}
D^{-1}\}\{D
E[S_{zz}|\mathcal{F}_n^*]^{-1}D\}a. \label{e:tildenjk-R}
\end{eqnarray}
Recall that  by (\ref{e:dafr}),    $DE[S_{zz}|\mathcal{F}_n^*]^{-1}D=O_p(1)$.
We show in (\ref{e:SuuC}) of  Lemma \ref{l:RegR} that
$$
D^{-1}S_{zzuu}^{(c)} D^{-1}= D^{-1}E[S_{zzuu}|\mathcal{F}_n^*]D^{-1}+o_p(1).
$$
Together  with (\ref{e:DA2}), this  implies
\begin{eqnarray*}
\sum_{t=1}^nE[\xi_{ t}^{ 2}\,|\mathcal{F}_{n,t-1}] &=&
v_n^{-2}\,a^\prime \{D\,E[S_{zz}|\mathcal{F}_n^*]^{-1}\,E[S_{zzuu}|\mathcal{F}_n^*]\,E[S_{zz}|\mathcal{F}_n^*]^{-1}D\}a+o_p(1)
\\
&=&v_n^{-2}(a^\prime D\Omega_nD\, a)+o_p(1)=1+o_p(1)
\end{eqnarray*}
which  proves (a).

Next we  prove (b). We  have
\begin{eqnarray*}
  \xi_t&=&v_n^{-1}d_nz_tu_t=v_n^{-1}(d_nD) (D^{-1}z_tu_t),
  \\
   \xi_t^2&\le&v_n^{-2}||d_nD ||^2||D^{-1}z_tu_t||^2.
    \nonumber
\end{eqnarray*}
By  definition of $d_n$, $||d_nD ||^2=||a^\prime D\,E[S_{zz}|\mathcal{F}_n^*]^{-1}D||^2.$  On the  other hand,
by (\ref{e:bn**}) of Corollary \ref{c:bn}, 
 for  any $a$, 
\begin{eqnarray*}
    a^\prime 
    D^{-1}E[S_{zzuu}|\mathcal{F}^*_n]D^{-1} a&\ge&     b_n||a||^2,\quad b_n^{-1}=O_p(1),
  \end{eqnarray*}
where $b_n$  is  $\mathcal{F}_n^*$  measurable,  and, thus, also $\mathcal{F}_{n,t-1}$  measurable. Then,
\begin{eqnarray*}&&v_n^2=a^\prime D\Omega_n D\,a=\{a^\prime D(E[S_{zz}|\mathcal{F}^*_n])^{-1}D\}\{D^{-1}E[S_{zzuu}|\mathcal{F}^*_n]D^{-1}\}\{D(E[S_{zz}|\mathcal{F}^*_n])^{-1}D \, a\}\nonumber\\
&&\quad \ge  ||a^\prime D(E[S_{zz}|\mathcal{F}^*_n])^{-1}D||^2b_n=||d_nD ||^2b_n, 
\nonumber \\
&& \xi_t^2\le b_n^{-1}||D^{-1}z_tu_t||^2. \nonumber
\end{eqnarray*}
Hence,
\begin{eqnarray*}
\sum_{t=1}^nE[\xi_{ t}^{ 2}I(\xi_{ t}^{ 2}\ge \epsilon)\,|\mathcal{F}_{n,t-1}] \le
\sum_{t=1}^nE\big[b_n^{-1}||D^{-1}z_tu_t||^2I\big(b_n^{-1}||D^{-1}z_tu_t||^2\ge \epsilon\big)\,|\mathcal{F}_{n,t-1}\big]
=o_p(1),
\end{eqnarray*}
by (\ref{e:LIZA2D}) of Lemma \ref{e:LEL}. 
This  completes the proof (b)  and the claim (\ref{e:hatb1-R}) of the theorem.

 \vskip.2cm \noindent
The claim (\ref{e:hatb1-R13})   follows from (\ref{e:hatb1-R}) by setting  $a=(a_1, ..., a_p)^\prime=(0, ...,0,1,0....)^\prime$ where  $a_k=1$  and  $a_j=0$  for  $j \ne  k$. Then $a^\prime D=v_k$ and $a^\prime D\Omega_n D a=v_k^2\omega_{kk}$, where $\omega_{kk}$ is the  $(k,k)$-th diagonal element of  $\Omega_n$. Then,
  $$
  \frac{a^\prime D( \widehat \beta -\beta)}{\sqrt{a^\prime D\Omega_n D a }}
  =\frac{( \widehat \beta -\beta)}{\sqrt{\omega_{kk}}}
  \rightarrow _d \mathcal{N}(0, 1)
  $$
  by (\ref{e:hatb1-R}).
  This  completes  the
proof  of the theorem.
   \hfill$\Box$

\vskip.2cm

\noindent
\noindent {\bf Proof of Corollary \ref{c:co1}}.
We   will  show  that
\begin{eqnarray}\label{e:show}
\frac{ \widehat \omega_{kk}}{ \omega_{kk}}=1+o_p(1)
 \end{eqnarray}
 which together  with (\ref{e:hatb1-R13})  implies (\ref{e:hatb1-Rno}):
\begin{eqnarray*}
 \frac{\widehat \beta_k -\beta_k}{\sqrt{\widehat \omega_{kk}}}&=& \big(\sqrt{\frac{{\omega_{kk}}}{{\widehat \omega_{kk}}}}\,\big)
 \frac{\widehat \beta_k -\beta_k}{\sqrt{ \omega_{kk}}}=(1+o_p(1))\frac{\widehat \beta_k -\beta_k}{\sqrt{ \omega_{kk}}}
 \rightarrow _d \mathcal{N}(0, 1).
\end{eqnarray*}
To  prove (\ref{e:show}), we  will verify that
 \begin{eqnarray}\label{e:omegaO}
 D\widehat  \Omega_n D= D  \Omega_n D+o_p(1)
\end{eqnarray}
which  implies  the  following property  for diagonal elements:
$$
v_k^2 \widehat \omega_{kk}=v_k^2 \omega_{kk}+o_p(1).
$$
 In (\ref{e:OMEGA})  of Lemma \ref{l:RegR}
  it is  shown  that
 \begin{eqnarray}\label{e:bada}
 a^\prime D\Omega_nD a \ge  b_n, \quad a^\prime D\Omega_nD a\le b_{n2}
 \end{eqnarray}
   for    any $a=(a_1, ...,a_p)^\prime$, $||a||=1$ where  $b_n,\,b_{n2}>0$  do  not depend  on $a, n$ and $b_n^{-1}=O_p(1),$
    $b_{n2}=O_p(1).$
Set $a=(0, ...,1,...0)^\prime$,  where $a_j=0$  for  $j \ne  k$  and  $a_k=1$. Then  $a^\prime D\Omega_nD a =v_k^2 \omega_{kk}$, and
by (\ref{e:bada}), $v_k^2 \omega_{kk}\ge b_n>0$. This proves  (\ref{e:show}):
 \begin{eqnarray*}
\frac{ \widehat \omega_{kk}}{ \omega_{kk}}=
\frac{v_k^2 \widehat \omega_{kk}}{v_k^2\omega_{kk}}=\frac{v_k^2 \omega_{kk}+o_p(1)}{v_k^2\omega_{kk}}=1+o_p(1).
\end{eqnarray*}
In addition, the bounds (\ref{e:bada})
 imply that $\sqrt{ \omega_{kk}}\asymp_p v_k^{-1}$:
 \begin{eqnarray*}
 v_k^{-1}\le b_n^{-1/2}\sqrt{ \omega_{kk}}=O_p(\sqrt{ \omega_{kk}}), \quad v_k\sqrt{ \omega_{kk}}=O_p(1), \quad \sqrt{ \omega_{kk}}=O_p( v_k^{-1}).
\end{eqnarray*}

\vskip.2cm
\noindent{\it Proof  of (\ref{e:omegaO})}. Set  $V_n=DD_g^{-1}$. By  (\ref{e:hkj3}) of Assumption  \ref{a:r3},   $V_n=O_p(1)$.
We have
\begin{eqnarray*}
 D\widehat  \Omega_n D&=&V_n\{D_gS_{zz}^{-1}D_g\} V_n\{D^{-1}S_{zz\widehat u\widehat u}D^{-1}\}V_n\{D_gS_{zz}^{-1}D_g\}V_n, \\
  D \Omega_n D&=&V_nW_{zz}^{-1}V_nW_{zzuu}V_nW_{zz}^{-1}V_n,  \quad \\
  & &W_{zz}^{-1}=D_gE[S_{zz}\, |\mathcal{F}_n^*]^{-1}D_g, \,\,\,
   W_{zzuu}=D^{-1}E[S_{zzuu}\, |\mathcal{F}_n^*]D^{-1}.
   \end{eqnarray*}
   By  (\ref {e:daz2-re}), (\ref{e:norm-re}), (\ref{e:RA-reU}) and (\ref{e:norm-reU+}) of Lemma \ref{l:RegR},
   \begin{eqnarray*}
   D_gS_{zz}^{-1}D_g&=&W_{zz}^{-1}+o_p(1), \,\,\,   W_{zz}^{-1}=O_p(1),  \\
  D^{-1}S_{zz u u}D^{-1}&=&W_{zzuu}+o_p(1),\,\,\,W_{zzuu}=O_p(1).
      \end{eqnarray*}
We  will show  that
 \begin{eqnarray}\label{e:Pro1}
 D^{-1}S_{zz\widehat u\widehat u} D^{-1}= D^{-1}S_{zzuu} D^{-1}+o_p(1).
 \end{eqnarray}
 This implies  (\ref{e:omegaO}):
 \begin{eqnarray*}
 D\widehat  \Omega_n D&=&V_n\{W_{zz}^{-1}+o_p(1)\}V_n \{W_{zzuu}+o_p(1)\}V_n\{W_{zz}^{-1}+o_p(1)\}V_n\\
 &=&V_nW_{zz}^{-1}V_nW_{zzuu}V_nW_{zz}^{-1}V_n+o_p(1)=  D \Omega_n D+o_p(1).
   \end{eqnarray*}
   \vskip.2cm
\noindent {\it Proof  of  (\ref{e:Pro1})}.
 By definition,
 \begin{eqnarray*}
&&||D^{-1}(S_{zz\widehat u\widehat u}-S_{zzuu})D^{-1}|| =  ||\sum_{t=1}^nD^{-1}z_tz^\prime _tD^{-1}  (\widehat u_t^2-u_t^2)||\nonumber\\
&&\qquad \le
  \sum_{t=1}^n||D^{-1}z_t||^2\, |\widehat u_t^2-u_t^2|\le i_n  \times  (\sum_{t=1}^n||D^{-1}z_t||^2), \quad   i_n=\max_{t=1,..., n} |\widehat u_t^2-u_t^2|.\hspace{1cm}
\nonumber 
  \end{eqnarray*}
Notice that
  $$\sum_{t=1}^n||D^{-1}z_t||^2\le ||D^{-1}D_g||^2\sum_{t=1}^n||D_g^{-1}z_t||^2=
  O_p(1),$$
   since  $||D_gD^{-1}||=O_p(1)$ by assumption (\ref{e:hkj3})  and
 $ \sum_{t=1}^n||D_g^{-1}z_t||^2=O_p(1)$
  by (\ref{e:RAM1x}) of Lemma \ref{l:RegR}.
Hence, to verify  (\ref{e:Pro1}), it suffices to show that
     \begin{eqnarray}
    i_n   = o_p(1).
\label{e:RAM2}
  \end{eqnarray}
     Recall the equality $\widehat u_t^2-u_t^2=(\widehat u_t-u_t)^2+2(\widehat u_t-u_t)u_t$.  Denote $q_n=||D(\beta -\widehat \beta)||$.
Then,
 \begin{eqnarray*}\widehat u_t-u_t&=&(\beta -\widehat\beta)^\prime z_t=\{(\beta -\widehat\beta)^\prime D\} \{D^{-1}z_t\},\nonumber\\
 | \widehat u_t-u_t|&\le &||D^{-1}z_t||\,q_n,\nonumber\\
 |\widehat u_t^2-u_t^2|&\le&(\widehat u_t-u_t)^2+2|(\widehat u_t-u_t)u_t|\le ||D^{-1}z_t||^2\,q_n^2+2||D^{-1}z_t||\,|u_t|\,q_n.
 \nonumber 
    \end{eqnarray*}
   Hence,
      \begin{eqnarray*}
    i_n\le (\max_{t=1,..., n} ||D^{-1}z_t||^2) \,q_n^2+2(\max_{t=1,..., n} ||D^{-1}z_tu_t||)\,q_n=o_p(1),
  \end{eqnarray*}
 where    $q_n=O_p(1)$  by Theorem \ref{t:r1},
 and $$\max_{t=1,..., n} ||D^{-1}z_t||^2=o_p(1), \quad \max_{t=1,..., n} ||D^{-1}z_tu_t||=o_p(1)$$
by (\ref{e:LIZA2}) of  Lemma \ref{e:LEL}.
  This implies (\ref{e:RAM2}) and  completes  the  proof of the  corollary.
\hfill $\Box$

\vskip.2cm

\noindent {\bf Proof of Corollary \ref{c:co1Power}}. Let $\beta_k$  be the true  value of the $k$-th  component of the parameter $\beta$, and  suppose  that
$\beta_k\ne\beta_k^0$.
Write
\begin{eqnarray*}
t_n=\frac{\widehat \beta_k -\beta_k^0}{\sqrt{\widehat \omega_{kk}}}=
\frac{\widehat \beta_k -\beta_k}{\sqrt{\widehat \omega_{kk}}}+
\frac{\beta_k-\beta_k^0}{\sqrt{\widehat \omega_{kk}}}=:t_{n,1}+t_{n,2}.
\end{eqnarray*}
By  (\ref{e:hatb1-Rno}) of Corollary \ref{c:co1}, $t_{n,1}
\rightarrow
_d \mathcal{N}(0, 1)$  and $\sqrt{ \omega_{kk}}\asymp_pv_k^{-1}$. Hence,
$$
t_{n,1}=O_p(1), \quad t_{n,2}\asymp_pv_k \rightarrow_p  \infty.
$$
Then,
$
t_n=t_{n,1}+t_{n,2}=O_p(1)+t_{n,2}\asymp_pv_k \rightarrow_p  \infty,
$ which proves the claim of   Corollary \ref{c:co1Power}. \hfill $\Box$

\vskip.4cm
\noindent{\bf Proof of  Lemma  \ref{l:le}}.
\noindent{\it Proof of  (\ref{e:hkj})}.
It  suffices to show  that
\begin{eqnarray}\label{e:impl+}
i_n=v_{gk}^{-2}\max_{1\le t \le n}(g^2_{kt}+\mu^2_{kt})=o_p(1).
 \end{eqnarray}
Notice  also  that
$z^2_{kt}=\mu^2_{kt}+2\mu_{kt}g_{kt}\eta_{kt}+g^2_{kt}\eta_{kt}^2$,
\begin{eqnarray}\label{e:impl+1}
E[z^2_{kt}\, |\mathcal{F}_n^*]=\mu^2_{kt}+2\mu_{kt}g_{kt}E[\eta_{kt}\, |\mathcal{F}_n^*]+g^2_{kt}E[\eta_{kt}^2\, |\mathcal{F}_n^*]
=\mu^2_{kt}+g^2_{kt}.
\end{eqnarray}
In  addition, by assumption (\ref{e:lec1++}) of lemma, $v_{gk}^{-2}=(\sum_{t=1}^n g_{kt}^2)^{-1}=O_p(n^{-1})$. Thus,
\begin{eqnarray*}
i_n=O_p(1)i_{n,1}, \quad i_{n,1}=n^{-1}\max_{1\le t \le n}E[z^2_{kt}\, |\mathcal{F}_n^*].
\end{eqnarray*}
We will show  that $Ei_{n,1}=o(1)$ which implies (\ref{e:impl+}).
Observe that  for  any  $L\ge 1$, $$z^2_{kt}\le  L+z^2_{kt}I(z^2_{kt}\ge L)\le L+L^{-1}z^4_{kt}.$$
By  assumption (\ref{e:lec1}), $E[z^4_{kt}]\le c<\infty$  where $c$   does not depend  on $t,n$.
Hence,
\begin{eqnarray*}
i_{n,1}&\le&n^{-1}L+n^{-1}L^{-1}\max_{t=1, ..., n}E[z^4_{kt}\, |\mathcal{F}_n^*]\le n^{-1}L+n^{-1}L^{-1}\sum_{t=1}^nE[z^4_{kt}\, |\mathcal{F}_n^*],\\
Ei_{n,1}&\le& \mbox{$n^{-1}L+n^{-1}L^{-1}\sum_{t=1}^nE[z^4_{kt}]\le n^{-1}L+L^{-1}c\rightarrow 0, \quad  n,L\rightarrow \infty$}
\end{eqnarray*}
which implies  $i_n=o_p(1)$  and  proves (\ref{e:hkj}).

\vskip.2cm
\noindent{\it Proof of  (\ref{e:zzuu})}. It  suffices to verify that
\begin{eqnarray}\label{e:impl+11}
i_n=v_{k}^{-2}\max_{1\le t \le n}(g^2_{kt}+\mu^2_{kt})h_t^2=o_p(1).
 \end{eqnarray}
By assumption (\ref{e:lec1++}) of lemma, $v_{k}^{-2}=O_p(n^{-1})$.
This  together  with (\ref{e:impl+1})
implies  that
\begin{eqnarray*}
i_n=O_p(1)i_{n,2}, \quad i_{n,2}=n^{-1}\max_{1\le t \le n}E[z^2_{kt}h_t^2\, |\mathcal{F}_n^*].
\end{eqnarray*}
We will show  that $Ei_{n,2}=o(1)$ which implies $i_{n,2}=o_p(1)$ and proves (\ref{e:impl+11}).

Similarly as  above,  for  any  $L\ge 1$, setting $L_0= \log L$, for  $\delta>0$ we  obtain
\begin{eqnarray*}z^2_{kt}h_t^2&\le&  L+z^2_{kt}h_t^2I(z^2_{kt}h_t^2\ge L)\\
&\le&L+L_0^{-1} z^4_{kt} I(h_t^2\le
L_0^{-1}z^2_{kt})  +L_0h_{t}^4 I(h_t^2> L_0^{-1}z^2_{kt})
  I(h_t^4L_0\ge L)\\
  &\le& L+L_0^{-1}z^4_{kt}+
  h_{t}^4L_0 \big(\frac{h_t^4}{LL_0^{-1}}\big)^\delta\\
  &\le& L+L_0^{-1}z^4_{kt}+
  h_{t}^{4+4\delta} A_L, \quad A_L=L^{-\delta}L_0^{1+\delta}.
  \end{eqnarray*}
By  assumption (\ref{e:lec1}), $E[z^4_{kt}]\le c$  and there  exists  $\delta>0$ such  that $E[|u_t|^{4+4\delta}]\le c$,
 where $c<\infty$   does not depend  on $t,n$. Hence,
$E[h^{4+4\delta}_{t}]=E[(E[u^2_{t}\, |\mathcal{F}_n^*])^{2+2\delta}]$ $\le
E[|u_t|^{4+4\delta}]\le c$. Notice  that $A_L\rightarrow 0$  as  $L\rightarrow  \infty$.
Therefore, as  $n, L\rightarrow\infty$,
\begin{eqnarray*}
Ei_{n,2}&\le& \mbox{$n^{-1}\sum_{t=1}^nE[z^2_{kt}h_t^2\, |\mathcal{F}_n^*]
$}\\
&\le& \mbox{$n^{-1}L+L_0^{-1}n^{-1}\sum_{t=1}^nE[z^4_{kt}]
+A_Ln^{-1}\sum_{t=1}^nE[h^{4+4\delta}_{t}]$}\\
&\le& \mbox{$ n^{-1}L+L_0^{-1}c+A_Lc\rightarrow 0, $}
\end{eqnarray*}
which implies  $i_n=o_p(1)$  and  proves  (\ref{e:zzuu}).

\vskip.2cm
\noindent{\it Proof of  (\ref{e:hkj3})}.
By assumption  (\ref{e:lec1})  of Lemma  \ref{l:le}, $E[z_{kt}^4]\le c$  and $E[u_{t}^4]\le c$  where $c<\infty$  does not depend on $t,k,n$.
By (\ref{e:impl+1}),
\begin{eqnarray}
\mu_{kt}^2&\le& E[z_{kt}^2\, |\mathcal{F}^*_n], \quad E[\mu_{kt}^2]\le E[z_{kt}^2]\le c,\nonumber\\
g_{kt}^2&\le& E[z_{kt}^2\, |\mathcal{F}^*_n], \quad E[g_{kt}^2]\le E[z_{kt}^2]\le c,\nonumber\\
E[\mu_{kt}^4]&\le& E[(E[z_{kt}^2\, |\mathcal{F}^*_n])^2]\le  E[(E[z_{kt}^4\, |\mathcal{F}^*_n])]\le E[z_{kt}^4]\le c,\nonumber\\
E[g_{kt}^4]&\le& E[(E[z_{kt}^2\, |\mathcal{F}^*_n])^2]\le  \ c,\nonumber\\
E[h_t^4]&=&E[(E[u_t^2\, |\mathcal{F}^*_n])^2]\le  E[(E[u_t^4\, |\mathcal{F}^*_n])]\le E[u_t^4]\le c,\nonumber\\
E[\mu_{kt}^2h_t^2]&\le& (E[\mu_{kt}^4]E[h_t^4])^{1/2}\le c,\nonumber\\
E[g_{kt}^2h_t^2]&\le& (E[g_{kt}^4]E[h_t^4])^{1/2}\le c,\label{e:ghc}
\end{eqnarray}
where $c<\infty$  does not depend on $t,n$.
Hence,
\begin{eqnarray*}
&&\mbox{$n^{-1}E[\sum_{t=1}^n\mu_{kt}^2]\le c, \quad \sum_{t=1}^n\mu_{kt}^2=O_p(n),$}\\
&&\mbox{$n^{-1}E[\sum_{t=1}^n\mu_{kt}^2h_t^2]\le c, \quad \sum_{t=1}^n\mu_{kt}^2h_t^2=O_p(n),$}\\
&&\mbox{$n^{-1}E[\sum_{t=1}^ng_{kt}^2h_t^2]\le c, \quad \sum_{t=1}^ng_{kt}^2h_t^2=O_p(n)$}\\
&&\mbox{$n^{-1}E[\sum_{t=1}^ng_{kt}^2]\le c, \quad \sum_{t=1}^ng_{kt}^2=O_p(n).$}
\end{eqnarray*}
By assumption $(\ref{e:lec1++})$,
$n/v_k^2=O_p(1)$  and $ n/v_{gk}^2=O_p(1)$. Thus,
\begin{eqnarray*}
&&\mbox{$v_{gk}^{-1}\sum_{t=1}^n\mu_{kt}^2=O_p(n/v_{gk}^2)=O_p(1),$}\\
&&\mbox{$v_{k}^{-1}\sum_{t=1}^n\mu_{kt}^2h_t^2=O_p(n/v_{k}^2)=O_p(1),$}\\
&&\mbox{$v_{gk}^{-1}v_{k}=v_{gk}^{-1}\sum_{t=1}^ng_{kt}^2h_t^2=O_p(n/v_{gk}^2)=O_p(1),$}\\
&&\mbox{$v_{k}^{-1}v_{gk}=v_{k}^{-1}\sum_{t=1}^ng_{kt}^2=O_p(n/v_{k}^2)=O_p(1),$}
\end{eqnarray*}
which  proves  (\ref{e:hkj3}).
This  completes  the proof of  the  lemma. \hfill $\Box$

\section{Proofs  of Theorem \ref{t:r1FT} and Corollaries \ref{c:co1TV} and \ref{c:co1TVPower}} \label{s:proTV}
\noindent {\bf Proof of Theorem \ref{t:r1FT}}.
Recall the notation introduced  in  Section \ref{s:FTV}. Set
$$
\widetilde y_j= b_{n,tj}^{1/2}y_j, \quad \widetilde z_j=b_{n,tj}^{1/2}z_j,\quad \widetilde u_j=b_{n,tj}^{1/2}u_j.
$$
Then we can write
\begin{eqnarray*}
\widetilde y_j=  \widetilde z_j^\prime \beta_t +  \widetilde u_j+r_j, \quad r_j=(\beta_j-\beta_t)^\prime z_j.
\end{eqnarray*}
Recall the   estimator $\widehat \beta_t$  given in  (\ref{e:OLS10}).  In Section \ref{s:FTV} we  introduced an auxiliary  regression model
with a fixed  parameter  $\beta=\beta_t$:
 \begin{eqnarray}  \label{e:r1TV1TV1}
 y_j^*&=&\beta^\prime \widetilde z_j + \widetilde u_j, \,\,\, \widetilde u_j =b_{n,tj}^{1/2}u_j, \quad j=1,..., n.
\end{eqnarray}
Recall  the  OLS   estimator $\widehat \beta$ of the  fixed  parameter  $\beta$ in this  model,  given in (\ref{e:OLSrTV+TV+}):
 $$
\widehat \beta =\big( \sum_{j=1}^n \widetilde z_j \widetilde z_j^\prime\big)^{-1}
\big( \sum_{j=1}^n \widetilde z_j  y_j^*\big)=\beta+\big( \sum_{j=1}^n \widetilde z_j \widetilde z_j^\prime\big)^{-1}
\big( \sum_{j=1}^n \widetilde z_j  \widetilde u_j\big).
$$
In (\ref{e:OLSrTV+DD})  we  showed  that following relation:
\begin{eqnarray}
\widehat \beta_t-\beta_t&=&
\widehat \beta-\beta+R_t,\quad R_t=\big( \sum_{j=1}^n \widetilde z_j \widetilde z_j^\prime\big)^{-1}
\big( \sum_{j=1}^n \widetilde z_j \widetilde z_j^\prime (\beta_j-\beta_t)\big)\label{e:OLSrTV+DD11}
\end{eqnarray}
The remainder    $R_t=(R_{1t}, ...., R_{pt})^\prime$    arises due to  time variation  in the parameter   $\beta_j$ and  is   negligible.
We will obtain an upper bound  for  this   term. The term $\widetilde \beta -\beta$
is  the  main component  We will analyse it  using the  results of Section \ref{s:OLS}. Overall, equation (\ref{e:OLSrTV+DD11}) shows  that   properties  of $\widehat \beta_t-\beta_t$  are  determined by the properties of
$\widehat \beta-\beta$, with  an  additional  negligible term $R_t$.

\noindent First we  will show   that the  components
of $\widehat \beta-\beta=(\widetilde \beta_{1}- \beta_{1}, ..., \widetilde \beta_{p}- \beta_{p})^\prime
$ and  $R_t$
satisfy the   following   properties. For  $k=1, ..., p$,
\begin{eqnarray}\label{e:clt1}
\widetilde \beta_{k}- \beta_{k}&=&O_p(H^{-1/2}), \quad \frac{\widetilde \beta_{k}- \beta_{k}}{\sqrt{\omega_{kk,t}}}\rightarrow _d \mathcal{N}(0,1), \quad \sqrt{\omega_{kk,t}}\asymp_p H^{-1/2},\\
\label{e:clt2}
R_{kt}&=&O_p\big((H/n)^{\gamma}\big).
\end{eqnarray}

\vskip.2cm
\noindent {\it Proof  of  (\ref{e:clt1})}. Recall that $\widetilde z_j=(\widetilde z_{1j}, ..., \widetilde z_{pj})^\prime$, and
\begin{eqnarray}\label{e:clt3}
\widetilde z_{kj}&=&\widetilde\mu _{kj}+\widetilde g_{kj}\eta_{kj}, \quad \widetilde u_j= \widetilde h_j \varepsilon_j,\\
& & \widetilde\mu _{kj}=b_{n,tj}^{1/2}\mu _{kj}, \quad \widetilde g_{kj}=b_{n,tj}^{1/2}g_{kj}, \quad  \widetilde h_j =b_{n,tj}^{1/2}h_j.\nonumber
\end{eqnarray}
By  Lemma  \ref{l:TVA}, under  assumptions  of  theorem,  $ \widetilde\mu _{kj}$  and the  scale  factors
$\{\widetilde g_{kj}, \widetilde h_j\}$ satisfy  Assumptions   \ref{a:r3} and  \ref{a:4R}(ii). Thus, by  Theorem \ref{t:r1},
\begin{eqnarray*}
\widetilde \beta_{k}- \beta_{k}=O_p( v_k^{-1}) =O_p(H^{-1/2}),
\end{eqnarray*}
where $ v_k^2\equiv v_{kt}^2=\sum_{j=1}^n \widetilde g_{kj}^2\widetilde h_j^2=\sum_{j=1}^n b_{n,tj}g_{kj}^2 h_j^2$ and
\begin{eqnarray}\label{e:vkvkk}
v_k^2\asymp_pH.
\end{eqnarray}
Indeed,
 $v_{kt}^{-2}=O_p(H^{-1})$
by  (\ref{e:lecTV++}) of Assumption  \ref{a:tv5}. On the  other  hand,  (\ref{e:ghc})  implies   that  $Ev_{kt}^2\le \sum_{j=1}^n b_{n,tj}E[g_{kj}^2 h_j^2]\le c\sum_{j=1}^n b_{n,tj}=O(H)$,  where the  last  relation  easily  follows  using definition of $b_{n,tj}$ and (\ref{e:kernel}). Hence $v_{kt}^2=O_p(H)$, which proves  (\ref{e:vkvkk}).

This complete the proof of  the  first claim in (\ref{e:clt1}), while   the  second  claim holds  by   (\ref{e:hatb1-R13})  of Theorem   \ref{t:r1-R}. The third  claim  holds  since  by (\ref{e:hatb1-Rno}) of  Corollary \ref{c:co1} and
(\ref{e:vkvkk}), 
\begin{eqnarray}\label{e:clt4}\sqrt{\omega_{kk,t}}\asymp_p v_k^{-1}=(\sum_{j=1}^n \widetilde g_{kj}^2\widetilde h_{j}^2)^{-1/2}\asymp_p H^{-1/2}.
\end{eqnarray}

\vskip.2cm
\noindent {\it Proof  of  (\ref{e:clt2})}.
Write
\begin{eqnarray*}
R_t &=&S_{\tilde z\tilde z,t}^{-1}S_{\tilde z\tilde z\beta,t},  \quad \mbox{where $S_{\tilde z\tilde z\beta,t}=\sum_{j=1}^n \tilde z_j\tilde z_j^\prime (\beta_j-\beta_t)$}.
\end{eqnarray*}
We  will  show  that
\begin{eqnarray}\label{e:BB1}
 ||S_{\tilde z\tilde z,t}^{-1}||= O_p(H^{-1}),  \quad ||S_{\tilde z\tilde z\beta,t}||=O_p\big(H(H/n)^\gamma\big),
\end{eqnarray}
which implies
$||R_t||\le ||S_{\tilde z\tilde z,t}^{-1}||\,||S_{\tilde z\tilde z\beta,t}||=O_p\big((H/n)^\gamma\big).
$
Then,   $|R_{kt}|\le ||R_t||=O_p\big((H/n)^\gamma\big)$  which proves (\ref{e:clt2}).

To verify (\ref{e:BB1}), recall notation of the $p\times p$  diagonal matrix
\begin{eqnarray*}
&&D_{\widetilde g}={\rm diag}( v_{\widetilde g 1},...,v_{\widetilde g p}), \quad \mbox{$v_{\widetilde g k}=\sum_{j=1}^n \widetilde g_{kj}^2, \quad k=1, ..., p.$}
\end{eqnarray*}
Notice  that
\begin{eqnarray*}
||S_{\tilde z\tilde z,t}^{-1}||&=&||D_{\widetilde g}^{-1}(D_{\widetilde g}S_{\tilde z\tilde z,t}^{-1}D_{\widetilde g})|D_{\widetilde g}^{-1}||\le ||D_{\widetilde g}^{-1}||^2||D_{\widetilde g}S_{\tilde z\tilde z,t}^{-1}D_{\widetilde g}||=O_p(H^{-1})
\end{eqnarray*}
because $||D_{\widetilde g}^{-1}||^2=  \sum_{k=1}^p v_{\widetilde g k}^{-2}=O_p(H^{-1})$  
by  Assumption  \ref{a:tv5}. On the  other  hand,

\noindent  $D_{\widetilde g}S_{\tilde z\tilde z,t}^{-1}D_{\widetilde g}=O_p(1)$
by (\ref{e:daz2-re})  and (\ref{e:norm-re}) of  Lemma \ref{l:RegR}.
This  proves  the  first claim in (\ref{e:BB1}).

\noindent Next, bound
\begin{eqnarray*}
E||S_{\tilde z\tilde z\beta,t}||&\le& E[\sum_{j=1}^n ||\tilde z_j||^2 ||\beta_j-\beta_t||]
\le \sum_{j=1}^n (E||\tilde z_j||^4)^{1/2} (E||\beta_j-\beta_t||^2)^{1/2}.
\end{eqnarray*}
We  have $||\widetilde z_j||^4=b_{n,tj}^2|| z_j||^4$.
Recall that  $E|| z_j||^4\le c$   by Assumption  \ref{a:tv5}, $E||\beta_j-\beta_t||^2\le c(|t-j|/n)^{2\gamma}$
by  Assumption \ref{a:sm},  and it is trivial to show   that  under  (\ref{e:kernel}),  $$\sum_{j=1}^n b_{n,tj} (|t-j|/H)^\gamma=O(H).$$
This  implies
\begin{eqnarray}
\label{e:sumg}E||S_{\tilde z\tilde z\beta,t}||&\le& CH(H/n)^\gamma \Big(H^{-1}\sum_{j=1}^n b_{n,tj} (|t-j|/H)^\gamma\Big) \le CH(H/n)^\gamma
\end{eqnarray}
which  proves  the  second  claim in  (\ref{e:BB1}).

We  now  are ready  to  prove the   claims (\ref{e:hatb1FT}) and  (\ref{e:hatb1FTCLT})  of  the  theorem.
First, together with (\ref{e:OLSrTV+DD11}), the properties
 (\ref{e:clt1})  and  (\ref{e:clt2}) establish   the  consistency   result (\ref{e:hatb1FT}):
$$
\widehat \beta_t-\beta_t= (\widetilde\beta-\beta)+R_t=O_p\big(H^{-1/2}+(H/n)^\gamma\big).
$$
To prove the   asymptotic normality  property  (\ref{e:hatb1FTCLT}), recall
assumption $H=o(n^{2\gamma/(2\gamma+1)})$.
Then \begin{eqnarray*}
\frac{\widehat \beta_{kt} -\beta_{kt}}{\sqrt{\omega_{kk,t}}}&=&\frac{\widetilde \beta_{kt} -\beta_{kt}}{\sqrt{\omega_{kk,t}}}+
\omega_{kk,t}^{-1/2}R_t=\frac{\widetilde \beta_{kt} -\beta_{kt}}{\sqrt{\omega_{kk,t}}}+o_p(1).
\end{eqnarray*}
because  by  (\ref{e:clt4}) and   (\ref{e:clt2}),
$$\omega_{kk,t}^{-1/2}B_t=  O_p\big(H^{1/2}\big)O_p\big((H/n)^\gamma\big)
=O_p\big(H^{1/2}(H/n)^\gamma\big)=o_p(1)$$
under assumption  $H=o(n^{2\gamma/(2\gamma+1)})$.
Then,
\begin{eqnarray*}
\frac{\widehat \beta_{kt} -\beta_{kt}}{\sqrt{\omega_{kk,t}}}
=\frac{\widetilde \beta_{kt} -\beta_{kt}}{\sqrt{\omega_{kk,t}}}+
o_p(1)\rightarrow_d \mathcal{N}(0,1)
\end{eqnarray*}
by (\ref{e:clt1})  which proves the   asymptotic normality  property  (\ref{e:hatb1FTCLT})  of  the theorem.
Noting that 
$\sqrt{\omega_{kk,t}}\asymp_p H^{-1/2}$, as  shown in
(\ref{e:clt4}),  this  completes  the  proof of the theorem.
\hfill $\Box$

\vskip.2cm
\noindent {\bf Proof  of  Corollary \ref{c:co1TV}}.  In the proof of  Theorem  \ref{t:r1FT}
we wrote the time-varying  regression model as a regression model
\begin{eqnarray}\label{e:regr1}
 \widetilde y_j=  \widetilde z_j^\prime \beta +  \widetilde u_j+r_j, \quad r_j=(\beta_j-\beta_t)^\prime \widetilde z_j
  \end{eqnarray} with a fixed  parameter $\beta=\beta_t$.
   We  showed that the   regressors  $\widetilde z_j$  and
  the  noise $\widetilde u_j$ satisfy assumptions of   Theorem \ref{t:r1-R} 
 and that the contribution of the term  $r_j$ is asymptotically negligible. That allowed us to  establish  the  asymptotic  normality  property  (\ref{e:hatb1FTCLT})  of  Theorem  \ref{t:r1FT} for  $\widehat \beta_{kt}$
 using results  of  Section \ref{s:OLS}.

Clearly, to  prove   Corollary  \ref{c:co1TV},  it  suffices to   verify the  second  claim  in (\ref{e:hatb1-RnoTV}),
 \begin{eqnarray*}
 \frac{\widehat \omega_{kk,t}}{ \omega_{kk,t}}=1+o_p(1).
\end{eqnarray*}
Proof of the  corresponding result in the case of   fixed  parameter in  Corollary \ref{c:co1} shows  that we  need to  verify
the validity of  (\ref{e:Pro1})  for
our regression model (\ref{e:regr1}), i.e. to  show   that
 \begin{eqnarray}\label{e:Pro1TV}
j_n= D^{-1}S_{\widetilde z\widetilde z\widehat u\widehat u} D^{-1}= D^{-1}S_{\widetilde z\widetilde z\widetilde u\widetilde u} D^{-1}+o_p(1),
 \end{eqnarray}
where  $\widehat u_j=\widetilde  y_j-  \widehat \beta^\prime \widetilde z_j$,  $\widehat \beta=\widehat \beta_t$,
  $D={\rm diag}(v_1, ...., v_k)^\prime$ and $v_k^2=\sum _{j=1}^n \widetilde g_{kj}^2\widetilde h_j^2$.

Set  $\widehat  u_j^*=(\beta_t-\widehat  \beta_t)^\prime\widetilde z_j+\widetilde u_j$.
Write
\begin{eqnarray*} 
j_n= D^{-1}S_{\widetilde z\widetilde z\widehat u^*\widehat u^*} D^{-1}+D^{-1}(S_{\widetilde z\widetilde z\widehat u\widehat u}-
S_{\widetilde z\widetilde z\widehat u^*\widehat u^*}) D^{-1}=j_{n1}+j_{n2}.
 \end{eqnarray*}
 By (\ref{e:Pro1}),
$j_{n1}=D^{-1}S_{\widetilde z\widetilde z\widetilde u\widetilde u} D^{-1}+o_p(1).$
 Hence, to  prove (\ref{e:Pro1TV}), we need to  show   that
 \begin{eqnarray}\label{e:Pro1TV4}
j_{n2}=o_p(1).
 \end{eqnarray}
By  Assumption \ref{a:tv5}, $||D^{-1}||=O_p(H^{-1/2})$.  Hence,
  \begin{eqnarray*}
||j_{n2}||\le ||D^{-1}||^2||S_{\widetilde z\widetilde z\widehat u\widehat u}-
S_{\widetilde z\widetilde z\widehat u^*\widehat u^*}||=O_p(1)||j_{n3}||,  \quad  j_{n3}=
 H^{-1}(S_{\widetilde z\widetilde z\widehat u\widehat u}-
S_{\widetilde z\widetilde z\widehat u^*\widehat u^*}).
 \end{eqnarray*}
We  will show  that  $j_{n3}=o_p(1)$  which implies (\ref{e:Pro1TV4}).
Notice  that
\begin{eqnarray}
\widehat u_j&=&\widetilde y_j-\widehat  \beta_t^\prime\widetilde z_j=(\beta_t-\widehat  \beta_t)^\prime \widetilde z_j+\widetilde u_j+r_j=\widehat    u_j^*+r_j,
\nonumber\\
\widehat u_j^2-\widehat    u_j^{*\, 2}&=&(\widehat u_j-\widehat    u_j^*)^2+2(\widehat u_j-\widehat    u_j^*)\widehat    u_j^*
\nonumber\\
&=&r_j^2+2r_j\widehat    u_j^*=r_j^2+2r_j(\beta_t-\widehat  \beta_t)^\prime \widetilde z_j+2r_j\widetilde    u_j.\label{e:bobound}
\end{eqnarray}
Using the inequality  $2|ab|\le  a^2+b^2$, we  can  bound  in (\ref{e:bobound}),
\begin{eqnarray*}
&&2|r_j(\beta_t-\widehat  \beta_t)^\prime \widetilde z_j|\le r_j^2+\big((\beta_t-\widehat  \beta_t)^\prime \widetilde z_j\big)^2
\le  r_j^2+||\beta_t-\widehat  \beta_t||^2||\widetilde  z_j||^2.
\end{eqnarray*}
Next we evaluate $|r_j\widetilde    u_j|$ in (\ref{e:bobound}).  Let   $L>1$ be  large  number. Then,
\begin{eqnarray*}
&&|r_j|\le  L^{-1}||\widetilde z_j||\, I\big(|r_j|\le  L^{-1} ||\widetilde z_j||\big)+|r_j|I\big(|r_j|>  L^{-1}||\widetilde z_j||\big)
\le L^{-1} ||\widetilde z_j||+ Lr_j^2||\widetilde z_j||^{-1},\\
&&|r_j\widetilde    u_j|\le   L^{-1} ||\widetilde z_j||\, |\widetilde    u_j|+ Lr_j^2||\widetilde z_j||^{-1}\, |\widetilde    u_j|.
\end{eqnarray*}
Hence,

\hspace{1cm}$
|\widehat u_j^2-\widehat    u_j^{*\, 2}|\le2  r_j^2+||\beta_t-\widehat  \beta_t||^2||\widetilde  z_j||^2+2 L^{-1} ||\widetilde z_j||\, |\widetilde    u_j|
+2 L||\widetilde z_j||^{-1}\, |\widetilde    u_j|r_j^2.
$

\noindent Since $r_j^2\le ||\beta_j-\beta_t||^2||\widetilde z_j||^2$, this yields
\begin{eqnarray*}
||\widetilde z_j||^2|\widehat u_j^2-\widehat    u_j^{*\, 2}|&\le&2 ||\beta_j-\beta_t||^2||\widetilde z_j||^4+||\beta_t-\widehat  \beta_t||^2||\widetilde  z_j||^4+2 L^{-1} ||\widetilde z_j||^3\, |\widetilde    u_j|
\\\qquad & &+2 L||\widetilde z_j||^3\, |\widetilde    u_j|\, ||\beta_j-\beta_t||^2.
\end{eqnarray*}
Recall that $\widetilde  z_j=b_{n,tj}^{1/2}z_j$  and $\widetilde  u_j=b_{n,tj}^{1/2}u_j$.  Denote $\theta_j =2|| z_j||^4
+2||z_j||^3 |u_j|.$
Then,
\begin{eqnarray*}
||\widetilde z_j||^2|\widehat u_j^2-\widehat    u_j^{*\, 2}|\le Lb_{n,tj}^2||\beta_j-\beta_t||^2\theta_j
+\big(||\beta_t-\widehat  \beta_t||^2+L^{-1}\big)b_{n,tj}^2\theta_j.
\end{eqnarray*}
Hence,
\begin{eqnarray}
|j_{n3}|&=&\mbox{$H^{-1}\big|\sum_{j=1}^n\widetilde z_j\widetilde z_j^\prime (\widehat u_j^2-\widehat    u_j^{*\, 2})\big|
\le H^{-1}\sum_{j=1}^n||\widetilde z_j||^2 |\widehat u_j^2-\widehat    u_j^{*\, 2}|$}
\nonumber\\
&\le& \mbox{$L\{H^{-1}\sum_{j=1}^nb_{n,tj}^2 ||\beta_j-\beta_t||^2\theta_j\}
+(||\beta_t-\widehat  \beta_t||^2+L^{-1})\{H^{-1}\sum_{j=1}^n b_{n,tj}^2\theta_j\}$}\nonumber\\
&\le&\mbox{$L\{\sum_{j=1}^n b_{n,tj}||\beta_j-\beta_t||^2\}\{H^{-1}\sum_{j=1}^nb_{n,tj}\theta_j\}$}
\nonumber\\
& &\qquad \qquad \mbox{$+(||\beta_t-\widehat  \beta_t||^2+L^{-1})\{H^{-1}\sum_{j=1}^n b_{n,tj}^2\theta_j\}$}\nonumber\\
&=&Lq_{n1}q_{n2}+(||\beta_t-\widehat  \beta_t||^2+L^{-1})q_{n3}.\label{e:q33}
\end{eqnarray}
By (\ref{e:hatb1FT})  of  Theorem \ref{t:r1FT}, $||\beta_t-\widehat  \beta_t||^2  =o_p(1)$,  and   $L^{-1}$
can be  made  arbitrarily  small  by  selecting  large  $L$.
We will show  that
\begin{eqnarray}\label{e:DADA11}
Eq_{n1}=o(1),  \quad Eq_{n2}=O(1), \quad Eq_{n3}=O(1).
\end{eqnarray}
Combining this  with (\ref{e:q33}), we  obtain
$$
|j_{n3}|=Lo_p(1)+\big(o_p(1)+L^{-1}\big)O_p(1),
$$
so that the  right hand  side  can be  made  arbitrarily  small by  selecting  a large  enough $L$  and  letting $n \rightarrow  \infty$.
This  proves  (\ref{e:Pro1TV4}).

To bound $Eq_{n1}$  observe  that by  Assumption \ref{a:sm}, $E ||\beta_t-\beta_j||^2\le C(|t-j|/n)^{2\gamma}$,
where  $0<\gamma\le  1$ and  
and  recall (\ref{e:sumg}).
Then,
\begin{eqnarray*}
Eq_{n1}\le \sum_{j=1}^nb_{n,tj}E||\beta_j-  \beta_t||^2&\le& C\big(H (\frac{H}{n})^{2\gamma}\big)\{H^{-1}\sum_{j=1}^nb_{n,tj}(\frac{|t-j|}{H})^{2\gamma}\}
\\
&\le&  C
H(H/n)^{2\gamma}=o(1)
\end{eqnarray*}
when $H=o(n^{2\gamma/(2\gamma+1)})$.  This  proves  (\ref{e:DADA11})  for  $Eq_{n1}$.

To bound $Eq_{n2}$ and $Eq_{n3}$, recall  that  by  Assumption \ref{a:tv5},
$
Ez_{kj}^4\le C$ and $Eu_{j}^4\le C$  which  implies  that  $E\theta_j\le  C$. Moreover, under (\ref{e:kernel}) it holds
$H^{-1}\sum_{j=1}^nb_{n,tj}=O(1)$ and  $b_{n,tj}^2\le Cb_{n,tj}$. Hence,
\begin{eqnarray*}
Eq_{n2}\le \mbox{$H^{-1}\sum_{j=1}^nb_{n,tj}E\theta_j\le CH^{-1}\sum_{j=1}^nb_{n,tj}=O(1),$}\\
Eq_{n3}\le \mbox{$H^{-1}\sum_{j=1}^nb_{n,tj}^2E\theta_j\le CH^{-1}\sum_{j=1}^nb_{n,tj}=O(1)$}.
\end{eqnarray*}
This   completes  the  proof of  (\ref{e:DADA11})
and  the  corollary.
\hfill $\Box$

\vskip.2cm
\noindent {\bf Proof of Corollary \ref{c:co1TVPower}}. Let $\beta_{kt}$  be the true  value of the $k$-th  component of the time-varying parameter $\beta_t$. Suppose  that
$|\beta_{kt}^0-  \beta_{kt}|\ge a>0$ for $t=t_n \in [1,...,n]$  as $n \rightarrow  \infty$.
Write
\begin{eqnarray*}
\tau_{n,t}=\frac{\widehat \beta_{kt} -\beta_{kt}^0}{\sqrt{\widehat \omega_{kk,t}}}=
\frac{\widehat \beta_{kt} -\beta_{kt}}{\sqrt{\widehat \omega_{kk,t}}}+\frac{\beta_{kt}-  \beta_{kt}^0}{\sqrt{\widehat \omega_{kk,t}}}
=:\tau_{n1,t}+\tau_{n2,t}.
\end{eqnarray*}
By  (\ref{e:hatb1-RnoTV}) of Corollary \ref{c:co1TV}, $\tau_{n1, t}
\rightarrow
_d \mathcal{N}(0, 1)$  and $\sqrt{ \omega_{kk,t}}\asymp_p H^{-1/2}$. Hence,
$$
\tau_{n1,t}=O_p(1), \quad \tau_{n2,t}\asymp_pH^{1/2} \rightarrow_p  \infty.
$$
Then,
$
\tau_{n,t}=\tau_{n1,t}+\tau_{n2,t}=O_p(1)+\tau_{n2,t}\asymp_pH^{1/2} \rightarrow_p  \infty,
$ which proves  the claim of  the Corollary \ref{c:co1TVPower}. \hfill $\Box$

\begin{lemma}\label{l:TVA} Suppose   that Assumption \ref{a:tv5} holds and  Assumptions \ref{a:r0}, \ref{a:ETA} are  satisfied. Then
$\{\widetilde\mu _{kj},\widetilde g_{kj}, \widetilde h_{j}\}$  in (\ref{e:clt3}) satisfy   Assumption \ref{a:r3} and  Assumption \ref{a:4R}(ii).
\end{lemma}
\noindent{\bf Proof  of  Lemma  \ref{l:TVA}}.  Notice that  assumptions  (\ref{e:kernel})
imply $\sum_{j=1}^nb_{n,tj}\asymp H$. Thus,  the  claim of Lemma  \ref{l:TVA}  follows  using the  same argument
as  in the  proof of
 Lemma~\ref{l:le}. \hfill $\Box$

\section{Proofs  of Theorems \ref{t:prop1}, \ref{t:prop2} and Theorem \ref{t:AR}} \label{s:proTVmiss}
\noindent{\bf Proof of  Theorem \ref{t:prop1}}.
Suppose  that   $y_t=\beta^\prime  z_t+u_t$ follows the regression   model (\ref{e:r1}).
In the presence of  missing  data,  estimation  of the  parameter  $\beta$  is  based  on a
regression model with the  fixed  parameter (\ref{e:r1mis}):
\begin{eqnarray}  \label{e:miss1}
\widetilde y_t=
\beta^\prime \widetilde z_t +\widetilde u_t, 
\end{eqnarray}
where the regressors
 $\widetilde z_t=(\widetilde z_{1t}, ..., \widetilde z_{pt})^\prime$  and the noise $\widetilde u_t$ take  the  form
\begin{eqnarray}  \label{e:miss2}
\widetilde z_{kt}&=& \widetilde \mu_{kt}+\widetilde g_{kt}\eta_{kt}, \quad
\widetilde \mu_{kt}=\tau_t \mu_{kt}, \quad \widetilde g_{kt}=\tau_t g_{kt},\\
\widetilde u_{t}&=& \widetilde h_{t}\varepsilon_{t}, \quad \widetilde h_{t}=\tau_t h_{t},\nonumber
\end{eqnarray}
and  $\tau_t$  is  the  missing data indicator.
Under  Assumptions \ref{a:tau} and  \ref{a:mis} of the theoren, $\{\widetilde \mu_{t}, \widetilde g_{t},\widetilde h_{t}\}$  are independent  of
 $\{\varepsilon_t, \eta_t\}$. Therefore, $(z_t,u_t)$   belongs to the regression space  described  in
(\ref{e:r2})  and  (\ref{e:rz2}) of Section \ref{s:OLS}.

 We  estimate  the fixed parameter $\beta$  using the  estimator defined in (\ref{e:OLSrTILDE}):
\begin{eqnarray} \label{e:OLSrTILDE miss}
\widehat \beta&=&\big( \sum_{t=1}^n \widetilde z_{t} \widetilde z_{t}^\prime\big)^{-1}\big( \sum_{t=1}^n
\widetilde z_{t}\widetilde  y_{t}\big).
\end{eqnarray}

\noindent We  will show that $(\widetilde z_t,\widetilde u_t)$  satisfy Assumptions \ref{a:r0}, \ref{a:ETA},
 \ref{a:r3} and \ref{a:4R} of  Theorem   \ref{t:r1-R} of  Section \ref{s:OLS}. Then,  the required  result  (\ref{e:hatb1-RnoMD}) for  $\widehat  \beta$
of  this theorem follows directly from the claims
(\ref{e:hatb1-Rno})
of Corollary \ref{c:co1}.

We  split Assumptions \ref{a:r0}, \ref{a:ETA},
 \ref{a:r3} and \ref{a:4R} into two groups:

 (a) Assumptions \ref{a:r0}, \ref{a:ETA}, and \ref{a:4R}(i), and

  (b)
 Assumptions  \ref{a:r3} and
\ref{a:4R}(ii).

Assumptions (a) imposed
on the stationary  processes $ \eta_t, \varepsilon_t$ are  part of
 Assumption    \ref{a:mis}  of Theorem \ref{t:prop1}.

It  remains  to  show the validity of the assumptions in  group (b), i.e.  that the means $\widetilde  \mu_t$ and the  scales
$\widetilde h_t, \widetilde g_t$ satisfy  Assumptions \ref{a:r3} and
\ref{a:4R}(ii).
 By Assumption  \ref{a:mis}, we have $Ez_{kt}^4\le c$ and $ E u_{t}^4\le c$. Moreover,
$g_{kt} \ge  c_1>0$  and $h_{t} \ge  c_1>0$
where  $c, c_1>0$  do not depend  on  $k,t$ and $n$.  For  $k=1, ..., p$, define:
\begin{eqnarray}  \label{e:Dkmis}
\widetilde v_k^2&=&\sum_{t=1}^n \widetilde g_{kt}^2\widetilde h_t^2,\quad \widetilde v_{gk}^2=\sum_{t=1}^n \widetilde g_{kt}^2.
\end{eqnarray}
Notice  that
$$
\widetilde v_k^2\ge c_1^4\sum_{t=1}^n \tau_t=c_1^4 N, \quad \widetilde v_{gk}^2\ge c_1^2\sum_{t=1}^n \tau_t=c_1^2 N,
$$
where $N$ is   the  size  of the subsample  (\ref{e:subs}). By  assumption  of the theorem,  $n/N=O_p(1)$.
Thus,
\begin{eqnarray} \label{e:tilde1}
E\widetilde z_{kt}^4&\le& Ez_{kt}^4\le c, \quad E|\widetilde u_{t}|^{4+\delta}\le E| u_{t}|^{4+\delta}\le c,\\
n/\tilde v_k^2&=&O(n/N)= O_p(1), \quad n/\tilde v_{gk}^2=O(n/N)= O_p(1),\label{e:tilde2}
\end{eqnarray}
which confirms the validity  of  Assumptions \ref{a:r3} and 
\ref{a:4R}(ii); see  Lemma \ref{l:le}. This  completes  the  proof of the theorem. \hfill $\Box$

\vskip.2cm
\noindent{\bf Proof of  Theorem \ref{t:prop2}}.
Now, suppose  that   $y_t=\beta^\prime_t  z_t+u_t$ follows the regression   model (\ref{e:r1TV})  with a  time-varying parameter $\beta_t$.
In the presence of  missing  data,  estimation  of the time-varying parameter  $\beta_t$  is  based  on a  model (\ref{e:r1misTV}):
\begin{eqnarray}  \label{e:miss1+}
\widetilde y_t=
\beta^\prime_t \widetilde z_t +\widetilde u_t. 
\end{eqnarray}
Here, the regressors
 $\widetilde z_t=(\widetilde z_{1t}, ..., \widetilde z_{pt})^\prime$  and the noise $\widetilde u_t$
 are the  same as  in (\ref{e:miss2}).   We  showed in the proof of the Theorem \ref{t:prop1}
  that $(z_t,u_t)$   belongs to the regression space  described  in
(\ref{e:r2})  and  (\ref{e:rz2}) of Section \ref{s:OLS}.

 The estimator of the time-varying parameter $\beta_t$ is given in (\ref{e:OLS10MV}):
\begin{eqnarray} \label{e:OLSrTILDE missTV}
\widehat \beta_t&=&\big( \sum_{j=1}^nb_{n,tj} \widetilde z_j \widetilde z_j^\prime\big)^{-1}\big( 
\sum_{j=1}^nb_{n,tj} \widetilde z_j \widetilde y_j\big).
\end{eqnarray}

\noindent We  will show that $(\widetilde z_t,\widetilde u_t)$  satisfy Assumptions
 \ref{a:r0}, \ref{a:ETA}, \ref{a:4R}(i), \ref{a:sm} and \ref{a:tv5}  of  Theorem   \ref{t:r1FT}. Then,  the  results  (\ref{e:hatb1FTMD}), (\ref{e:hatb1-R13TVMD}) and (\ref{e:omegaVMD}) for  $\widehat  \beta_t$ of  Theorem \ref{t:prop2} follow from the results (\ref{e:hatb1FT}) of Theorem \ref{t:r1FT} and (\ref{e:hatb1-RnoTV}) of Corollary \ref{c:co1TV}.

Observe that  Assumptions  \ref{a:r0}, \ref{a:ETA},  \ref{a:4R}(i)
on $\eta_t, \varepsilon_t$ are  part of Assumption \ref{a:tau} of
this  theorem,  which also includes Assumption \ref{a:sm} for  $\beta_t$.

 It remains  to show  that $\widetilde z_t, \widetilde u_t$  satisfy  Assumption \ref{a:tv5}. This requires  to prove the validity of (\ref{e:tilde1}) and (\ref{e:tilde2}) under Assumption \ref{a:mis} of this theorem, which we showed  in
 the proof of  Theorem \ref{t:prop1}.
\hfill $\Box$

\vskip.2cm
\noindent {\bf Proof  of   Theorem  \ref{t:AR}}. We  consider  a stationary  AR($p$) model  (\ref{e:AR2S}),
$$
y_t=\phi_0+\phi_1y_{t-1}+...+\phi_py_{t-p}+\varepsilon_t,  $$
where $\varepsilon_t$ is a  stationary $m.d.$  sequence  with respect to the information set ${\cal F}_t=\sigma(\varepsilon_s, s\le t)$. Write it
 as a regression model  (\ref{e:r1}),
 \begin{equation}
  \label{e:RAR1+}
 y_t=\beta^\prime z_t+u_t,  \quad u_t=\varepsilon_t
 \end{equation}
with fixed parameter
 $\beta=(\beta_1, ..., \beta_{p+1})^\prime=(\phi_0,
...., \phi_p)^\prime$  and   regressors $z_t=(z_{1t},z_{2t},...,z_{p+1,t})^
\prime=(1,y_{t-1},y_{t-2}, ..., y_{t-p})^\prime $. Under  assumption  (\ref{e:AREE}) of  theorem,
AR($p$) model has a  stationary  solution:
\begin{equation}
  \label{e:AREE++1}
y_t=\mu+\sum_{j=0}^\infty  a_j\varepsilon_{t-j},  \quad  \mbox{where} \,\,\sum_{j=0}^\infty  |a_j|<\infty, \,\,\mu=Ey_t,
\end{equation}
and
regressors
$$z_{kt}= \mu_{kt}+g_{kt}\eta_{kt},  \quad \mu_{kt}=E[y_{t-k}]=Ey_1, \quad g_{kt}=1, \quad \eta_{kt}=y_{t-k}-E[y_{t-k}],
$$ for  $k=2, ..., p+1$,  satisfy   regression  assumption   (\ref{e:rz2}).
From (\ref{e:AREE++1})  it follows   that the regressors  $\eta_t=(\eta_{1t}, ...., \eta_{pt})^\prime =(y_{t-1},y_{t-2}, ..., y_{t-p})^\prime
$ are  $\mathcal{F}_{t-1}=\sigma(\varepsilon_s, s\le t-1)$ measurable.
Moreover, under the assumptions  of  the theorem, $(\varepsilon_t, \eta_t)$  satisfy  Assumptions \ref{a:r0}, \ref{a:ETA},
 \ref{a:r3} and \ref{a:4R} of  Theorem   \ref{t:r1-R} in  Section \ref{s:OLS}.
Finally,  we  show  that  $Ey_t^8\le C<\infty$. Recall that  by the assumption of  theorem, $\varepsilon_t$ is a  stationary $m.d.$  sequence  such that  $E\varepsilon_t^8<\infty.$ It is  known  that if   $E|\varepsilon_t|^p<\infty$, for some  $p>2$, then
$$
E\big|\sum_{j=0}^\infty  a_j\varepsilon_{t-j}\big|^p\le C\big(\sum_{j=0}^\infty  a_j^2\big)^{p/2},
$$
where  $C<\infty$  does  not  depend  on $n$; see e.g., Lemma 2.5.2  in \cite{GKS2012}. Hence $E(y_t-\mu)^8<\infty$  and $E\eta_{kt}^8<\infty$ from $k=1, ..., p$.

Thus,  regressors   $z_t$  and   regression noise  $u_t=\varepsilon_t$  satisfy  Assumptions  \ref{a:r0}, \ref{a:ETA}, \ref{a:r3} and  \ref{a:4R}  of  Section \ref{s:OLS}.  Therefore,  the robust OLS estimator $\widehat \beta$  of $\beta$ has properties  derived   in  Corollary  \ref{c:co1}  which  implies Theorem \ref{t:AR}.
\hfill $\Box$

\section{Proofs  of Section \ref{s:OLS}: Auxiliary  lemmas}\label{AuxLem}
\noindent 
This  section  contains auxiliary lemmas  used  in the proofs of the  main results for  Section \ref{s:OLS}. For the  ease  of  referencing,  we  include  the statement of Lemma \ref{l:bba1-R}(i)  established  in \cite{glp2024Suppl}.

\begin{lemma}
\label{l:bba1-R} 
Assume  that  sequences $\{\beta_t\}$  and  $\{z_t\}$  are  mutually independent.

 \vskip.1cm \noindent {\rm(i)} If  $\{z_t\}$ is a  covariance stationary short memory sequence, then
\begin{eqnarray}\label{e:su12-R}
\sum_{t=1}^n\beta_tz_t = \Big(\sum_{t=1}^n\beta_t\Big)Ez_1+O_p\Bigl((\sum_{t=1}^n\beta_t^2)^{1/2}\Bigr).
\end{eqnarray}
\vskip.1cm \noindent {\rm(ii)}  If  $E|z_t|<\infty$, then
\begin{eqnarray}\label{e:su12-R+}
\big|\sum_{t=1}^n\beta_tz_t\big| = O_p\Big(\sum_{t=1}^n|\beta_t|\Big)(\max_{t=1, ..., n}E|z_t|).
\end{eqnarray}
\end{lemma}
\noindent{\bf Proof of  Lemma \ref{l:bba1-R}.} The   claim  (i)  of  Lemma \ref{l:bba1-R} was derived in
 (\cite{glp2024Suppl},  Lemma A5).
To  prove (ii), denote  $s_n=\sum_{t=1}^n|\beta_t|$. Then,
\begin{eqnarray*}
\mbox{$E\big[s_n^{-1}\sum_{t=1}^n|\beta_t|\, |z_t|\big]$}&=&\mbox{$\sum_{t=1}^nE[s_n^{-1}|\beta_t|]\, E[|z_t|]$}\\
&\le&\mbox{$ (\max_{t=1, ..., n}E|z_t|)E[s_n^{-1}\sum_{t=1}^n|\beta_t|]=\max_{t=1, ..., n}E|z_t|,$}\\
\mbox{$s_n^{-1}\sum_{t=1}^n|\beta_t|\, |z_t|$}&=& \mbox{$O_p\Big(\max_{t=1, ..., n}E|z_t|\Big).$}
\end{eqnarray*}
This  implies
\begin{eqnarray*}
\mbox{$\big|\sum_{t=1}^n\beta_tz_t\big| \le s_n
\big\{s_n^{-1}\sum_{t=1}^n|\beta_t|\, |z_t|\big\}=s_nO_p\big(\max_{t=1, ..., n}E|z_t|\big).$}
\end{eqnarray*}
This  completes  the  proof of (\ref{e:su12-R+})  and the  lemma.  \hfill $\Box$
\vskip.2cm \noindent
Recall notation
 \begin{eqnarray*}
S_{zz}&=&\mbox{$\sum _{t=1}^nz_tz^\prime_t, \quad S_{zzuu}=\sum _{t=1}^nz_tz^\prime_tu_t^2, \quad S_{zu}=\sum _{t=1}^nz_tu_t,$}\\
\nonumber&&D={\rm diag}(v_1, ..., v_p),  \quad \mbox{$v_k=(\sum_{t=1}^ng_{kt}^2h_t^2)^{1/2}$},
\\
\nonumber&&D_g={\rm diag}(v_{g1}, ..., v_{gp}),  \quad  \mbox{$v_{gk}=(\sum_{t=1}^ng_{kt}^2)^{1/2}$}.
 \end{eqnarray*}
Recall definition $\mathcal{F}_n^*=\sigma(\mu_t, g_t, t=1, ..., n)$ and $\mathcal{F}_{n,t-1}$  in (\ref{e:MD*}).
Denote
\begin{eqnarray*}
 W_{zz}&=&D_g^{-1} E[S_{zz}|\mathcal{F}^*_n]  D_g^{-1}, \quad W_{zzuu}=D^{-1} E[S_{zzuu}|\mathcal{F}^*_n]  D^{-1},\\
 &&\Omega_n=(E[S_{zz}|\mathcal{F}^*_n])^{-1}(E[S_{zzuu}|\mathcal{F}^*_n])(E[S_{zz}|\mathcal{F}^*_n])^{-1}.
\end{eqnarray*}

\begin{lemma}\label{l:RegR}
\vskip.2cm \noindent
Suppose that   $z_t$ and $u_t$   satisfy  Assumptions  \ref{a:r0},  \ref{a:ETA} and \ref{a:r3}.  Then  the  following holds.

\vskip.2cm \noindent{\rm (i)}
There  exists  $b_n>0$ such that $b_n^{-1}=O_p(1)$ and such  that  for  any $a=(a_{1}, ..., a_p)^\prime $, $||a||=~1$,
\begin{eqnarray}\label{e:norm-re}
 a^\prime W_{zz}a&\ge&  b_n, 
 \quad  ||W_{zz}^{-1}||_{sp}\le b_n^{-1}, \\
 || W_{zz}||&\le& b_{2n}=O_p(1).\label{e:norm-re+}
\end{eqnarray}
Moreover,
\begin{eqnarray}
\label{e:RA-re}
  D_g^{-1}S_{zz}D_g^{-1}&=&
  W_{zz}+o_p(1),\\
\label{e:daz2-re}
 D_gS_{zz}^{-1}D_g&=&
 W_{zz}^{-1}  +o_p(1),\\
\label{e:daz2-re+}
 D^{-1}S_{zu}&=&
O_p(1),\\
\sum_{t=1}^n||D_g^{-1}z_t||^2&=&O_p(1).
  \label{e:RAM1x}\end{eqnarray}
\vskip.2cm \noindent (ii) In addition, if  Assumption \ref{a:4R} holds, then there exists  $b_n>0$ such that $b_n^{-1}=O_p(1)$ and such  that  for  any $a=(a_{1}, ..., a_p)^\prime $, $||a||=1$,
\begin{eqnarray}\label{e:norm-reU}
 a^\prime W_{zzuu}a&\ge &  b_n,  \quad ||W_{zzuu}^{-1}||_{sp}\le b_n^{-1}, \\
 || W_{zzuu}||&\le &b_{2n}=O_p(1),\label{e:norm-reU+}\\
  \label{e:OMEGA}
 a^\prime D\Omega_nD a&\ge &b_{n},  \quad
 a^\prime D\Omega_nD a \le b_{2n}=O_p(1).
\end{eqnarray}
Moreover,
\begin{eqnarray}
\label{e:RA-reU}
  D^{-1}S_{zzuu}D^{-1}&=&
  W_{zzuu}+o_p(1),\\
\label{e:daz2-reU}
 DS_{zzuu}^{-1}D&=&
 W_{zzuu}^{-1}  +o_p(1),\\
 \label{e:SuuC}
  D^{-1}S_{zzuu}^{(c)}D^{-1}=
  W_{zzuu}+o_p(1),&&  S_{zzuu}^{(c)}=\sum_{t=1}^n z_tz^\prime_tE[u_t^2\,|\mathcal{F}_{n,t-1}].
 \end{eqnarray}
\end{lemma}
\vskip.2cm
\noindent
Before the proof  of  lemma, we   will state  the  following corollary.
Denote 
 \begin{eqnarray}\label{e:c*}
c_{*,n}=\sum_{t=1}^n|| D_g^{-1}\mu_t||^2,  \qquad c_{**,n}=\sum_{t=1}^n|| D^{-1}\mu_th_t||^2.
 \end{eqnarray}
 Notice  that  under (\ref{e:hkj3}) of Assumption \ref{a:r3},
  \begin{eqnarray}\label{e:c*1}
c_{*,n}=\sum_{k=1}^p\{v_{gk}^{-2}\sum_{t=1}^n\mu_{kt}^2\}=O_p(1),\quad
 c_{**,n}=\sum_{k=1}^p\{v_{k}^{-2}\sum_{t=1}^n\mu_{kt}^2h_t^2\}=O_p(1).
 \end{eqnarray}
\begin{corollary}\label{c:bn}
In Lemma  \ref{l:RegR}, the  claims (\ref{e:norm-re}) and (\ref{e:norm-reU})  hold  with  $b_n$ as  below:
\begin{eqnarray}\label{e:bn*}
    a^\prime W_{zz}a &\ge&  b_n = \begin{cases}
                        c^{-1}:\qquad  \qquad \quad \quad \text{Case 1 (intercept not included), } \\
                      c^{-1} (1+c_{*,n})^{-1}: \quad  \text{Case 2 (intercept included),}
                    \end{cases}
\\
\label{e:bn**}
    a^\prime W_{zzuu}a&\ge&     b_n = \begin{cases}
                        c^{-1} (1+c_{**,n})^{-4}:\quad   \text{Case 1 (intercept not included), } \\
                         c^{-1} (1+c_{**,n})^{-9}: \quad  \text{Case 2 (intercept included),}
                    \end{cases}
\end{eqnarray}
where   $c>0$    does not  depend  on $n$, $b_n^{-1}=O_p(1)$  and  $b_n$  is $\mathcal{F}^*_n$  measurable.
\end{corollary}
\vskip.2cm \noindent {\bf Proof of Lemma  \ref{l:RegR}(i)}.
{\it Proof of (\ref{e:norm-re})}.
Set  $I_{gt}={\rm diag}(g_{1t},..., g_{pt})$. By definition,
 \begin{eqnarray}z_t&=&\mu_t+I_{gt}\eta_t=\mu_t+\widetilde  z_t,\quad \widetilde  z_t=I_{gt}\eta_t.
\label{e:Igt}
 \end{eqnarray}
Then
 \begin{eqnarray}
  z_t z^\prime_t&=&(\mu_t+\widetilde z_t) (\mu_t+\widetilde z_t)^\prime = \widetilde z_t\widetilde z_t^\prime + \mu_t\mu_t^\prime
+ \mu_t\widetilde z_t^\prime +\widetilde z_t\mu_t^\prime ,\nonumber\\
E[ z_t z^\prime_t |\mathcal{F}^*_n]&=& E[ \widetilde z_t \widetilde z^\prime_t|\mathcal{F}^*_n] + \mu_t\mu_t^\prime
+\mu_tE[\widetilde z^\prime_t|\mathcal{F}^*_n]+E[\widetilde z_t|\mathcal{F}^*_n]\mu_t^\prime\nonumber\\
&=& E[ \widetilde z_t \widetilde z^\prime_t|\mathcal{F}^*_n] + \mu_t\mu_t^\prime
+\mu_te_t^\prime+e_t\mu_t^\prime\nonumber\\
&=& E[ \widetilde z_t \widetilde z^\prime_t|\mathcal{F}^*_n] + (\mu_t+e_t)(\mu_t+e_t)^\prime
-e_te_t^\prime,
\label{e:secon}
 \end{eqnarray}
where $ e_t=E[\widetilde z_t|\mathcal{F}^*_n]=I_{gt}E[\eta_t]$.
Using (\ref{e:secon}), we   can write
\begin{eqnarray}
&&a^\prime W_{zz}a= \sum_{t=1}^n a^\prime D_g^{-1}E[ z_t z^\prime_t |\mathcal{F}^*_n]D_g^{-1}a
\nonumber  \\
&& =
\sum_{t=1}^n  a^\prime D_g^{-1}E[\widetilde z_t \widetilde z^\prime_t |\mathcal{F}^*_n]D_g^{-1}a
+\sum_{t=1}^n ( a^\prime D_g^{-1}\mu_t)^2
 +2\sum_{t=1}^n ( a^\prime D_g^{-1}\mu_t)(e_t^\prime D_g^{-1}a)\label{e:aWa-reRR0}\\
&& =
\sum_{t=1}^n  a^\prime D_g^{-1}E[\widetilde z_t \widetilde z^\prime_t |\mathcal{F}^*_n]D_g^{-1}a
 +\sum_{t=1}^n ( a^\prime D_g^{-1}(\mu_t+e_t))^2-\sum_{t=1}^n( a^\prime D_g^{-1}e_t)^2.
 \label{e:aWa-reRR}
\end{eqnarray}
We split  the  proof  into  two cases when regression  model (\ref{e:r1}) does not include intercept and  when  intercept is   included.

Case 1 (no intercept): $e_t=I_{gt}E[\eta_t]=(0,...,0)^\prime$. 

Case  2 (intercept included): $e_t=I_{gt}E[\eta_t]=I_{gt}(1,0,...,0)^\prime=(g_{1t},0,...,0)^\prime$, $g_{1t}=1$.

\vskip.2cm
\noindent {\bf Case 1}.  Let   $e_t=0$.
 Then (\ref{e:aWa-reRR0})  implies
\begin{eqnarray}
a^\prime W_{zz}a &\ge& \sum_{t=1}^n  a^\prime D_g^{-1}E[\widetilde z_t \widetilde z^\prime_t |\mathcal{F}^*_n]D_g^{-1}a.
 \label{e:aWa-reRR1}
\end{eqnarray}
In this instance,
$$
E[ \widetilde z_t \widetilde z^\prime_t |\mathcal{F}^*_n]=  I_{gt}E[\eta_t\eta_t^\prime]I_{gt}
=I_{gt}\Sigma I_{gt},
$$
where  $
E[\eta_t\eta_t^\prime]=\Sigma =(\sigma_{jk})_{j,k=1, ...,p}$.
By  Assumption  \ref{a:ETA}(ii), the matrix $\Sigma$ is  positive  definite.
 Therefore, there  exists  $b>0$  such   that for any   $\alpha=(\alpha_{1}, ..., \alpha_p)^\prime $, 
 $$
\alpha^\prime \Sigma \alpha
\ge b||\alpha||^2.
 $$
Hence, setting  $\gamma_{kt}=v_{gk}^{-1}g_{kt}$, we  derive
\begin{eqnarray*}
&&\sum_{t=1}^n a^\prime D_g^{-1}E[ z_t z^\prime_t |\mathcal{F}^*_n]D_g^{-1}a =  \sum_{t=1}^n \{a^\prime D_g^{-1}I_{gt}\}\Sigma \{I_{gt}D_g^{-1}a\}
\nonumber\\
&&\qquad \quad \ge b\sum_{t=1}^n ||a^\prime D_g^{-1} I_{gt}||^2=
 b\sum_{t=1}^n
\big[ \sum_{k=1}^p a_k^2\gamma_{kt}^2\big] \notag \\
&&\qquad \quad  =b\sum_{k=1}^p a_k^2 \, (\sum_{t=1}^n \gamma_{kt}^2)=b\sum_{k=1}^p a_k^2 =b||a||^2=b, 
\end{eqnarray*}
since $\sum_{t=1}^n \gamma_{kt}^2=1$ and $||a||=1$.
With (\ref{e:aWa-reRR1}) this proves  the  first  claim  in~(\ref{e:norm-re}):
\begin{eqnarray}
a^\prime W_{zz}a&\ge & b.
  \label{e:LIZ}
\end{eqnarray}
 Matrix  $W_{zz}$ is  symmetric and, thus, it has  real  eigenvalues.  The bound  (\ref{e:LIZ})
 implies   that the  smallest  eigenvalue  of   $W_{zz}$ has  property $ \lambda_{ min}\ge b_n>0$. Therefore $W_{zz}$  is  positive definite, and the  largest  eigenvalue $\theta_{\max}$ of  $W_{zz}^{-1}$ has  property $\theta_{\max}=\lambda_{min}^{-1}\le 1/b_n$, which implies  that $||W_{zz}^{-1}||_{sp}\le 1/b_n$.   This proves  the   second  claim  in (\ref{e:norm-re}).

\vskip.2cm\newpage
\noindent {\bf Case 2} (intercept included): $e_t=I_{gt}E[\eta_t]=I_{gt}(1,0,...,0)^\prime=(g_{1t},0,...,0)^\prime$.
Recall that in presence of intercept, $g_{1t}=1$ and  $\eta_{1t}=1$. 

\vskip.2cm
 \noindent  {\it  Proof of (\ref{e:norm-re})}. Set  $a=(a_1, ..., a_p)^\prime$, $\widetilde a=(a_2, ..., a_p)^\prime$. Recall that
\begin{eqnarray} \label{e:Raf00}
 1=||a||^2 =a_1^2+....+a_p^2=a_1^2+||\widetilde a||^2.
\end{eqnarray}
We  will show   that there  exists  $b>0$ such that  for  any $a$ and  $n \ge 1$,
\begin{eqnarray} \label{e:Raf1}
a^\prime W_{zz}a&\ge& b||\widetilde a||^2,\\
 \label{e:Raf2} a^\prime W_{zz}a&\ge&  b||\widetilde a||^2+\{a_1^2-2|a_1|\,||\widetilde a||c_{*,n}^{1/2}\},
\end{eqnarray}
where $c_{*,n}$ is  defined  as in (\ref{e:c*}). 
These bounds imply (\ref{e:norm-re}). Indeed, suppose  that $
||\widetilde a||>  (1-b)|a_1|/(2c_{*,n}^{1/2})$. By
(\ref{e:Raf00}), this is  equivalent to
\begin{eqnarray*}
||\widetilde a||^2>\frac{(1-b)^2a_1^2}{4c_{*,n}}=\frac{(1-b)^2(1 -||\widetilde a||^2)}{4c_{*,n}},\quad
||\widetilde a||^2>\frac{(1-b)^2}{(1-b)^2+4c_{*,n}}.
\end{eqnarray*}
Then, by (\ref{e:Raf1}),
\begin{eqnarray*}
a^\prime W_{zz}a&\ge& b||\widetilde a||^2= \frac{b(1-b)^2}{(1-b)^2+4c_{*,n}}.
\end{eqnarray*}
On the other  hand,  if  $
||\widetilde a||\le  (1-b)|a_1|/(2c_{*,n}^{1/2})$, then in  (\ref{e:Raf2}),
 $$a_1^2-2|a_1|\,||\widetilde a||c_{*,n}^{1/2}\ge a_1^2-(1-b)a_1^2=
 b\,a_1^2$$
 which together with (\ref{e:Raf2}) implies
\begin{eqnarray*}
a^\prime W_{zz}a\ge b||\widetilde a||^2 +a_1^2b=b( ||\widetilde a||^2 +a_1^2)=b||a||^2=b.
\end{eqnarray*}
Therefore,
\begin{eqnarray*}
a^\prime W_{zz}a\ge \min \Big( \frac{b(1-b)^2}{(1-b)^2+4c_{*,n}}, \,b\Big)=\frac{b(1-b)^2}{(1-b)^2+4c_{*,n}}.
\end{eqnarray*}
This  implies  that   there  exists  $c>0$  such  that
\begin{eqnarray}\label{e:gasa1}
a^\prime W_{zz}a\ge b_n= c^{-1}(1+c_{*,n})^{-1},
\end{eqnarray}
where
$
b_n^{-1}=c(1+c_{*,n})=O_p(1)
$
by (\ref{e:c*1}). This verifies   the  first   claim in (\ref{e:norm-re}).

\vskip.2cm
\noindent {\it Proof of (\ref{e:Raf1})}. Below  we  will show  that there  exists   $b>0$  such that
\begin{eqnarray} \label{e:Raf6}
i_n=\sum_{t=1}^n  a^\prime D_g^{-1}E[\widetilde z_t \widetilde z^\prime_t |\mathcal{F}^*_n]D_g^{-1}a&\ge &a_1^2+b||\widetilde a||^2.
\end{eqnarray}
In addition, observe  that in  Case  2,
\begin{eqnarray} \label{e:data1}
e_t^\prime D_g^{-1}a=a_1v_{g1}^{-1}g_{1t},  \quad \sum_{t=1}^n( a^\prime D_g^{-1}e_t)^2=a_1^2v_{g1}^{-2}\sum_{t=1}^ng_{1t}^2=a_1^2.
\end{eqnarray}
Then from (\ref{e:aWa-reRR}), using (\ref{e:Raf6}) and (\ref{e:data1}) we  arrive  at  (\ref{e:Raf1}):
\begin{eqnarray*}
a^\prime W_{zz}a&\ge&
\sum_{t=1}^n  a^\prime D_g^{-1}E[\widetilde z_t \widetilde z^\prime_t |\mathcal{F}^*_n]D_g^{-1}a
 -\sum_{t=1}^n( a^\prime D_g^{-1}e_t)^2\nonumber\\
 &\ge&\{a_1^2+b||\widetilde a||^2\}-a_1^2=b||\widetilde a||^2.\nonumber 
\end{eqnarray*}
\noindent {\it Proof of (\ref{e:Raf2})}. By (\ref{e:aWa-reRR0})  and  (\ref{e:Raf6}),
\begin{eqnarray}\label{e:Ga1}
a^\prime W_{zz}a
&\ge&
\sum_{t=1}^n  a^\prime D_g^{-1}E[\widetilde z_t \widetilde z^\prime_t |\mathcal{F}^*_n]D_g^{-1}a
 -2\big|\sum_{t=1}^n ( a^\prime D_g^{-1}\mu_t)(e_t^\prime D_g^{-1}a)\big|\\
 &\ge&\{a_1^2+b||\widetilde a||^2\}-2|q_n|,  \quad q_n=\sum_{t=1}^n ( a^\prime D_g^{-1}\mu_t)(e_t^\prime D_g^{-1}a).\nonumber
\end{eqnarray}
By  Cauchy inequality  and  (\ref{e:data1}),
\begin{eqnarray*}
|q_n|\le \big\{\sum_{t=1}^n ( a^\prime D_g^{-1}\mu_t)^2\sum_{t=1}^n(e_t^\prime D_g^{-1}a)^2\big\}^{1/2}
=|a_1|\big(\sum_{t=1}^n ( a^\prime D_g^{-1}\mu_t)^2\big)^{1/2}.
\end{eqnarray*}
Since $\mu_{1t}=0$, then
$
| a^\prime D_g^{-1}\mu_t|\le ||\widetilde a||\,|| D_g^{-1}\mu_t||.
$
 Hence, using notation $c_{*,n}$  introduced  in (\ref{e:c*}), we obtain
\begin{eqnarray*}
&&\sum_{t=1}^n ( a^\prime D_g^{-1}\mu_t)^2\le  ||\widetilde a||^2(\sum_{t=1}^n|| D_g^{-1}\mu_t||^2)= ||\widetilde a||^2c_{*,n},
\end{eqnarray*}
which together  with (\ref{e:Ga1}) and (\ref{e:Raf6})  proves  (\ref{e:Raf2}): 
\begin{eqnarray*}\
a^\prime W_{zz}a
&\ge&
\{a_1^2+b||\widetilde a||^2\}
 -2|a_1|||\widetilde a||c_{*,n}^{1/2}=b||\widetilde a||^2+\{a_1^2-2|a_1| \, ||\widetilde a||c_{*,n}^{1/2}\}.
\end{eqnarray*}
\vskip.2cm \noindent {\it  Proof of  (\ref{e:Raf6})}.
Recall, that  in presence of  intercept, $\eta_t=(1,\eta_{2t}, ..., \eta_{pt})^\prime$  and $E[\eta_{kt}]
=0.$
Denote $\tilde  \eta=(\eta_{2t}, ...., \eta_{pt})^\prime$ and $\widetilde \Sigma =E[\tilde \eta\tilde \eta^\prime]$.
 Then
$$
E[ \widetilde z_t \widetilde z^\prime_t |\mathcal{F}^*_n]=  I_{gt}E[\eta_t\eta_t^\prime]I_{gt}
=I_{gt}{\rm diag}(1, \widetilde\Sigma )I_{gt}={\rm diag}\big(g_{1t}^2, \tilde I_{gt}\widetilde\Sigma\tilde I_{gt}\big),
$$
where
${\rm diag}(1, \widetilde\Sigma)$ is a  block diagonal matrix
and $\widetilde I_{gt}={\rm diag}(g_{2t}, ..., g_{pt})$.
By  assumption,   the matrix $\widetilde \Sigma$ is positive definite. Denote  $\widetilde D_g={\rm diag}(v_{g2}, ..., v_{gp})$. Then,
\begin{eqnarray*}\nonumber
i_n&=&\sum_{t=1}^n  a^\prime D_g^{-1}E[\widetilde z_t \widetilde z^\prime_t |\mathcal{F}^*_n]D_g^{-1}a
\\&=&
 a_1^2\{v_{g1}^{-2}\sum_{t=1}^n g^2_{1t}\}+\sum_{t=1}^n\widetilde
a^\prime \widetilde D_g^{-1}\widetilde I_{gt}\widetilde \Sigma \widetilde I_{gt}\widetilde D_g^{-1}\widetilde a\nonumber\\
&=&i_{n,1}+i_{n,2}.\nonumber
\end{eqnarray*}
Observe that $i_{n,1}=a_1^2$ since $v_{g1}^{-2}\sum_{t=1}^n g^2_{1t}=1$.
Recall that $||\widetilde a||\le 1$. Hence,
by (\ref{e:LIZ}),
\begin{eqnarray*}
i_{n,2}\ge b ||\widetilde a||^2, \quad i_{n}\ge a_1^2+b ||\widetilde a||^2
\end{eqnarray*}
for some  $b>0$ which  does not depend  on $n$  and  $a$.
This  implies (\ref{e:Raf6}).

Summarizing, note that
 by (\ref{e:LIZ})  and (\ref{e:gasa1}),
\begin{equation}\label{e:gasa4}
    a^\prime W_{zz}a \ge      b_n = \begin{cases}
                        c^{-1}:\qquad  \qquad \qquad  \text{Case 1 (intercept not included), }\\
                        c^{-1}(1+c_{*,n})^{-1}: \quad  \text{Case 2 (intercept included),}
                    \end{cases}
\end{equation}
where $c>0$  does not depend  on $n$.
Notice that $b_n^{-1}\le c(1+c_{*,n})=O_p(1)$  by (\ref{e:c*1}).
This proves the   first  claim in
(\ref{e:norm-re}).

Proof of  the  second  claim in (\ref{e:norm-re}) is  the  same  as   in Case 1.

\vskip.2cm
\noindent{\it Proof of  (\ref{e:norm-re+})}. Observe that
 \begin{eqnarray*}
 ||W_{zz}||&\le&|| E\big[ (\sum _{t=1}^n D_g^{-1}z_tz^\prime_t D_g^{-1} \big   |\mathcal{F}^*_n]  ||
 \le E\big[ ||\sum _{t=1}^n D_g^{-1}z_tz^\prime_t D_g^{-1})|| \big   |\mathcal{F}^*_n]\\\
&\le& \sum _{t=1}^nE[||D_g^{-1} z_t||^2 \,|\mathcal{F}^*_n]\le c(1+c_{*,n})=O_p(1)
\end{eqnarray*}
by  (\ref{e:LIZA0}) of Lemma \ref{e:LEL}. This proves (\ref{e:norm-re+}).
\vskip.2cm

\noindent{\bf  Proof  of  (\ref{e:RA-re}), (\ref{e:daz2-re}), (\ref{e:daz2-re+}) and (\ref{e:RAM1x})}.
  Denote  by $\delta_{jk}$ the $jk$-th element of  the  matrix
 \begin{eqnarray}\label{e:ZAS!}
D_g^{-1}S_{zz}D_g^{-1}-W_{zz}
&=&\sum _{t=1}^nD_g^{-1}\{z_tz^\prime_t-E[z_tz^\prime_t|\mathcal{F}_n^*]\}D_g^{-1}=\big(\delta_{jk}\big).
 \end{eqnarray}
 To  prove   (\ref{e:RA-re}), it remains to show  that
\begin{eqnarray}\label{e:prr1+}
\delta_{jk}=o_p(1).
 \end{eqnarray}

\vskip.2cm \noindent{\bf Case 1}:  $e_t=0$. Then, by (\ref{e:secon}), we  have
$$
z_tz^\prime_t-E[z_tz^\prime_t|\mathcal{F}_n^*]=I_{gt}(\eta_t\eta_t^\prime-E[\eta_t\eta_t^\prime])I_{gt}
+\mu_t\eta_t^\prime I_{gt}+I_{gt}\eta_t\mu_t^\prime.
$$
Therefore, setting $\gamma_{jt}=v_{gj}^{-1}g_{jt}$, we can write
\begin{eqnarray}
\delta_{jk}&=&
\sum _{t=1}^n\gamma_{jt}\gamma_{kt}(\eta_{jt}\eta_{kt}-E[\eta_{jt}\eta_{kt}])+\sum _{t=1}^n\{v_{gj}^{-1}\mu_{jt}\gamma_{kt}\}\eta_{kt}+\sum _{t=1}^n \{v_{gk}^{-1}\mu_{kt}\gamma_{jt}\}\eta_{jt}\nonumber\\
&=&
S_{n,1}+S_{n,2}+S_{n,3},\label{e:def1-r+1}\\
\delta_{jk}^2&\le&3(S_{n,1}^2+S_{n,2}^2+S_{n,3}^2)\nonumber.
 \end{eqnarray}
 By  assumption,  sequences  $\{w_{1t}= \eta_{jt}\eta_{kt}-E[\eta_{jt}\eta_{kt}]\}$,
 $\{w_{2t}= \eta_{kt}\}$ and  $\{w_{3t}= \eta_{jt}\}$
 are covariance stationary short  memory
sequences with  zero mean,  and  the weights
 $\{b_{1t}= \gamma_{jt}\gamma_{kt}\}$ are independent of $\{w_{1t}\}$,
 $\{b_{2t}=v_{gj}^{-1}\mu_{jt}\gamma_{kt}\}$ are  independent of $\{w_{2t}\}$
 and
 $\{b_{3t}= v_{gk}^{-1}\mu_{kt}\gamma_{jt}\}$ are independent of $\{w_{3t}\}$,
 Thus, applying Lemma  \ref{l:bba1-R} to $S_{n,i}, \,i=1,2,3$, we  obtain
 \begin{eqnarray*}
\delta_{jk}^2&=&O_p\Big(\sum_{t=1}^n(b_{1t}^2+b_{2t}^2+b_{3t}^2)\Big).
 \end{eqnarray*}
 Denote  $r_{jn}=\max_{t=1,...,n}\gamma_{jt}^2$.  Then,
  \begin{eqnarray*}
 \sum_{t=1}^n(b_{1t}^2+b_{2t}^2+b_{3t}^2)&\le& r_{jn}\sum_{t=1}^n\gamma_{kt}^2+r_{kn}(v^{-2}_{gj}\sum_{t=1}^n\mu_{jt}^2)
 +r_{jn}(v^{-2}_{gk}\sum_{t=1}^n\mu_{kt}^2).
   \end{eqnarray*}
  Notice that $\sum_{t=1}^n\gamma_{kt}^2=1$. Observe  that $r_{jn}=o_p(1)$
 by (\ref{e:hkj})  and $v^{-2}_{gj}\sum_{t=1}^n\mu_{jt}^2=O_p(1)$
 by  (\ref{e:hkj3}) of  Assumption \ref{a:r3}.
 This implies $\delta_{jk}^2=o_p(1)$  which
  proves   (\ref{e:prr1+}).

\vskip.2cm \noindent{\bf Case 2}.  Let  $e_t=(1, 0,...,0)^\prime$.

\noindent To  prove (\ref{e:RA-re}), it suffices to show that $\delta_{jk}$, $j,k=1, ..., p$ in (\ref{e:ZAS!}) have  property  (\ref{e:prr1+}):
$\delta_{jk}=o_p(1)$. Recall  that in presence of intercept we  have
$z_t=(1, z_{2t}, ..., z_{pt})^\prime$.

First, observe  that for  $j,k=2, ...,p$,
$\delta_{jk}$ are the same  as  in (\ref{e:def1-r+1}) and
whence $\delta_{jk}=o_p(1)$ by
 (\ref{e:prr1+}).
Second, $\delta_{11}=0$  since $z_{1t}=1$. Finally,
for $k=2, ..., p$, we  have
\begin{eqnarray*}
z_{1t}z_{kt}=z_{kt}=\mu_{kt}+g_{kt}\eta_{kt},\\
E[z_{1t}z_{kt}|\mathcal{F}_n^*]=E[z_{kt}|\mathcal{F}_n^*]=\mu_{kt}.
\end{eqnarray*}
Then,
\begin{eqnarray*}
\delta_{1k}&=&\sum _{t=1}^nv_{g1}^{-1}\{z_{1t}z_{kt}-E[z_{1t}z_{kt}|\mathcal{F}_n^*]\}v_{gk}^{-1}\\
&=&v_{g1}^{-1}\sum _{t=1}^n\{v_{gk}^{-1}g_{kt}\}\eta_{kt}=n^{-1/2}\sum _{t=1}^n\gamma_{kt}\eta_{kt}.
\end{eqnarray*}
By assumption, $\{\eta_{kt}\}$ is a covariance stationary  short  memory   sequence  with $E[\eta_{kt}]=0$,  and $\{\eta_{kt}\}$ and $\{\gamma_{kt}\}$  are mutually independent. Therefore,  by  Lemma \ref{l:bba1-R},
$$
\delta_{1k}=n^{-1/2}
O_p\Bigl(( \sum_{t=1}^n\gamma_{kt}^2)^{1/2} \Bigr)=n^{-1/2}O_p(1)=o_p(1)
$$
which  proves  (\ref{e:prr1+}). This  completes the  proof of  (\ref{e:RA-re}) in Case 2.

\vskip.2cm
\noindent{\it Proof  of   (\ref{e:daz2-re})}. It follows  using  the  same  argument as  in Case  1.

\vskip.2cm
\noindent{\it Proof  of   (\ref{e:daz2-re+})}. To prove that $D^{-1}S_{zu}=O_p(1)$, write
 \begin{eqnarray*}
D^{-1}S_{zu}
=\sum _{t=1}^nD^{-1}z_tu_t=\sum _{t=1}^nD^{-1}(\mu_t+I_{gt}\eta_t)h_t\varepsilon_t=(\nu_{1}, ..., \nu_p)^\prime.
 \end{eqnarray*}
 It suffices to show  that
\begin{eqnarray}\label{e:prr1L}
\nu_{k}=O_p(1).
 \end{eqnarray}
We  have
\begin{eqnarray*}
\nu_{k}&=&
\sum _{t=1}^n\{v_{k}^{-1}\mu_{kt}h_t\}\varepsilon_t+\sum _{t=1}^n \{v_{k}^{-1}g_{kt}h_t\}\eta_{kt}\varepsilon_t\\
&=&
S_{n,1}+S_{n,2}, \nonumber\\
\nu_{k}^2&\le &
2S^2_{n,1}+2S^2_{n,2}.
 \end{eqnarray*}
 By Assumptions, \ref{a:r0} and \ref{a:ETA},  the sequences  $\{w_{1t}=\varepsilon_t\}$, $\{w_{2t}=\eta_{kt}\varepsilon_t\}$ are covariance stationary short  memory sequences with  zero mean,  the weights
 $\{b_{1t}=v_{k}^{-1}\mu_{kt}h_t\}$ are independent of $\{w_{1t}\}$, and
 $\{b_{2t}=v_{k}^{-1}g_{kt}h_t\}$ are  independent of $\{w_{2t}\}$.

Thus, applying  Lemma  \ref{l:bba1-R} to  each  of the sum $S_{n,1}, S_{n,2}$, we  obtain
 \begin{eqnarray*}
\nu_{k}^2&=&O_p\Big(\sum_{t=1}^n(b_{1t}^2+b_{2t}^2)\Big).
 \end{eqnarray*}
 Notice that,
  \begin{eqnarray*}
 \sum_{t=1}^n(b_{1t}^2+b_{2t}^2)&=& v_{k}^{-2}\sum_{t=1}^n\mu_{kt}^2h^2_t+ v_{k}^{-2}\sum_{t=1}^ng_{kt}^2h^2_t
 =v_{k}^{-2}\sum_{t=1}^n\mu_{kt}^2h^2_t+1=O_p(1)
   \end{eqnarray*}by (\ref{e:hkj3}) of  Assumption \ref{a:r3}
   which proves (\ref{e:prr1L}).

\vskip.2cm
\noindent{\it Proof of  (\ref{e:RAM1x})}.  Observe that by (\ref{e:RA-re}) and (\ref{e:norm-re+})
  of Lemma \ref{l:RegR}, $D_g^{-1}(\sum_{t=1}^nz_tz_t^\prime) D_g^{-1}=O_p(1)$. Therefore,
  $$\sum_{t=1}^n||D_g^{-1}z_t||^2={\rm trace}\Big(D_g^{-1}(\sum_{t=1}^nz_tz_t^\prime) D_g^{-1}\Big)=O_p(1).$$
This proves (\ref{e:RAM1x})
and   completes  the  proof of  the  part (i)  of  the  lemma.

\vskip.2cm \noindent {\bf Proof of Lemma  \ref{l:RegR} (ii)}.
\noindent {\it Proof of  (\ref{e:norm-reU})}.
We   can write
\begin{eqnarray*}
a^\prime W_{zzuu}a&=& \sum_{t=1}^n a^\prime D^{-1}E[ z_t z^\prime_t u_t^2|\mathcal{F}^*_n]D^{-1}a
\nonumber  \\
&=& E\big[ \big(\sum_{t=1}^n ||a^\prime D^{-1}z_th_t||^2 \varepsilon_t^2\big)|\mathcal{F}^*_n\big].
\end{eqnarray*}
Let  $\delta>0$   be  a   small number  which will be  selected  below.  Then,
\begin{eqnarray*}
\varepsilon_t^2&=&\{\varepsilon_t^2I(\varepsilon_t^2\ge  \delta)+\delta I(\varepsilon_t^2< \delta)\}
+(\varepsilon_t^2-\delta)I(\varepsilon_t^2<  \delta)\\
&\ge&\delta- \delta I(\varepsilon_t^2<  \delta).
\end{eqnarray*}
Thus,
\begin{eqnarray}
a^\prime W_{zzuu}a&\ge & \delta \big\{ E\big[ \big(\sum_{t=1}^n ||a^\prime D^{-1}z_th_t||^2 \big)|\mathcal{F}^*_n\big]-
 E\big[ \big(\sum_{t=1}^n ||a^\prime D^{-1}z_th_t||^2 I(\varepsilon_t^2<  \delta)|\mathcal{F}^*_n\big]\big\}
 \nonumber\\
 &=&\delta\{q_{1,n}-q_{2,n}\}.\label{e:aWa-reRRx1}
\end{eqnarray}
We  will show  that   there  exist  $b_n>0$  and  $\delta=\delta_n>0$ such that  $b_n^{-1}=O_p(1)$,  $\delta_n^{-1}=O_p(1)$
and for  any  $a=(a_1, ..., a_p)^\prime$, $||a||=1$
and  $n\ge 1$,
\begin{eqnarray}
q_{1,n}&\ge&  b_n,
 \label{e:aWa-reRRx2}\\
 q_{2,n}&\le&  b_n/2.
 \label{e:aWa-reRRx3}
\end{eqnarray}
Using these  bounds  in (\ref{e:aWa-reRRx1}), we obtain
\begin{eqnarray} \label{e:Lela1}
a^\prime W_{zzuu}a\ge b_n^*= \delta_n\{b_n- (b_n/2)\}= \delta_n b_n/2, \quad 1/b_n^*=O_p(1).
\end{eqnarray}

\vskip.2cm
\noindent First we  prove (\ref{e:aWa-reRRx2}).   Setting
\begin{eqnarray*}
Z_t&=&
 \{h_t\mu_t\}+ \{h_tI_{gt}\} \eta_t=\mu_t^*+I_{g^*t}\eta_t, \quad \mbox{where $\mu^*_t=h_t\mu_t$,
$g^*_t=h_t g_t$},\\
D_{g*}&=&(v_{g^*1}, ..., v_{g^*p})^\prime, \quad \mbox{$ v_{g^*k}=(\sum _{t=1}^n g^{*\, 2}_{kt})^{1/2}$},
\end{eqnarray*}
 we  can  write
$$
q_{1,n}=\sum_{t=1}^n a^\prime  D_{g^*}^{-1}E[ Z_t Z^\prime_t|\mathcal{F}^*_n]D_{g^*}^{-1}a= a^\prime \, W_{ZZ} \, a.
$$
Observe that the variables $Z_t=\mu_t^*+I_{g^*t}\eta_t$  satisfy
assumptions of  Lemma  \ref{l:RegR}(i). Hence by
(\ref{e:gasa4}),
\begin{eqnarray}\label{e:gasa2}
 a^\prime W_{ZZ}a \ge  b_n = \begin{cases}
                        c^{-1}:\qquad  \qquad  \qquad \,\,\, \text{Case 1 (intercept not included), }\\
                        c^{-1}(1+c_{**,n})^{-1}: \quad  \text{Case 2 (intercept included),}
                    \end{cases}
\end{eqnarray}
where $c>0$  does not depend  on $n$.
Notice that $b_n^{-1}\le c(1+c_{**,n})=O_p(1)$  by (\ref{e:c*1}).
This proves
(\ref{e:aWa-reRRx2}).

To prove  (\ref{e:aWa-reRRx3}), recall that $||a||=1$. Bound
$$
q_{n,2}\le ||a||^2q_{n,2}^*=q_{n,2}^*,  \quad q_{n,2}^*=
\sum_{t=1}^n E[||D^{-1}z_th_t||^2I(\varepsilon_t^2<  \delta)|\mathcal{F}^*_n\big].
$$
In  (\ref{e:LIZA0d})  of Lemma \ref{e:LEL} we  show  that
$q_{n,2}^*\le c_1(1+c_{**,n}) \delta ^{1/4}$,   where $c_1>0$  does  not depend  on   $n$  and $c_{**,n}$
is  defined  in (\ref{e:c*}).
Thus, selecting
\begin{eqnarray*}
\delta_n&=&\Big(\frac{b_n/2}{c_1(1+c_{**,n})}\Big)^{4},
\end{eqnarray*}
we  obtain
$
q_{n,2}\le c_1(1+c_{**,n}) \delta_n ^{1/4}=  b_n/2,
$
which proves the   bound (\ref{e:aWa-reRRx3}). Notice  that  $\delta_n \le (2cc_1)^{-4}$ can be  made small  by
selecting large  $c$  in  (\ref{e:gasa2}).

\noindent In turn,  by (\ref{e:Lela1}),
\begin{eqnarray*} 
a^\prime W_{zzuu}a&\ge& 
 (b_n/2)\delta_n =(b_n/2)\Big(\frac{(b_n/2)}{c_1(1+c_{**,n})}\Big)^{4}
\end{eqnarray*}
where $b_n$  is defined in (\ref{e:gasa2}).  This  implies
\begin{equation}\label{e:gasa2++}
  a^\prime W_{zzuu}a\ge      b_n^*  = \begin{cases}
                        c^{-1}(1+c_{**,n})^{-4}:\quad   \text{Case 1 (intercept not included), }\\
                        c^{-1}(1+c_{**,n})^{-9}: \quad  \text{Case 2 (intercept included)}
                    \end{cases}
\end{equation}
for some $c>0$  which  does not  depend on $n$.
Notice that $b_n^*$  is   $\mathcal{F}^*_n$ measurable, and
 $(b_n^*)^{-1}\le   c(1+c_{**,n})^{9}=O_p(1)$  by (\ref{e:c*1}).
This proves the  first claim in (\ref{e:norm-reU}). The  second   claim  follows  using the  same  argument as  in the  proof of (\ref{e:norm-re}).

\vskip.2cm \noindent
{\it Proof of  (\ref{e:norm-reU+})}.
 Observe that
 \begin{eqnarray*}
 ||W_{zzuu}||&\le&|| E\big[ (\sum _{t=1}^n D^{-1}z_tz^\prime_t u_t^2 D^{-1} )\big   |\mathcal{F}^*_n]  ||
 \le E\big[ ||\sum _{t=1}^n D^{-1}z_tu_t^2z^\prime_t D^{-1}|| \big   |\mathcal{F}^*_n]\\
&\le& \sum _{t=1}^nE[||D^{-1} z_tu_t||^2 \,|\mathcal{F}^*_n]\le b_{n3}=c(1+c_{**,n})=O_p(1)
\end{eqnarray*}
by  (\ref{e:LIZA0}) of Lemma \ref{e:LEL} which implies (\ref{e:norm-reU+}).

\vskip.2cm
\noindent{\it  Proof of  (\ref{e:OMEGA})}.  
Write  $D\Omega_n D= W_{zz}^{-1}W_{zzuu}W_{zz}^{-1}$, $(D\Omega _nD)^{-1}= W_{zz}W_{zzuu}^{-1}W_{zz}$.
By (\ref{e:norm-re}), (\ref{e:norm-re+}), (\ref{e:norm-reU}) and (\ref{e:norm-reU+}),
 \begin{eqnarray}
 \label{e:aWLLG}
 ||D\Omega_n D||_{sp}&\le& ||D\Omega_n D||\le ||W_{zz}^{-1}|| \, ||W_{zzuu}||\, ||W_{zz}^{-1}||\le b_{n4}=O_p(1),
\\
||(D\Omega_n D)^{-1}||_{sp}&\le&||(D\Omega_n D)^{-1}||\le  ||W_{zz}|| \,||W_{zzuu}^{-1}|| \, ||W_{zz}||\le b_{n5}=O_p(1).
\label{e:aWLLGG}
\end{eqnarray}
We will show  that
 \begin{eqnarray}\label{e:first}a^\prime D\Omega_n D a\ge b_n:=b_{n5}^{-1}.
\end{eqnarray}
Since $b_n^{-1}=b_{n5}=O_p(1)$   this proves  the  first claim in (\ref{e:OMEGA}). To verify (\ref{e:first}),
notice that
 the   smallest  eigenvalue $\lambda_{min}$
 of the matrix $D\Omega D$
and the  largest  eigenvalue   $\theta_{max}$  of the inverse matrix $(D\Omega_n D)^{-1}$ are related by the
equality $\theta_{max}=\lambda_{min}^{-1}$.  By (\ref{e:aWLLGG}),  $\theta_{max}\le b_{n5}$.  Thus, for $||a||=1$,
  $$
a^\prime D\Omega_n D a\ge \lambda_{min}=\theta_{max}^{-1}\ge b_n:= b_{n5}^{-1},
  $$
  where $b_n^{-1}=b_{n5}=O_p(1)$  which proves (\ref{e:first}).
  Finally, by (\ref{e:aWLLG}), for  $||a||=1$,
$a^\prime D\Omega_n D a\le  ||D\Omega_n D||_{sp}\le b_{n4}=O_p(1)$  which  proves the  second bound  in (\ref{e:OMEGA}).

\vskip.2cm
\noindent{\bf  Proof  of  (\ref{e:RA-reU}), (\ref{e:daz2-reU}) and (\ref{e:SuuC})}.
 Write
 \begin{eqnarray*}
D^{-1}S_{zzuu}D^{-1}-W_{zzuu}
&=&\sum _{t=1}^nD^{-1}\{z_tz^\prime_tu_t^2-E[z_tz^\prime_tu_t^2|\mathcal{F}_n^*]\}D^{-1}=\big(\delta_{jk}\big).
 \end{eqnarray*}
 To  prove   (\ref{e:RA-reU}), it suffices to  verify  that
\begin{eqnarray}\label{e:prr1+1}
\delta_{jk}=o_p(1).
 \end{eqnarray}
Recall  that $z_t=\mu_t+\widetilde z_t$ and $u_t=h_t\varepsilon_t$, where $E\varepsilon_t^2=1$. Hence,
 \begin{eqnarray*}
 E[ u_t^2 |\mathcal{F}^*_n]&=&h_t^2,\\
 E[ \widetilde z_t u_t^2 |\mathcal{F}^*_n]&=&h_t^2I_{gt}E[ \eta_t \varepsilon_t ^2]
=I_{gt}h_t^2\bar e, \quad  \bar e=E[ \eta_1 \varepsilon_1 ^2].
 \end{eqnarray*}
By (\ref{e:secon}),
 \begin{eqnarray*}
  z_t z^\prime_tu_t^2 &=&\widetilde z_t\widetilde z_t^\prime u_t^2 + \mu_t\mu_t^\prime u_t^2
+ \mu_t\widetilde z_t^\prime u_t^2 +\widetilde z_t\mu_t^\prime u_t^2,\nonumber\\
E[ z_t z^\prime_t u_t^2|\mathcal{F}^*_n]&=& E[ \widetilde z_t \widetilde z^\prime_tu_t^2|\mathcal{F}^*_n] + \mu_t\mu_t^\prime E[ u_t^2|\mathcal{F}^*_n]
+\mu_tE[\widetilde z^\prime_tu_t^2|\mathcal{F}^*_n]+E[\widetilde z_tu_t^2|\mathcal{F}^*_n]\mu_t^\prime\nonumber\\
&=& E[ \widetilde z_t \widetilde z^\prime_t u_t^2|\mathcal{F}^*_n] + \mu_t\mu_t^\prime h_t^2E[\varepsilon_t^2]
+\{h_t\mu_t\}\bar e^\prime\{h_tI_{gt}\} + \{h_tI_{gt}\}\bar e\{h_t\mu_t^\prime\}.\hspace{0.5cm}
 \end{eqnarray*}
 Then,
\begin{eqnarray*}
&&z_tz^\prime_tu_t^2-E[z_tz^\prime_tu_t^2|\mathcal{F}_n^*]=h_tI_{gt}(\eta_t\eta_t^\prime \varepsilon_t^2-E[\eta_t\eta_t^\prime \varepsilon_t^2])h_tI_{gt}+\mu_t\mu_t^\prime h_t^2(\varepsilon_t^2-E[\varepsilon_t^2])
\\&&\qquad +h_t\mu_t (\eta_t^\prime \varepsilon_t^2- E[\eta_t^\prime \varepsilon_t^2])h_tI_{gt}+h_tI_{gt}(\eta_t\varepsilon_t^2- E[\eta_t^\prime \varepsilon_t^2])h_t\mu_t^\prime.
\end{eqnarray*}
Therefore, setting $\gamma_{jt}=v_j^{-1}g_{jt}h_t$, it follows that
\begin{eqnarray*}
\delta_{jk}&=&
\sum _{t=1}^n\gamma_{jt}\gamma_{kt}(\eta_{jt}\eta_{kt}\varepsilon_t^2-E[\eta_{jt}\eta_{kt}\varepsilon_t^2])+\sum _{t=1}^n\{v_j^{-1}\mu_{jt}h_t\}\gamma_{kt}(\eta_{kt} \varepsilon_t^2- E[\eta_{kt} \varepsilon_t^2])\\
& & +\sum _{t=1}^n \{v_k^{-1}\mu_{kt}h_t\}\gamma_{jt}(\eta_{jt} \varepsilon_t^2- E[\eta_{jt} \varepsilon_t^2])+
\sum _{t=1}^n\{v_j^{-1}\mu_{jt}h_t\} \{v_k^{-1}\mu_{kt}h_t\}(\varepsilon_t^2- E[ \varepsilon_t^2])\nonumber\\
&=&
r_{n,jk}^{(1)}+r_{n,jk}^{(2)}+r_{n,jk}^{(3)}+r_{n,jk}^{(4)}.\nonumber
 \end{eqnarray*}
 To  prove   (\ref{e:prr1+1}), it suffices to show  that
\begin{eqnarray}\label{e:prr1++}
r_{n,jk}^{(i)}=o_p(1),\quad i=1,...,4.
 \end{eqnarray}
By  Assumption \ref{a:4R}, $\{\eta_{jt}\eta_{kt}\varepsilon_t^2\}$,
$\{\eta_{kt} \varepsilon_t^2\}$  and $\{\varepsilon_t^2\}$
are covariance  stationary  short memory zero mean sequences, and these  sequences   are  mutually   independent  of
the  weights  $\{\gamma_{jt}\gamma_{kt}\}$, $\{v_j^{-1}\mu_{jt}h_t\gamma_{kt}\}$
and $\{(v_j^{-1}\mu_{jt}h_t)(v_k^{-1}\mu_{kt}h_t)\}$. Moreover,
definition of  $v_k$  and $\gamma_{kt}$ and  (\ref{e:hkj3})  of Assumption  \ref{a:r3} imply that
$$
\sum_{t=1}^n \gamma^2_{kt}=1,  \quad  v_k^{-2}\sum_{t=1}^n\mu_{kt}^2h^2_t=O_p(1)
$$
and  by  (\ref{e:zzuu})
  of Assumption  \ref{a:4R},
$$
\max_{t=1, ..., n}\gamma^2_{kt}=o_p(1),  \quad v_k^{-2}\max_{t=1, ..., n}\mu_{kt}^2h^2_t=o_p(1).
$$
Thus, (\ref{e:prr1++}) follows by using Lemma  \ref{l:bba1-R} and applying a similar argument as in the proof of (\ref{e:RA-re}).
This  completes  the  proof of  (\ref{e:RA-reU}).

The claim (\ref{e:daz2-reU})  follows  using (\ref{e:RA-reU}) and  property
$W_{zzuu}^{-1}=O_p(1)$ of
 (\ref{e:norm-reU}):
\begin{eqnarray*}DS_{zzuu}^{-1}D&=&\big(D^{-1}S_{zzuu}D^{-1}\big)^{-1}=
\big(W_{zzuu}+o_p(1)\big)^{-1}=W_{zzuu}^{-1}\big(1+W_{zzuu}^{-1}\times o_p(1)\big)^{-1}\\
&=&W_{zzuu}^{-1}\big(1+o_p(1)\big)^{-1}
=W_{zzuu}^{-1} +o_p(1).
\end{eqnarray*}
\vskip.2cm
\noindent
\noindent{\it Proof of (\ref{e:SuuC})}. Write
\begin{eqnarray}
\label{e:SuuC1}
  D^{-1}S_{zzuu}^{(c)}D^{-1}&=&D^{-1}S_{zzuu}D^{-1}+D^{-1}(S_{zzuu}^{(c)}-S_{zzuu})D^{-1}.
  \end{eqnarray}
  By  (\ref{e:RA-reU}), $D^{-1}S_{zzuu}D^{-1}= W_{zzuu}+o_p(1).$  We  will  show  that
 \begin{eqnarray}
\label{e:SuuC2}
 D^{-1}(S_{zzuu}^{(c)}-S_{zzuu})D^{-1}=o_p(1),
  \end{eqnarray}
  which together  with (\ref{e:SuuC1})  implies (\ref{e:SuuC}):
 $ D^{-1}S_{zzuu}^{(c)}D^{-1}=W_{zzuu}+o_p(1).$
 We  have, $u_t^2-E[u_t^2|\mathcal{F}_{n,t-1}]=h_t^2(\varepsilon_t^2-\sigma_t^2)$, where
$\sigma_t^2=E[\varepsilon_t^2\,|\mathcal{F}_{t-1}]$. Write
 \begin{eqnarray*}
D^{-1}(S_{zzuu}-S_{zzuu}^{(c)})D^{-1}
&=&\sum _{t=1}^nD^{-1}z_tz^\prime_t(u_t^2-E[u_t^2|\mathcal{F}_{n,t-1}])D^{-1}=\big(\delta_{jk}\big).
 \end{eqnarray*}
Then  (\ref{e:SuuC2})  follows  if we show that
\begin{eqnarray}\label{e:ccc1+}
\delta_{jk}=o_p(1).
 \end{eqnarray}
We  have $z_t=\mu_t+\widetilde z_t$ and $u_t=h_t\varepsilon_t$. So,
 \begin{eqnarray*}
 &&z_t z^\prime_t =\widetilde z_t\widetilde z_t^\prime + \mu_t\mu_t^\prime
+ \mu_t\widetilde z_t^\prime+\widetilde z_t\mu_t^\prime ,\nonumber\\
 &&z_tz^\prime_t(u_t^2-E[u_t^2|\mathcal{F}_{n,t-1}])=z_tz^\prime_th_t^2(\varepsilon_t^2-\sigma_t^2)\nonumber\\
&&\quad =
h_tI_{gt}\eta_t\eta_t^\prime I_{gt}h_t(\varepsilon_t^2-\sigma_t^2)
+\mu_t\mu_t^\prime h_t^2(\varepsilon_t^2-\sigma_t^2)
\\&& \hspace{4.55cm} +h_t\mu_t \eta_t^\prime I_{gt}h_t(\varepsilon_t^2-\sigma_t^2)
+I_{gt}\eta_t\mu_t^\prime h_t^2(\varepsilon_t^2-\sigma_t^2).\nonumber
\end{eqnarray*}
Hence, denoting $\gamma_{jt}=v_j^{-1}g_{jt}h_t$,  we obtain
\begin{eqnarray*}
\delta_{jk}&=&
\sum _{t=1}^n\gamma_{jt}\gamma_{kt}\{\eta_{jt}\eta_{kt}(\varepsilon_t^2-\sigma_t^2)\}+\sum _{t=1}^n\{v_j^{-1}\mu_{jt}h_t\}\gamma_{kt}
\{\eta_{kt}(\varepsilon_t^2-\sigma_t^2)\}
\\
& & +\sum _{t=1}^n \{v_k^{-1}\mu_{kt}h_t\}\gamma_{jt}\{\eta_{jt}(\varepsilon_t^2-\sigma_t^2)\}+
\sum _{t=1}^n\{v_j^{-1}\mu_{jt}h_t\} \{v_k^{-1}\mu_{kt}h_t\}\{\varepsilon_t^2-\sigma_t^2\}\nonumber\\
&=&
r_{n,jk}^{(1)}+r_{n,jk}^{(2)}+r_{n,jk}^{(3)}+r_{n,jk}^{(4)}.\nonumber
 \end{eqnarray*}
 Observe,  that sequences  $\{w_{1t}=\eta_{jt}\eta_{kt}(\varepsilon_t^2-\sigma_t^2)\}$,
 $\{w_{2t}=\eta_{kt}(\varepsilon_t^2-\sigma_t^2)\}$,
 $\{w_{3t}=\eta_{jt}(\varepsilon_t^2-\sigma_t^2)\}$,
 $\{w_{4t}=\varepsilon_t^2-\sigma_t^2\}$
 are  sequences  of uncorrelated   random  variables  with  zero mean and  constant variance.
 For  example,  by assumption,  $\eta_{jt}\eta_{kt}$  are $\mathcal{F}_{t-1}$  measurable. Then, for  $t\ge s$,
\begin{eqnarray*}
 E[w_{1t}]&=&E\big[E[w_{1t}|\mathcal{F}_{t-1}]\big]=E\big[\eta_{jt}\eta_{kt}E[(\varepsilon_t^2-\sigma_t^2)|\mathcal{F}_{t-1}]\big]=0,\\
  E[w_{1t}w_{1s}]&=&E\big[\eta_{jt}\eta_{kt}\eta_{js}\eta_{ks}(\varepsilon_s^2-\sigma_s^2)E[(\varepsilon_t^2-\sigma_t^2)|\mathcal{F}_{t-1}]\big]=0,\\
   E[w_{1t}^2]&=&E\big[\eta_{jt}^2\eta^2_{kt}E[(\varepsilon_t^2-\sigma_t^2)^2|\mathcal{F}_{t-1}]\big]\\
   &\le &E\big[\eta_{jt}^2\eta^2_{kt}E[\varepsilon_t^4|\mathcal{F}_{t-1}]\big]=E\big[E[\eta_{jt}^2\eta^2_{kt}\varepsilon_t^4|\mathcal{F}_{t-1}]\big]=E\big[\eta_{j1}^2\eta^2_{k1}\varepsilon_1^4\big]<\infty.
\end{eqnarray*}
Then  using the   same  argument as  in the proof  of (\ref{e:prr1++}) it  follows
\begin{eqnarray*} 
r_{n,jk}^{(i)}=o_p(1),\quad i=1,...,4.
 \end{eqnarray*}
which  proves  (\ref{e:ccc1+})  and  completes  the proof of   (\ref{e:SuuC}).

 This   completes  the  proof of  the part  (ii) and of  the  lemma.
\hfill $\Box$

\newpage

\vskip.2cm\noindent {\bf Proof of  Corollary \ref{c:bn}}.  The  claim (\ref{e:bn*})   is  shown in  (\ref{e:gasa4}), and the  claim  (\ref{e:bn**}) is  shown in  (\ref{e:gasa2++}). \hfill $\Box$

\vskip.2cm
\begin{lemma}\label{e:LEL}Under Assumptions  of  Theorem \ref{t:r1}, the  exists $c>0$  such  that
\begin{eqnarray}
\sum _{t=1}^n  E\big[||D_g^{-1}z_t||^2 \ |\mathcal{F}^*_n\big]\le c(1+c_{*,n}),&& \hspace{-10pt} \sum _{t=1}^n  E\big[||D^{-1}z_tu_t||^2 \ |\mathcal{F}^*_n\big]\le c(1+c_{**,n}),
\label{e:LIZA0}\\
\sum_{t=1}^n E\big[||D^{-1}z_th_t||^2 I(\varepsilon_t^2<  \delta)|\mathcal{F}^*_n\big] &\le&  c(1+c_{**,n})\delta^{1/4},
\label{e:LIZA0d}
\end{eqnarray}
for  sufficiently  small $\delta>0$,
where $c$ does not  depend  on  $n$ and $\delta $ and $c_{*,n}=O_p(1)$,

\noindent $c_{**,n}=O_p(1)$.

In addition, under assumptions  of  Theorem \ref{t:r1-R},
\begin{eqnarray}
\max_{t=1,..., n}||D^{-1}z_tu_t||^2=o_p(1),&&   \max_{t=1,..., n}||D_g^{-1}z_t||^2=o_p(1),\label{e:LIZA2}\\
\sum_{t=1}^nE\big[b_n^{-1}||D^{-1}z_tu_t||^2I\big(b_n^{-1}||D^{-1}z_tu_t||^2 &\ge& \epsilon\big)\,|\mathcal{F}_{n,t-1}\big]
=o_p(1) \,\,\, \mbox{for any  $\epsilon>0$},
\label{e:LIZA2D}
\end{eqnarray}
where  $b_n$  is $\mathcal{F}_{n}^*$  measurable, $b_n^{-1}=O_p(1)$
and $\mathcal{F}_{n,t-1}$  is  defined  as  in (\ref{e:MD*}).
\end{lemma}
\noindent {\bf Proof of  Lemma \ref{e:LEL}}.
{\it Proof of  (\ref{e:LIZA0})}. Denote
\begin{eqnarray*}
 b_{1t}=||D_g^{-1}\mu_t||^2+||D_g^{-1}I_{gt}||^2, \quad  \theta_{1t}=1+ ||\eta_t||^2,\\
 b_{2t}=||D_g^{-1}\mu_t h_t||^2+||D_g^{-1}I_{gt}h_t||^2, \quad  \theta_{2t}=\varepsilon_t^2+ ||\eta_t||^2\varepsilon_t^2.\nonumber
\end{eqnarray*}
By (\ref{e:Igt}),
\begin{eqnarray}
\nonumber
||D_g^{-1}z_t||^2=||D_g^{-1}\mu_t+D_g^{-1}I_{gt}\eta_t||^2&\le&2 (||D_g^{-1}\mu_t||^2+||D_g^{-1}I_{gt}||^2||\eta_t||^2)\hspace{0.7cm}\\
&\le& 2b_{1t}\theta_{1t},\label{e:IgtLL} \\
||D_g^{-1}z_tu_t||^2=||D_g^{-1}\mu_t h_t\varepsilon_t+D_g^{-1}I_{gt}\eta_t h_t\varepsilon_t||^2&\le&2 b_{2t}\theta_{2t}. \nonumber \end{eqnarray}
 By  Assumption \ref{a:ETA}(i)  and Assumption \ref{a:4R}(i),

\noindent \hspace{2.3cm} $
E[\theta_{1t} \ |\mathcal{F}^*_n]=E[\theta_{1t}]=E[\theta_{11}], \quad E[\theta_{2t} \ |\mathcal{F}^*_n]=E[\theta_{2t}]=E[\theta_{21}].
$

 \noindent This  implies
\begin{eqnarray}
 E\big[||D_g^{-1}z_t||^2 \ |\mathcal{F}^*_n\big]&\le&
2 b_{1t}E[\theta_{11}],\label{e:gasil1}\\
 E\big[||D^{-1}z_tu_t||^2 \ |\mathcal{F}^*_n\big]&\le&
2 b_{2t}E[\theta_{21}], \nonumber\\
\mbox{$\sum _{t=1}^n E\big[||D_g^{-1}z_t||^2 \ |\mathcal{F}^*_n\big]$}&=&\mbox{$2 E[\theta_{11}](\sum _{t=1}^n b_{1t})$}, \nonumber\\
\mbox{$\sum _{t=1}^n E\big[||D_g^{-1}z_tu_t||^2 \ |\mathcal{F}^*_n\big]$}&=&\mbox{$2E[\theta_{21}](\sum _{t=1}^n b_{2t})$}. \nonumber
 \end{eqnarray}
 Notice  that
 \begin{eqnarray}&&\mbox{$\sum _{t=1}^nb_{1t}=\sum _{t=1}^n||D_g^{-1}\mu_t||^2+\sum _{t=1}^n||D_g^{-1}I_{gt}||^2=c_{*,n}+p,$}\nonumber\\
 &&\mbox{$\sum _{t=1}^nb_{2t}=\sum _{t=1}^n||D^{-1}\mu_th_t||^2+\sum _{t=1}^n||D^{-1}I_{gt}h_t||^2=c_{**,n}+p,$}
  \label{e:bounds18}
 \end{eqnarray}
 by  definition (\ref{e:c*}) of $c_{*,n}$  and $c_{**,n}$  and  because
  \begin{eqnarray*}
  &&\mbox{$\sum _{t=1}^n||D_g^{-1}I_{gt}||^2= \sum_{k=1}^pv_{gk}^{-2}(\sum _{t=1}^ng_{kt}^2)=p,$}\\
  &&\mbox{$ \sum _{t=1}^n||D^{-1}I_{gt}h_t||^2= \sum_{k=1}^pv_{k}^{-2}(\sum _{t=1}^ng_{kt}^2h_t^2)=p.$}\nonumber
 \end{eqnarray*}
 Moreover, $c_{*,n}=O_p(1)$, $c_{**,n}=O_p(1)$ by (\ref{e:c*1}).
Clearly, (\ref{e:gasil1})  and (\ref{e:bounds18})  prove~(\ref{e:LIZA0}).

\vskip.2cm
\noindent {\it Proof of  (\ref{e:LIZA0d})}. Denote

\hspace{3.5cm}$\theta_{2t}(\delta)= I(\varepsilon_t^2<  \delta)+||\eta_t||^2I(\varepsilon_t^2<  \delta).
$

\noindent Recall, that  by  assumption, $\varepsilon_t$ is a  stationary  sequence,
and  by  Assumption  \ref{a:ETA}(i), $E[||\eta_t||^4]=E[||\eta_1||^4]$.
 Then,
\begin{eqnarray*}
E[\theta_{2t}(\delta)]&\le&
 E[I(\varepsilon_t^2<  \delta)]+(E[||\eta_t||^4)^{1/2}(E[I(\varepsilon_t^2<  \delta)])^{1/2}\\
 &=& E[I(\varepsilon_1^2<  \delta)]+(E[||\eta_1||^4)^{1/2}(E[I(\varepsilon_1^2<  \delta)])^{1/2}.
\end{eqnarray*}
We will show   that  for  sufficiently small $\delta>0$,

\hspace{4.5cm}
$
\mbox{$E[I(\varepsilon_1^2<  \delta)]\le C\delta^{1/2}.$}
$

\noindent Indeed,   by  Assumption  \ref{a:r0}, the variable $\varepsilon_1$  has probability distribution density $f(x)$ and $f(x)\le c<\infty$  when $|x|\le  x_0$   for  some  $x_0>0$. Without restriction of  generality assume that $\delta\le x_0$. Then,
$$
\mbox{$E[I(\varepsilon_1^2<  \delta)]=\int I(|x|\le \delta^{1/2})f(x)dx
\le c\int I(|x|\le \delta^{1/2})dx\le C\delta^{1/2}.$}
$$
Therefore,   $E[\theta_{2t}(\delta)]\le C\delta^{1/4}$, and
 as  in (\ref{e:gasil1}), we  obtain
\begin{eqnarray*}
&&E\big[||D^{-1}z_th_t||^2I(\varepsilon_t^2<  \delta) \ |\mathcal{F}^*_n\big]\le
2 b_{2t}
E[\theta_{2t}(\delta)]
\le C \delta^{1/4}b_{2t},\\
 &&\sum_{t=1}^n E\big[||D^{-1}z_th_t||^2I(\varepsilon_t^2<  \delta) \, |\mathcal{F}^*_n\big]\le
 C\delta^{1/4}(\sum_{t=1}^n b_{2t}) 
 \le C\delta^{1/4}(p+c_{**,n}),
    \end{eqnarray*}
   which proves  (\ref{e:LIZA0d}).

\vskip.2cm
\noindent{\it Proof of  (\ref{e:LIZA2})}. We  will  prove the first claim (the  proof of  the  second claim is  similar).
By (\ref{e:IgtLL}),
$ 
||D^{-1}z_tu_t||^2\le2 b_{2t}\theta_{2t}.
$ 
\noindent Let  $K>0$  be  a  large  number. Then, $\theta_{2t}\le K+ \theta_{2t}I(\theta_{2t}\ge  K)$. Therefore,
  \begin{eqnarray}
\max_{t=1, ..., n}||D^{-1}z_tu_t||^2&\le&  2K (\max_{t=1, ..., n}b_{2t})+2\sum_{t=1}^nb_{2t}\,\theta_{2t}I(\theta_{2t}\ge  K).
 \label{e:dafg}
 \end{eqnarray}
 By (\ref{e:zzuu})  of  Assumption \ref{a:4R} and (\ref{e:bounds18}),
  \begin{eqnarray}
 \max_{t=1, ..., n}b_{2t}=o_p(1), \qquad  \sum_{t=1}^nb_{2t}=O_p(1).
 \label{e:dafg2}
 \end{eqnarray}
 Since  $\{b_t\}$  and  $\{\theta_{2t}\}$  are   mutually  independent, then by (\ref{e:su12-R+})  of  Lemma \ref{l:bba1-R},
 \begin{eqnarray}
\sum_{t=1}^nb_{2t}\,\theta_{2t}I(\theta_{2t}\ge  K)=O_p\big( \sum_{t=1}^nb_{2t}\big)\Delta_{n,K}, \quad \Delta_{n,K}=\max_{t=1, ..., n}E[\theta_{2t}I(\theta_{2t}\ge  K)].
 \label{e:dafg1}
 \end{eqnarray}
 We  will show  that
   \begin{eqnarray}  \Delta_{n,K}\le   \Delta_{K},
 \label{e:delta}
 \end{eqnarray}
 where $\Delta_{K}\rightarrow 0$, $K \rightarrow  \infty$ and  $\Delta_{K}$ does not depend  on $n$. Together  with (\ref{e:dafg}) this implies
  \begin{eqnarray*}
\max_{t=1, ..., n}||D^{-1}z_tu_t||^2\le K o_p(1)+O_p(1) \Delta_K=o_p(1), \quad  n,K\rightarrow \infty.
 \end{eqnarray*}
Next we  prove (\ref{e:delta}). Set  $L=K^{1/4}$.  Then, letting $\varepsilon_{L,t}^{2+}=\varepsilon_t^2I(\varepsilon_t^2>L)$, we  obtain
  \begin{eqnarray*} \theta_{2t}&=&\varepsilon_t^2 (||\eta_t||^2+1)\le \{\varepsilon_{L,t}^{2+}
  + LI(\varepsilon_t^2\le L)\}(||\eta_t||^2+1),\\
\theta_{2t} I(\theta_{2t}\ge  K)&\le&
\varepsilon_{L,t}^{2+}(||\eta_t||^2+1)+
L(||\eta_t||^2+1)I\big(L(||\eta_t||^2+1)\ge K\big),\\
E[\theta_{2t} I(\theta_{2t}\ge  K)]&\le&(E[
(\varepsilon_{L,t}^{2+})^2])^{1/2}(E[(||\eta_t||^2+1)^2])^{1/2}+
L E[(||\eta_t||^2+1)^4] (K/L)^{-1}\\
&\le&(E[
(\varepsilon_{L,1}^{2+})^2])^{1/2}(E[(||\eta_1||^2+1)^2])^{1/2}+
(L^2/K) E[(||\eta_1||^2+1)^2]\\
 &=:&\Delta_K \rightarrow 0, \quad  K\rightarrow \infty
 \end{eqnarray*}
 since, as  $K\rightarrow  \infty$,  $L^2/K=K^{-1/2}  \rightarrow 0$, $E[(\varepsilon_{L,1}^{2+})^2]
\rightarrow 0$ and $E[||\eta_1||^4<\infty$. This  implies (\ref{e:delta}).

\vskip.2cm
\noindent {\it Proof of (\ref{e:LIZA2D})}. Denote  by  $i_n$  the  left hand  side  of (\ref{e:LIZA2D}).
By (\ref{e:IgtLL}), $||D^{-1}z_tu_t||^2\le 2 b_{2t}\theta_{2t}.$ Let  $K>0$  be a  large  number. Then,
\begin{eqnarray*}
&&b_n^{-1}||D^{-1}z_tu_t||^2I(b_n^{-1}||D^{-1}z_tu_t||^2\ge \epsilon)\le
2b_n^{-1} b_{2t}\theta_{2t}I\big(2b_n^{-1} b_{2t}\theta_{2t}\ge \epsilon\big)\\
&&\quad \le
2b_n^{-1} b_{2t}KI\big(2b_n^{-1} b_{2t}K\ge \epsilon\big)I(\theta_{2t}\le K)
+2b_n^{-1} b_{2t}\theta_{2t}I(\theta_{2t}> K)\\
&&\quad \le 
\epsilon_n^{-1}K^2(2b_n^{-1} b_{2t})^2+
2b_n^{-1} b_{2t}\theta_{2t}I(\theta_{2t}> K).
\end{eqnarray*}
Observe, that $b_n^{-1} b_{2t}$  is $\mathcal{F}_{n,t-1}$  measurable.  Then,

\hspace{2cm}$
i_n\le 
\epsilon_n^{-1}K^2(2b_n^{-1})^2 \sum_{t=1}^nb_{2t}^2+
2b_n^{-1}\sum_{t=1}^n b_{2t}\theta_{2t}I(\theta_{2t}> K).
$

\noindent Together   with (\ref{e:dafg1}), (\ref{e:delta}) and  (\ref{e:dafg2}), this implies:
\begin{eqnarray*}
i_n
&\le&\mbox{$
\epsilon_n^{-1}K^2(2b_n^{-1})^2 (\max_{t=1,..., n}b_{2t})(\sum_{t=1}^nb_{2t})+
2b_n^{-1}(\sum_{t=1}^n b_{2t})\Delta_K$}\\
&\le&
\epsilon_n^{-1}K^2O_p(1)o_p(1)+O_p(1)\Delta_K=o_p(1), \quad n,K\rightarrow  \infty.
\end{eqnarray*}
This proves (\ref{e:LIZA2D})
 and completes  the  proof of  the  lemma. \hfill $\Box$

\section{Additional Monte Carlo simulation results}
In this section, we further evaluate the finite-sample performance of our robust OLS estimation method using two examples of regression models with fixed parameters, where the regressors $z_t$ and regression noise $u_t$ exhibit complex, non-standard structures.

\vskip.2cm \noindent {\bf Example  1}.
As in the Monte Carlo section of the main paper, we generate arrays of samples from a regression model with a fixed parameter and an intercept, using a sample size of $n = 1500$ and $1000$ replications. We first consider the following model:
\begin{eqnarray}
y_t=\beta_1+\beta_{2}z_{2t}+\beta_{3}z_{3t}+u_t, \quad u_t=h_t \varepsilon_t, \nonumber\\
\quad \beta=(\beta_1,\beta_2,\beta_3)^\prime= (0.5, 0.4,
0.3)^\prime.
\label{SMC:OLSy}
\end{eqnarray}
We  specify  the  scale factor $h_t$ in the regression noise $u_t=h_t\varepsilon_t$  as  a deterministic  trend  $h_t=0.4(t/n)$, and  a stationary martingale difference noise   $\varepsilon_t$ is generated  from a GARCH($1,1$) process
\begin{eqnarray}  \label{S:GARCHnoise}
\varepsilon_t =\sigma_t e_t, \quad \sigma^2_t =
1+0.7\sigma^2_{t-1}+0.2\varepsilon^2_{t-1}, \quad e_t\sim i.i.d.\,\mathcal{N}(0,1).
\end{eqnarray}

\noindent Define the regressors  as $z_{1t}=1$ and  $z_{kt}=\mu_{kt}+g_{kt}\eta_{kt}$ for $k=2,3$,
where
\begin{eqnarray}\label{SMC:Model_5}
\mu_{2t}=0.5\sin(\pi t/n)+1, &&  g_{2t}=\Big|\dfrac{1}{2\sqrt{n}}\sum\limits^t_{j=1} \nu_{j}
\Big|+0.25, \quad \nu_j \sim {\rm i.i.d.} {\cal N}(0,1),\nonumber\\
\mu_{3t}=0.5\sin(0.5\pi t/n)+1, && g_{3t}=0.5\sin(3\pi t/n)+1, \nonumber\\
\eta_{kt}= 0.5\eta_{k,t-1}+\xi_{kt},&& \xi_{2t}=\varepsilon_{t-1},\quad \xi_{3t}=\varepsilon_{t-2}.
\end{eqnarray}

\begin{figure}[H]
\centering
\begin{minipage}[t]{0.45\linewidth}
    \centering
\includegraphics[width=\textwidth]{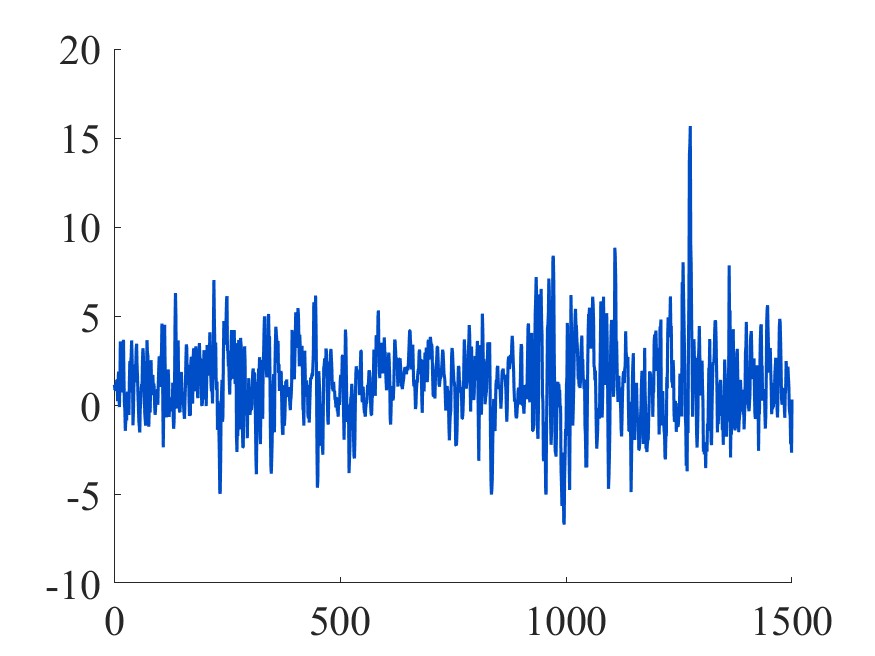}
 \subcaption{$y_t$}
	  \end{minipage}
\begin{minipage}[t]{0.45\linewidth}
    \centering
\includegraphics[width=\textwidth]{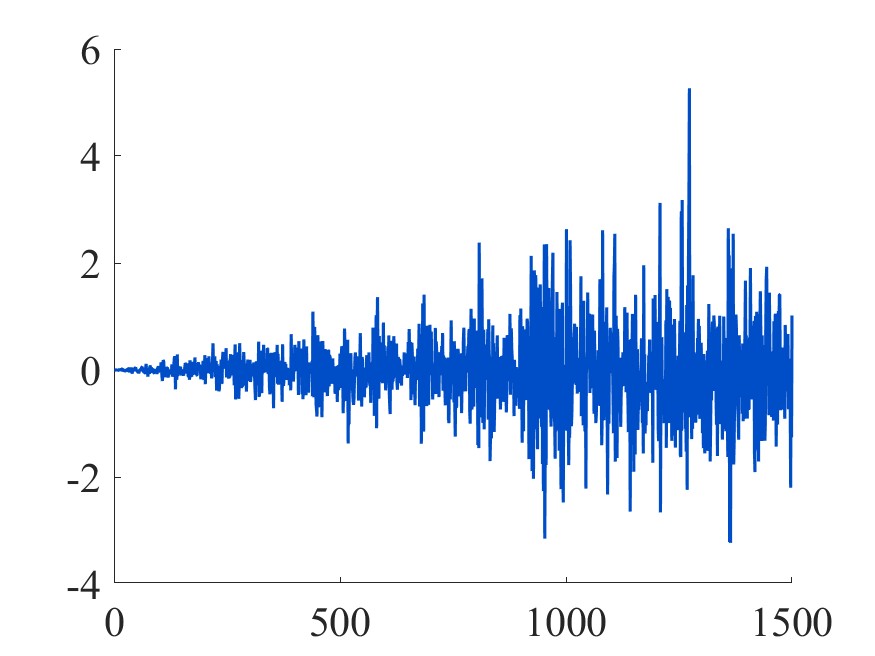}
\subcaption{{$u_{t}$}}
	  \end{minipage}\\
\begin{minipage}[t]{0.45\linewidth}
    \centering
\includegraphics[width=\textwidth]{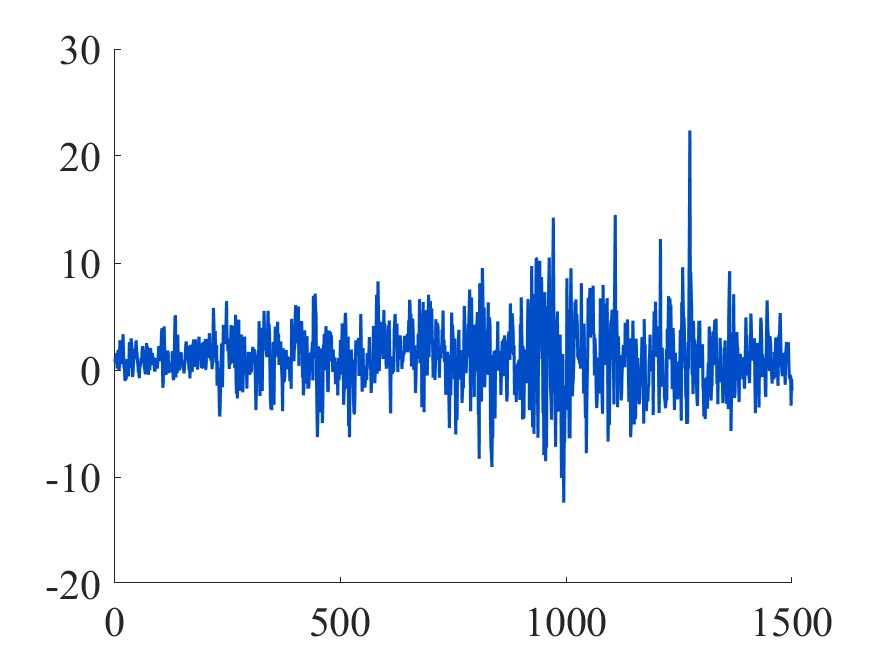}
 \subcaption{$z_{2t}$}
	  \end{minipage}
	  \begin{minipage}[t]{0.45\linewidth}
    \centering
\includegraphics[width=\textwidth]{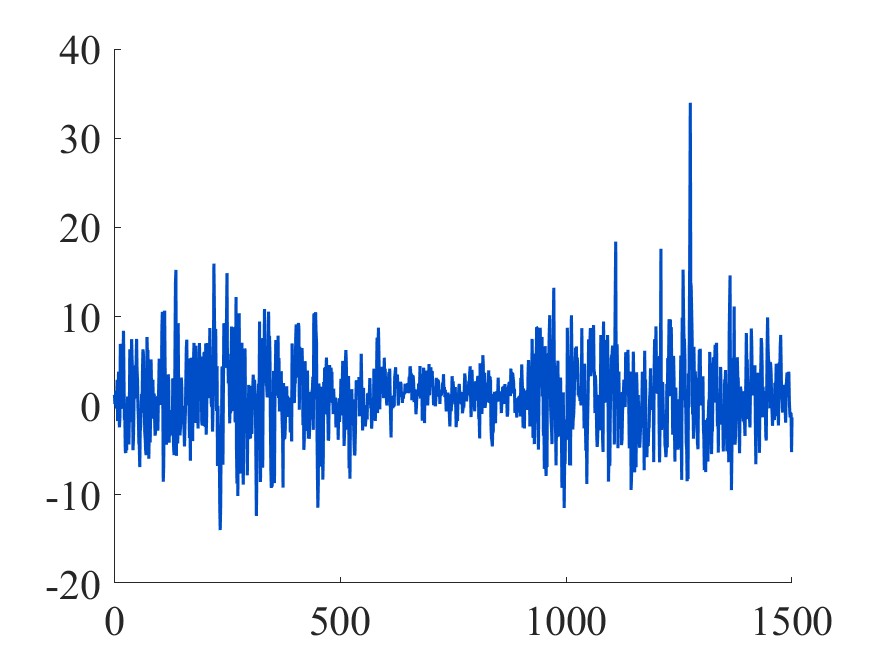}
 \subcaption{$z_{3t}$}
	  \end{minipage}
\caption{Plots of $y_t$, $u_{t}$, $z_{2t}$ and $z_{3t}$ in Model
(\ref{SMC:Model_5}), $n=1500$.}
\label{WSDA3YZ}
\end{figure}

Figure \ref{WSDA3YZ} displays plots of a sample of variables $y_t$, $z_t$, and $u_t$ for $t = 1, \dots, 1500$ generated by Model (\ref{SMC:OLSy})-(\ref{SMC:Model_5}), which exhibit clear patterns of non-stationary behavior.
The Monte Carlo simulation results  for sample size $n = 1500$ based on 1000 replications are reported in the Table \ref{SMC:tab:OLSm5}.
 Since the regressors $z_t$ and regression noise $u_t$ in this model satisfy the assumptions of Corollary \ref{c:co1}, as expected, the Monte Carlo simulation results confirm excellent performance of the robust OLS estimator. In particular, the empirical coverage of the $95\%$ confidence intervals is close to the nominal $95\%$, whereas the standard OLS estimator exhibits significant coverage distortions.

\begin{table}[H]
\caption{Robust  OLS estimation in Model (\ref{SMC:Model_5}), $n=1500$.}
\label{SMC:tab:OLSm5}\centering
\begin{tabular}{cccccc}
\hline Parameters & Bias & RMSE & CP & CP$_{st}$ & SD \\
\hline $\beta_1$ & -0.00036  & 0.02738 & 94.3 & 89.8 & 0.02738  \\
$\beta_2$ & 0.00050  & 0.01681  & 93.8 & 79.5 & 0.01680  \\
$\beta_3$ & -0.00003  & 0.00682  & 95.6 & 85.5 & 0.00682  \\
\hline &  &  &  &  &
\end{tabular}
\end{table}

\vskip.2cm \noindent {\bf Example  2}. Next, we provide an example of a regression model in which the components $\beta_1, \beta_2, \beta_3$  of the fixed regression parameter are estimated at different rates.
Consider regression model (\ref{SMC:OLSy}) with    $\varepsilon_t$,  $\eta_{2t},\eta_{3t}$ defined  as  in Example 1. Set  $h_t \equiv 1$, and let
 the means  $\mu_{kt}$  and  scale  factors  $g_{kt}$, $k=2,3$ be defined as  follows:
 \begin{eqnarray}
\mu_{2t}=[0.5\sin(10\pi t/n)+1]\, \sqrt{g_{2t}}, &&  g_{2t}=t, \nonumber\\
\mu_{3t}=[0.5\sin(5\pi t/n)+1] \, g_{3t}, && g_{3t}=t^{\gamma},\quad \gamma=\dfrac{1}{2}, 0,-\dfrac{1}{4}, -\dfrac{1}{2}. \label{SMC:Model6}
\end{eqnarray}
This model satisfies the assumptions of Corollary \ref{c:co1} (see also Remark \ref{r:2.1} in the main paper).
Therefore, the  corresponding $t$-statistics  for $k=1,2,3$ have the following property:
\begin{eqnarray}\label{e:nono}
&& \frac{\widehat \beta_k -\beta_k}{\sqrt{\widehat \omega_{kk}}}\rightarrow
_d \mathcal{N}(0, 1), \quad \sqrt{ \widehat \omega_{kk}}\asymp_pv_k^{-1},
\end{eqnarray}
where, the robust   standard errors $\sqrt{ \widehat \omega_{kk}}$  are inversely   proportional to the consistency   rate
$$
v_k= \big(\sum_{j=1}^n g_{jt}^2\big)^{1/2}.
$$
  In this  model, the  intercept $\beta_1$   associated  with the regressor  $z_{1t}=1$ is  estimated  at   the  consistency  rate  $v_1=\sqrt n$;   the  parameter $\beta_2$   linked  with  the regressor  $z_{2t}$ (with $g_{2t}=t$)
   at the   rate  $v_2\sim n^{3/2}$,
 and  the  parameter $\beta_3$  linked  with the regressor  $z_{3t}$ (with $g_{3t}=t^\gamma$)  at the   rate $v_3\sim n^{\gamma+1/2}$. 
The rate $v_3$ is   super-fast,  $n$,   when  $\gamma=1/2$;   standard, $n^{1/2}$,  when  $\gamma=0$;   super-slow,  $n^{1/4}$,  when  $\gamma=-1/4$;  and  logarithmic,  $\log n$,  when  $\gamma=-1/2$.
Monte Carlo results reported in Table \ref{SMC:tab:OLS6g} confirm the validity of the normal approximation (\ref{e:nono}) in finite samples ($n=1500$, based on $1000$ replications).
 In particular, the coverage of the robust $95\%$ confidence intervals is close to the nominal level for all three parameters
$\beta_1, \beta_2, \beta_t$  and for  all
 values  of   $\gamma$ considered in the construction of the regressor $z_{3t}$.
In contrast, the coverage rates based on the standard OLS method exhibit noticeable distortions, especially for
 $\beta_{2t}$ and $\beta_{3t}$.

 As expected, smaller values of
$\gamma$ are associated with slower consistency rates
$v_3$, wider confidence intervals, and larger standard deviations for the estimator of
$\beta_3$.

\begin{table}[H]
\caption{Robust  OLS estimation in Model (\ref{SMC:Model6}), $n=1500$.}
\label{SMC:tab:OLS6g}\centering
\begin{tabular}{ccccccc}
\hline $\gamma$ &Parameters & Bias & RMSE & CP & CP$_{st}$ & SD \\
\hline
$ $ & $\beta_1$ & -0.00331   & 0.08888  & 94.1 & 93.3 & 0.08882  \\
$1/2$ & $\beta_2$ & 2.4E-06  & 0.00004   & 95.3 & 86.4 & 0.00004   \\
$ $ & $\beta_3$ & -0.00008  & 0.00128  & 94.5 & 85.6 & 0.00128  \\
\hline
$ $ & $\beta_1$ & -0.00406    & 0.08976  & 94.5 & 93.4 & 0.08967  \\
$0$ & $\beta_2$ & 2.4E-06  & 0.00004   & 95.8 & 86.9 & 0.00004  \\
$ $ & $\beta_3$ & 0.00272   & 0.03384   & 94.9 & 85.9 & 0.03373  \\
\hline
$ $ & $\beta_1$ & -0.00397  & 0.08884  & 94.6 & 93.9 & 0.08875  \\
$-1/4$ & $\beta_2$ & 2.6E-06  & 0.00004   & 95.6 & 85.9 & 0.00004   \\
$ $ & $\beta_3$ & 0.01275   & 0.14219   & 95 & 87.2 & 0.14162   \\
\hline
$ $ & $\beta_1$ & -0.00319   & 0.08628  & 95 & 94.7 & 0.08622   \\
$-1/2$ & $\beta_2$ & 3.0E-06  & 0.00004   & 95.5 & 86.3 & 0.00004   \\
$ $ & $\beta_3$ & 0.04468   & 0.43022   & 95.1 & 91.3 & 0.42790  \\
\hline
\end{tabular}
\end{table}

\vskip.2cm \noindent Table \ref{SMC:tab:OLS6n} reports the estimation results for the parameters $\beta_1, \beta_2, \beta_3$   for  sample  sizes  $n=200, 800, 1500, 3000$, when the regressor $z_t$ is  generated   with $\gamma=-1/4$  and $\beta_3$  is estimated  with the super-slow  rate  $v_3=n^{1/4}$. The coverage rates for the robust OLS method are close to the nominal level in all cases. As expected, as $n$  increases, the standard errors of all three parameter estimates decrease; however, for $\beta_3$, which is estimated  with  the super-slow rate  $ n^{1/4}$, the reduction in the standard deviation is relatively slow.

\begin{table}[H]
\caption{Robust  OLS estimation in Model (\ref{SMC:Model6}), $\gamma=-1/4$, $n=1500$.}
\label{SMC:tab:OLS6n}\centering
\begin{tabular}{ccccccc}
\hline $n$ &Parameters & Bias & RMSE & CP & CP$_{st}$ & SD \\
\hline
$ $ & $\beta_1$    & -0.01307  & 0.23584  & 94.9  & 95.1  & 0.23548  \\
$200$ & $\beta_2$  & 0.00008  & 0.00073  & 92.5  & 86.6  & 0.00073  \\
$ $ & $\beta_3$   & 0.03138  & 0.22902  & 94.6  & 91.7  & 0.22686  \\
\hline
$ $ & $\beta_1$     & -0.00740  & 0.12187  & 95.1  & 94.6  & 0.12164  \\
$800$ & $\beta_2$    & 0.00001  & 0.00010  & 94.3  & 86.3  & 0.00009  \\
$ $ & $\beta_3$    & 0.01302  & 0.16151  & 94.8  & 88.1  & 0.16098  \\
\hline
$ $ & $\beta_1$ & -0.00397  & 0.08884  & 94.6 & 93.9 & 0.08875  \\
$1500$ & $\beta_2$ & 2.6E-06  & 0.00004   & 95.6 & 85.9 & 0.00004   \\
$ $ & $\beta_3$ & 0.01275   & 0.14219   & 95 & 87.2 & 0.14162   \\
\hline
$ $ & $\beta_1$    & -0.00319  & 0.06599  & 93.2  & 92.2  & 0.06592  \\
$3000$ & $\beta_2$    & 0.00000  & 0.00001  & 94.1  & 83.3  & 0.00001  \\
$ $ & $\beta_3$    & 0.00913  & 0.12430  & 94.8  & 84.3  & 0.12396  \\
\hline
\end{tabular}
\end{table}

\begin{figure}[H]
\centering
\begin{minipage}[t]{0.42\linewidth}
    \centering
\includegraphics[width=\textwidth]{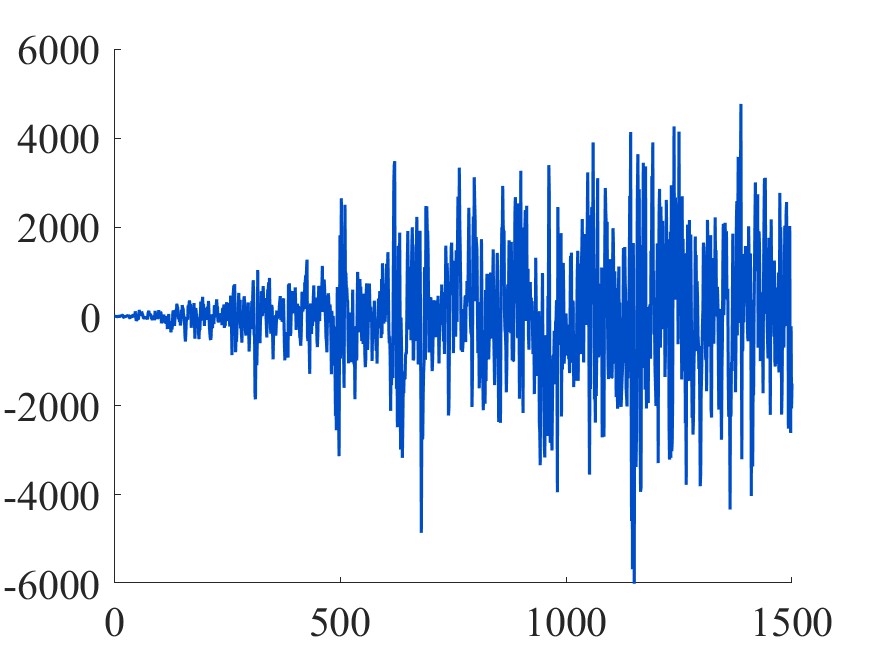}
 \subcaption{$y_t$ $(\gamma=1/2)$}
	  \end{minipage}
\begin{minipage}[t]{0.42\linewidth}
    \centering
\includegraphics[width=\textwidth]{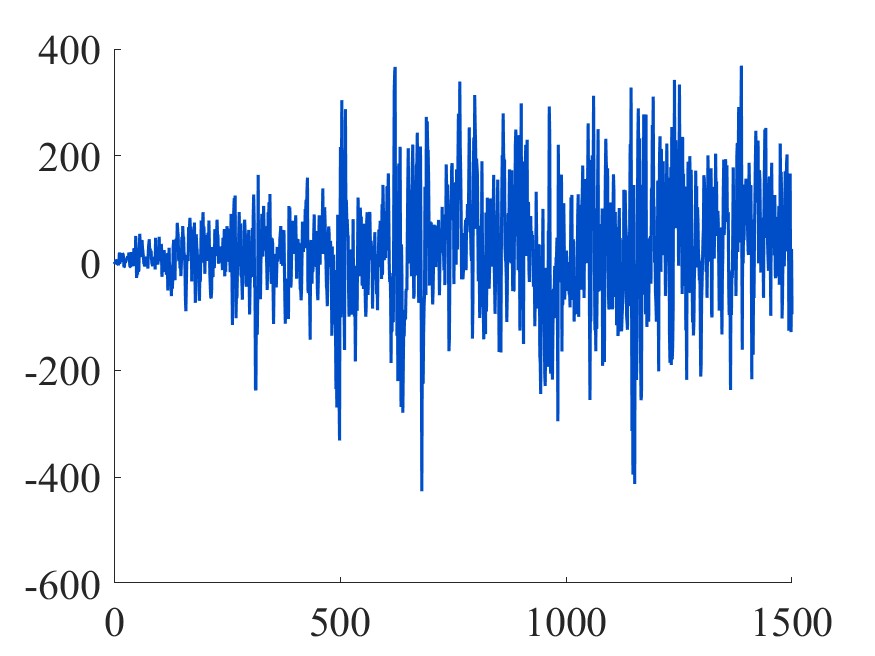}
\subcaption{{$z_{3t}$ $(\gamma=1/2)$}}
	  \end{minipage}\\
\begin{minipage}[t]{0.42\linewidth}
    \centering
\includegraphics[width=\textwidth]{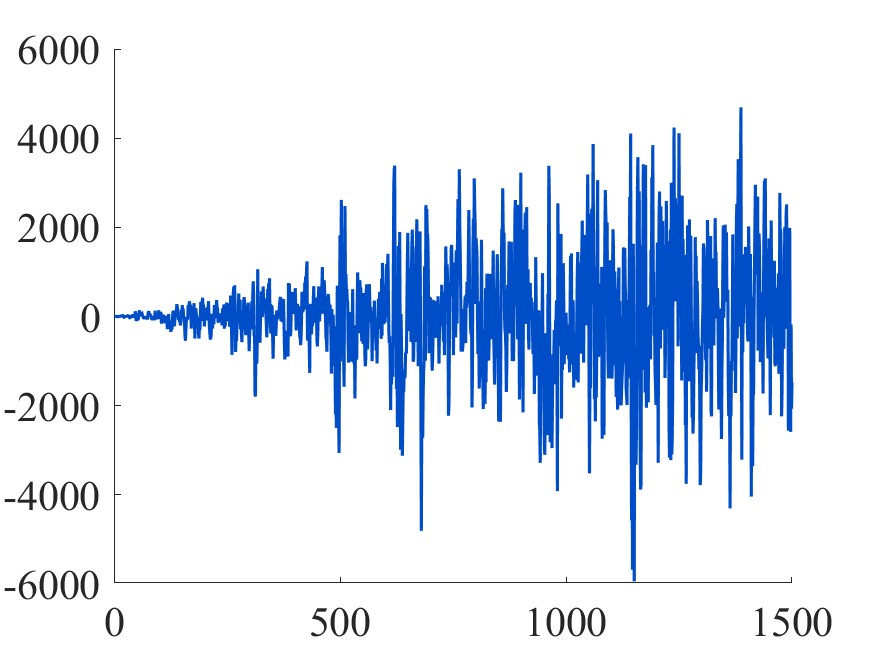}
 \subcaption{$y_t$ $(\gamma=0)$}
	  \end{minipage}
	  \begin{minipage}[t]{0.42\linewidth}
    \centering
\includegraphics[width=\textwidth]{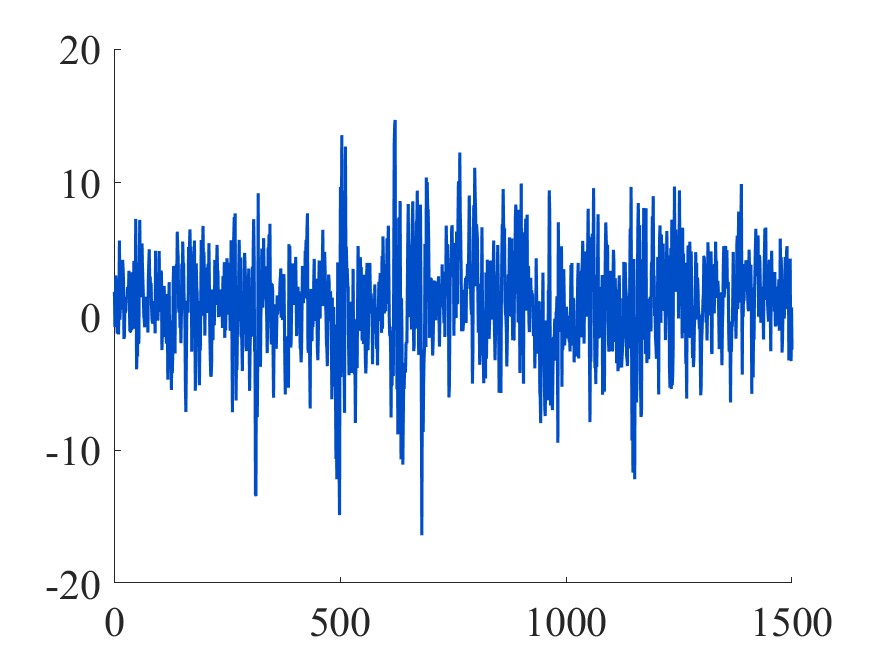}
 \subcaption{$z_{3t}$ $(\gamma=0)$}
	  \end{minipage}\\
\begin{minipage}[t]{0.42\linewidth}
    \centering
\includegraphics[width=\textwidth]{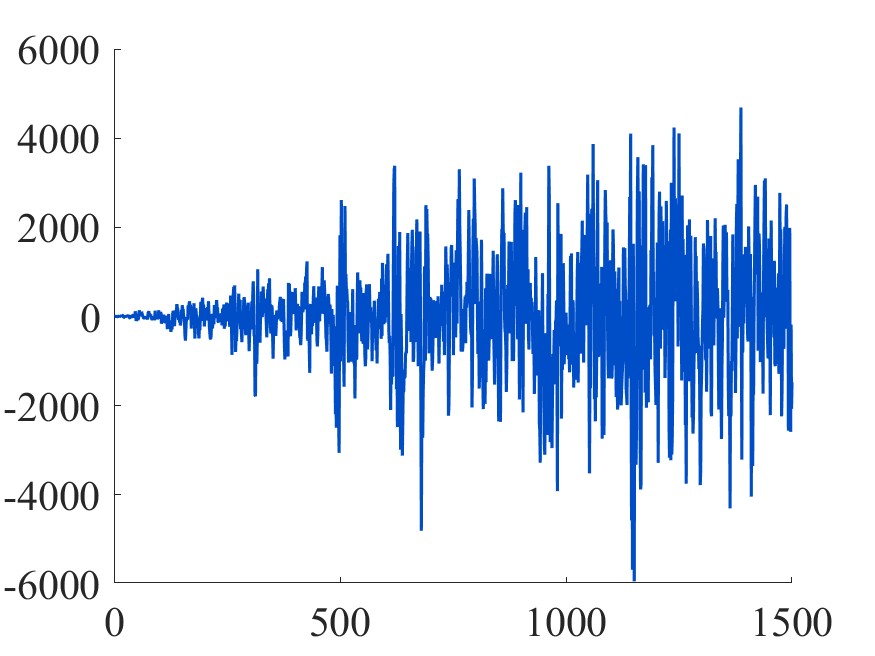}
 \subcaption{$y_t$ $(\gamma=-1/4)$}
	  \end{minipage}
	  \begin{minipage}[t]{0.42\linewidth}
    \centering
\includegraphics[width=\textwidth]{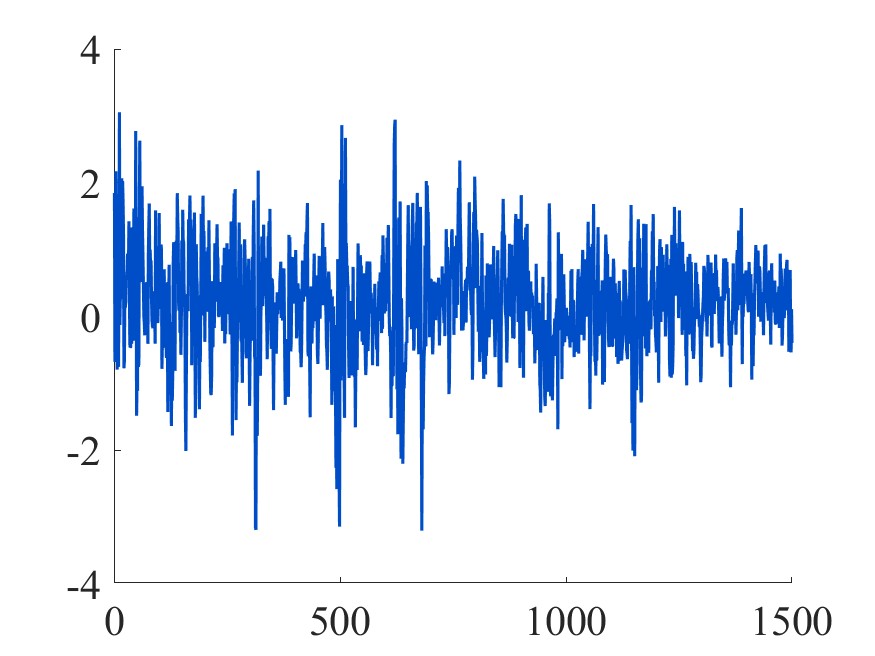}
 \subcaption{$z_{3t}$ $(\gamma=-1/4)$}
	  \end{minipage}\\
	  \begin{minipage}[t]{0.42\linewidth}
    \centering
\includegraphics[width=\textwidth]{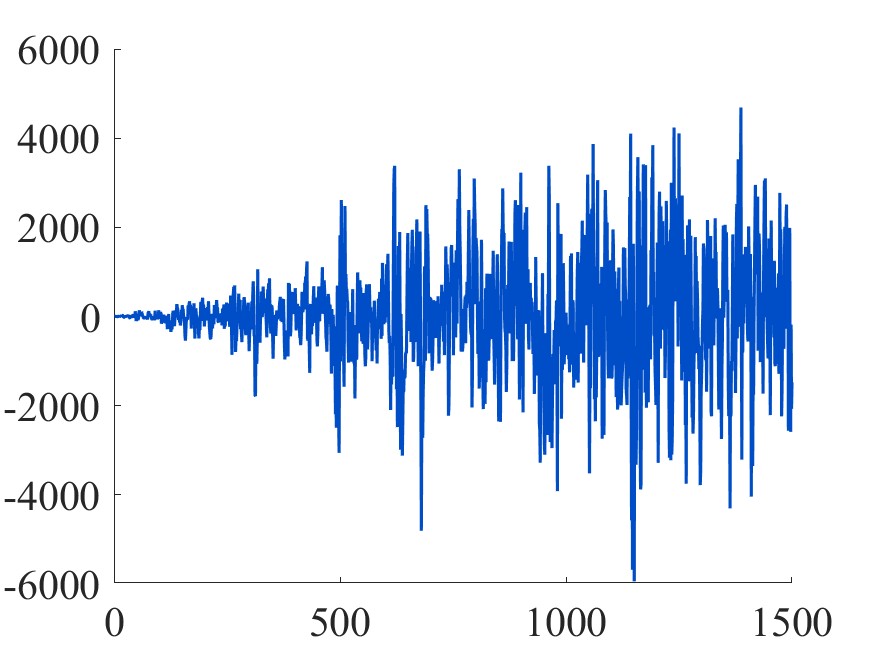}
 \subcaption{$y_{t}$ $(\gamma=-1/2)$}
	  \end{minipage}
	  \begin{minipage}[t]{0.42\linewidth}
    \centering
\includegraphics[width=\textwidth]{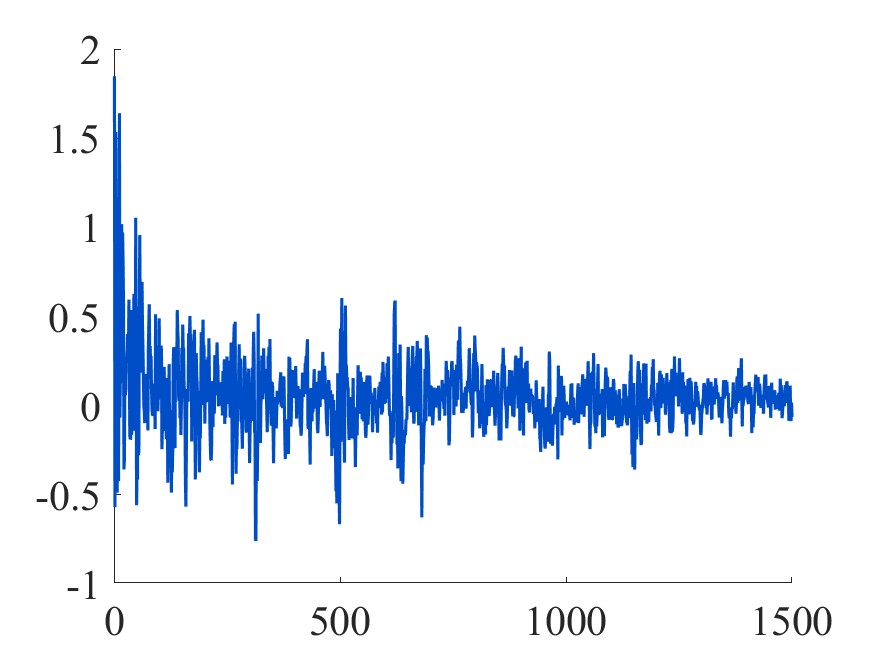}
 \subcaption{$z_{3t}$ $(\gamma=-1/2)$}
	  \end{minipage}\\
\caption{Plots of $y_t$, $z_{3t}$ of a  single sample of  the  model  (\ref{SMC:Model6}) for  $\gamma=1/2,0,-1/4,-1/2$ .}
\label{SDTAYZ}
\end{figure}

\vskip.2cm \noindent Figure \ref{SDTAYZ} displays plots of a single  sample of the variables $y_t$ and $z_{3t}$ for $t = 1, \dots, 1500$ generated by Model (\ref{SMC:Model6}) for  $\gamma=1/2,0,-1/4,-1/2$.
These samples exhibit clear patterns of non-stationary behavior.

\newpage

\end{document}